\documentclass[aps,prx,twocolumn,unsortedaddress,floatfix,superscriptaddress]{revtex4-2}
\usepackage{amsthm}
\usepackage{amsfonts}
\usepackage{siunitx}
\usepackage{amsmath}
\usepackage{amssymb}
\usepackage{graphicx}
\usepackage{verbatim}
\usepackage[colorlinks]{hyperref}
\usepackage{tikz}
\usepackage{tabularx}
\usepackage{braket}
\usepackage{xcolor}
\usepackage{sansmath}
\usepackage{csquotes}

\makeatletter
\def\@ssect@ltx#1#2#3#4#5#6[#7]#8{%
  \def\H@svsec{\phantomsection}%
  \@tempskipa #5\relax
  \@ifdim{\@tempskipa>\z@}{%
    \begingroup
      \interlinepenalty \@M
      #6{%
       \@ifundefined{@hangfroms@#1}{\@hang@froms}{\csname @hangfroms@#1\endcsname}%
       {\hskip#3\relax\H@svsec}{#8}%
      }%
      \@@par
    \endgroup
    \@ifundefined{#1smark}{\@gobble}{\csname #1smark\endcsname}{#7}%
  }{%
    \def\@svsechd{%
      #6{%
       \@ifundefined{@runin@tos@#1}{\@runin@tos}{\csname @runin@tos@#1\endcsname}%
       {\hskip#3\relax\H@svsec}{#8}%
      }%
      \@ifundefined{#1smark}{\@gobble}{\csname #1smark\endcsname}{#7}%
      \addcontentsline{toc}{#1}{\protect\numberline{}#8}%
    }%
  }%
  \@xsect{#5}%
}%
\makeatother

\definecolor{linkcolor}{RGB}{0,83,166}
\hypersetup{
  colorlinks = true,
  allcolors = {linkcolor}
}

\definecolor{myteal}{RGB}{23,190,187}
\definecolor{myblue}{RGB}{42,125,225}
\definecolor{myorange}{RGB}{243,120,32}
\definecolor{mygray}{RGB}{134,134,134}

\newcommand{\affildw}{D-Wave Systems, Burnaby, British Columbia, Canada}
\newcommand{\affilsfu}{Department of Physics, Simon Fraser University, Burnaby, British Columbia, Canada}

\begin{document}
\newcommand{\mytitle}{Coherent quantum annealing in a programmable 2000-qubit Ising chain}
\title{\mytitle}

\author{Andrew D.~King}
\affiliation{\affildw}
\author{Sei Suzuki}
\affiliation{Department of Liberal Arts, Saitama Medical University, Moroyama, Saitama 350-0495, Japan}
\author{Jack Raymond}
\affiliation{\affildw}
\author{Alex Zucca}
\affiliation{\affildw}
\author{Trevor Lanting}
\affiliation{\affildw}
\author{Fabio Altomare}
\affiliation{\affildw}
\author{Andrew J.~Berkley}
\affiliation{\affildw}
\author{Sara Ejtemaee}
\affiliation{\affildw}
\author{Emile Hoskinson}
\affiliation{\affildw}
\author{Shuiyuan Huang}
\affiliation{\affildw}
\author{Eric Ladizinsky}
\affiliation{\affildw}
\author{Allison MacDonald}
\affiliation{\affildw}
\author{Gaelen Marsden}
\affiliation{\affildw}
\author{Travis Oh}
\affiliation{\affildw}
\author{Gabriel Poulin-Lamarre}
\affiliation{\affildw}
\author{Mauricio Reis}
\affiliation{\affildw}
\author{Chris Rich}
\affiliation{\affildw}
\author{Yuki Sato}
\affiliation{\affildw}
\author{Jed D.~Whittaker}
\affiliation{\affildw}
\author{Jason Yao}
\affiliation{\affildw}
\author{Richard Harris}
\affiliation{\affildw}
\author{Daniel A.~Lidar}
\affiliation{Departments of Electrical and Computer Engineering, Chemistry, Physics \& Astronomy, and Center for Quantum Information Science \& Technology (CQIST), University of Southern California, Los Angeles, CA, USA.}
\author{Hidetoshi Nishimori}
\affiliation{Institute of Innovative Research, Tokyo Institute of Technology, Yokohama, Kanagawa 226-8503, Japan}
\affiliation{Graduate School of Information Sciences, Tohoku University, Sendai, Miyagi 980-8579, Japan}
\affiliation{Interdisciplinary Theoretical and Mathematical Sciences, RIKEN, Wako, Saitama 351-0198, Japan}
\author{Mohammad H.~Amin}
\affiliation{\affildw}
\affiliation{\affilsfu}

\date{\today}

\maketitle

\begin{figure}\includegraphics[scale=1]{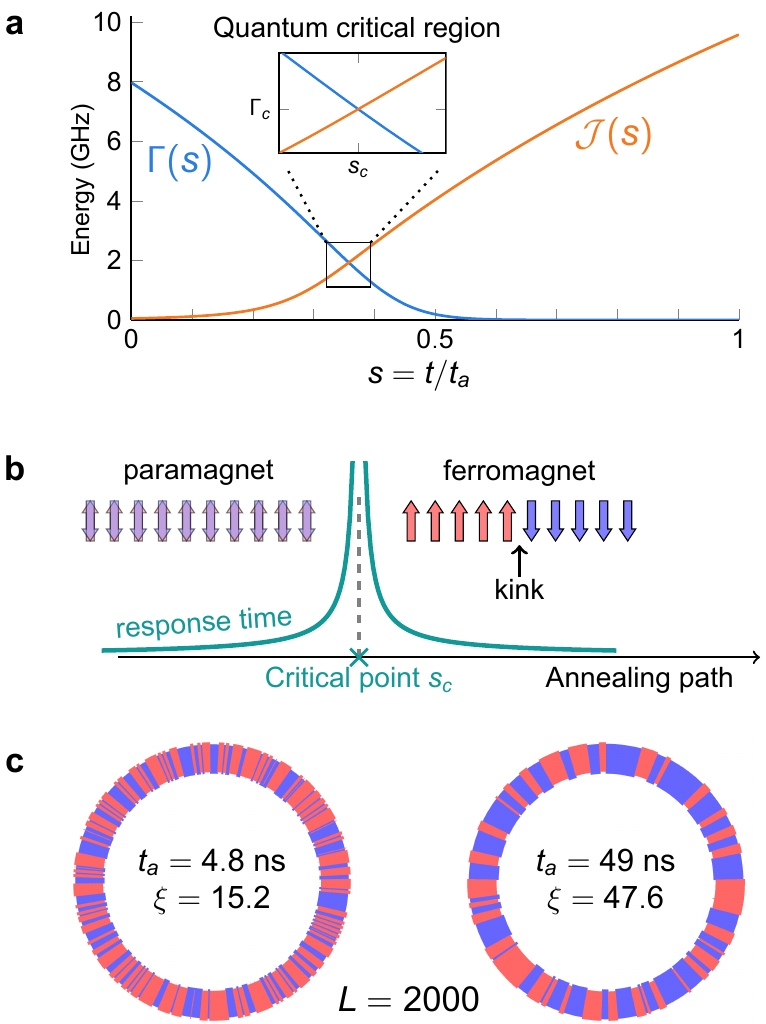}
  \caption{{\bf Quantum phase transition in an annealed Ising chain.}  {\bf a,} Quantum annealing of the transverse-field Ising chain.  Using a dimensionless annealing parameter $s$ to control Hamiltonian (\ref{eq:ham}) with $J=1$, the system is tuned through its QPT at $\Gamma(s_c)=\mathcal J(s_c)$ for $s_c\approx 0.36$.  The critical point separates a quantum paramagnet ($s<s_c$) from an ordered ground state ($s>s_c$). {\bf b,} Response time diverges at the quantum critical point, as a function $\tau \propto |s-s_c|^{-z\nu }$ for critical exponents $z$ and $\nu$.  Consequently, a finite-time traversal of the QPT results in kinks separating ordered domains after annealing. {\bf c}, Example QA output states for a chain of $L=2000$ qubits with $J=-1.4$, whose alternating domains of up (red) and down (blue) spins have correlation length $\xi =1/\bar n$, where $\bar n$ is the average kink density.}\label{fig:1}
\end{figure}

{\bf Quantum simulation has emerged as a valuable arena for demonstrating and understanding the capabilities of near-term quantum computers \cite{Kandala2017, Zhang2017, Keesling2018}.  Quantum annealing \cite{Kadowaki1998,Johnson2011} has been used successfully in simulating a range of open quantum systems, both at equilibrium \cite{Harris2018, King2018, Nishimura2020} and out of equilibrium \cite{Gardas2018, Bando2020, Weinberg2020}.  However, in all previous experiments, annealing has been too slow to simulate a closed quantum system coherently, due to the onset of thermal effects from the environment.  Here we demonstrate coherent evolution through a quantum phase transition in the paradigmatic setting of the 1D transverse-field Ising chain, using up to 2000 superconducting flux qubits in a programmable quantum annealer.  In large systems we observe the quantum Kibble-Zurek mechanism with theoretically predicted kink statistics, as well as characteristic positive kink-kink correlations, independent of system temperature.  In small chains, excitation statistics validate the picture of a Landau-Zener transition at a minimum gap.  In both cases, results are in quantitative agreement with analytical solutions to the closed-system quantum model.  For slower anneals we observe anti-Kibble-Zurek scaling in a crossover to the open quantum regime.  
These experiments demonstrate that large-scale quantum annealers can be operated coherently, paving the way to exploiting coherent dynamics in quantum optimization, machine learning, and simulation tasks.}

Quantum phase transitions (QPTs) describe the sudden macroscopic change of a system's ground state driven by quantum fluctuations \cite{Dutta2015}. An important aspect of phase transitions is the divergence of the correlation length $\xi$ at the critical point, resulting in universal behavior: macroscopic properties become independent of Hamiltonian details. The growth of the correlation length happens within the response time $\tau$, which also diverges at the critical point due to critical slowing down. For a finite system, the correlation length is limited by the system size. Therefore, a slow quench through a QPT, i.e., within a time longer than $\tau$, can transition the system adiabatically into its new ground state \cite{Albash2016}.  Outside the adiabatic regime, the correlation length remains shorter than the system size, leading to defects, i.e., boundaries between domains with different order. The average distance between defects is set by the correlation length, which itself is a function of quench velocity. The defect density scales polynomially with the speed at which the critical point is traversed. This phenomenon, known as the Kibble-Zurek mechanism (KZM) \cite{Zurek1985}, has its origins in early universe cosmology but has since been observed in various experimental platforms such as Bose-Einstein condensates \cite{Anquez2016}, Rydberg atoms \cite{Keesling2018,Browaeys2020}, and trapped ions \cite{Cui2020}.

The quantum Ising chain is a popular testbed for studying the KZM   \cite{Zurek2005,Dziarmaga2005,Cincio2009,Arceci2018, DelCampo2018, Keesling2018} in part because it can be solved exactly using fermionization via the Jordan-Wigner transformation \cite{Dziarmaga2005}.  We implement this model using a programmable superconducting quantum annealer \cite{Johnson2011}.  The Hamiltonian of this system is given by
\begin{equation}\label{eq:ham}
  H(s) = -\Gamma(s)\sum_{i=1}^L \sigma_i^x  + \mathcal J(s) \sum_{i=1}^L J\sigma_i^z\sigma_{i+1}^z,
\end{equation}
where $\sigma_i^z$ and $\sigma_i^x$ are Pauli operators on the $i$th qubit, and $J$ is a dimensionless programmable coupling. For anneal time $t_a$ the annealing parameter $s=t/t_a$ ranges from $0$ to $1$, controlling the transverse field $\Gamma(s)$ and Ising energy scale $\mathcal J(s)$ according to the schedule depicted in Fig.~\ref{fig:1}a  \footnote{For detailed modeling, we determine the schedule terms $\Gamma(s)$ and $\mathcal J(s)$ for each programmed value of $J$, based on a radio-frequency SQUID flux qubit model (see Supplementary Materials).}.  We use periodic boundary conditions ($\sigma^\alpha_{L+1} =\sigma^\alpha_1$) and program all couplers with the same value $J$, which can be either positive (antiferromagnetic) or negative (ferromagnetic).

\begin{figure*}
    \includegraphics[width=\linewidth]{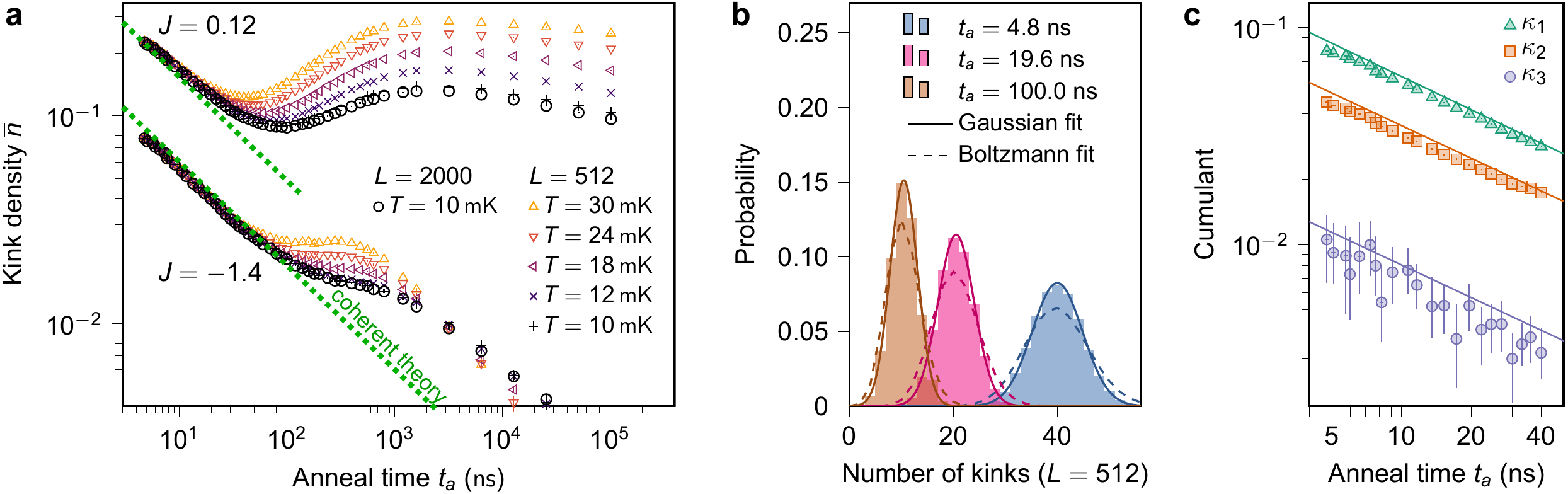}
    \caption{{\bf Kink density scaling and distribution.}  {\bf a}, Shown are QA data for weak coupling $(J=0.12)$ and strong coupling ($J=-1.4$, used for {\bf b--c}) for a range of temperatures and anneal times.  The weak coupling regime exhibits anti-Kibble-Zurek behavior, with a local minimum in $\bar n $.  For strong coupling and fast anneals, $\bar n $ is unaffected by temperature and agrees quantitatively with closed-system coherent quantum theory (dotted green lines, following Eq.~(\ref{eq:bconversion})).  {\bf b}, A best-fit thermal (Boltzmann) model is significantly broader than measurement results, which are better described by a Gaussian model, as expected given the predicted binomial form.  {\bf c}, First three cumulants of the kink distribution. Lines indicate coherent theory.  All error bars represent $95\%$ statistical confidence intervals.}\label{fig:2}
\end{figure*}

In the paramagnetic phase, when $s \approx 0$, the system is dominated by quantum fluctuations and the ground state is an approximately uniform superposition of computational basis states (eigenstates of the $\sigma_i^z$). At the end of the anneal, when $s=1$, the system is diagonal in the computational basis, with frozen dynamics.  This ordered phase has a ferromagnetic ground state; these two phases are separated by a quantum critical point (QCP) at $s=s_c$ such that $\Gamma(s_c)=\mathcal J(s_c)|J|$ (see Fig.~\ref{fig:1}b).

To probe kink density scaling in the thermodynamic limit, we anneal chains of $L=512$ and $L=2000$  qubits for varying $t_a$, at operating temperatures between $\SI{10}{mK}$ and $\SI{30}{mK}$, and for several values of $J$ ranging in magnitude from $0.12$ to $1.4$.  Fig.~\ref{fig:1}c shows examples of experimental data from the quantum annealer (QA) for $t_a=\SI{4.8}{ns}$ and $t_a=\SI{49}{ns}$ with $J=-1.4$.  As expected from the KZM, the longer anneal exhibits fewer kinks.

We define the kink operator  
\begin{equation}
    K_i = \big[1 + \textrm{sign}(J)\sigma^z_i \sigma^z_{i+1} \big]/2. \label{eq:kinkop}
\end{equation}
At the end of the anneal, when all qubits are measured in the computational basis, $K_i = 1$ if there is a kink between qubits $i$ and $i{+}1$, and $K_i = 0$ otherwise. We define the kink density operator as
\begin{equation}
    n = \frac 1{L} \sum_{i=1}^L K_i . \label{eq:kinkdensity}
\end{equation}
The average kink density $\bar n = \langle n \rangle$ is obtained by running the experiment many times and averaging over the outcomes. Measurements of $\bar n$ are summarized in Figure~\ref{fig:2}a.  To test the ability of $L=512$ to represent the thermodynamic limit, we confirmed consistency with $L=2000$ at $\SI{10}{mK}$.  For $t_a\geq \SI{1}{\micro s}$, $\bar n$ decreases monotonically as a function of $t_a$, consistent with previous experiments in the same regime \cite{Bando2020}.  For the previously unexplored region $t_a < \SI{1}{\micro s}$, $\bar n$ is non-monotonic, particularly for high temperature and weak coupling.  This ``anti-Kibble-Zurek'' behavior is a result of coupling to a thermal environment, which generates additional excitations and thus increases $\bar n$; such behavior has been seen in classical simulations of open-system quantum Ising chains \cite{Arceci2018,Bando2020} as well as 2D systems in a quantum annealer outside the coherent regime \cite{Weinberg2020}.

For the shortest anneals, kink densities at all temperatures collapse on a common curve.  This temperature independence is evidence of coherent evolution, wherein the system traverses the QCP faster than the environment's response time.  In this case the system is unable to exchange energy with the environment.  The exactly solvable coherent (closed-system) quantum model predicts \cite{Dziarmaga2005} (see Supplementary Materials (SM))
\begin{equation}\label{eq:bconversion}
  \bar n = \frac{t_a^{-1/2}}{2\pi\sqrt{2b}}, 
  \qquad b = {\Gamma(s_c) /\hbar \over  \mathcal J'(s_c)/\mathcal J(s_c) {-} \Gamma'(s_c)/\Gamma(s_c)}.
\end{equation}
This theoretical kink density (dashed lines in Fig.~\ref{fig:2}a) is in quantitative agreement with the experimental measurements in the fast-anneal regime, with no fitting parameters. 

Kink distributions in the quantum Ising chain have been characterized theoretically beyond just average densities.  The number of kinks follows a binomial distribution \cite{DelCampo2018}, and when the number of kinks is large, this distribution is well approximated by a Gaussian distribution.  This clearly differentiates the data from a Boltzmann distribution describing thermal equilibrium  (see Fig.~\ref{fig:2}b).  Unlike a Gaussian distribution, the binomial kink distribution is expected to skew slightly away from zero, and therefore have a positive third cumulant.  Moreover, the first three cumulants of the kink distribution, $\kappa_1 {=} \bar n$, $\kappa_2 {=} \langle (n{-}\bar n)^2\rangle$, and $\kappa_3 {=} \langle (n{-}\bar n)^3\rangle$, are expected to be proportional to $t_a^{-1/2}$, at fixed ratios \cite{DelCampo2018}
\begin{eqnarray}
  \kappa_2/\kappa_1 &=& 2-\sqrt{2} \approx 0.586,
  \label{eq:k2k1} \\
  \kappa_3/\kappa_1 &=& 4(1-3/\sqrt{2}+2/\sqrt{3}) \approx 0.134.
  \label{eq:k3k1}
\end{eqnarray}
Measurements of these cumulants are shown in Fig.~\ref{fig:2}c.  Lines in the figure are derived from theory, showing good agreement with the experimental data.

Although single-point QA statistics agree with the closed-system quantum model, some aspects of the kink distribution can be reproduced by classical models \cite{Mayo2021}.  For example, the scaling exponent $-1/2$ (Eq.~\eqref{eq:bconversion}) is identical to that of a purely classical diffusion/annihilation model \cite{Krebs1995}.  We therefore investigate two-point statistics \cite{Roychowdhury2021,Nowak2021}.  We define the normalized kink-kink correlator as

\begin{figure}
  \includegraphics[width=\linewidth]{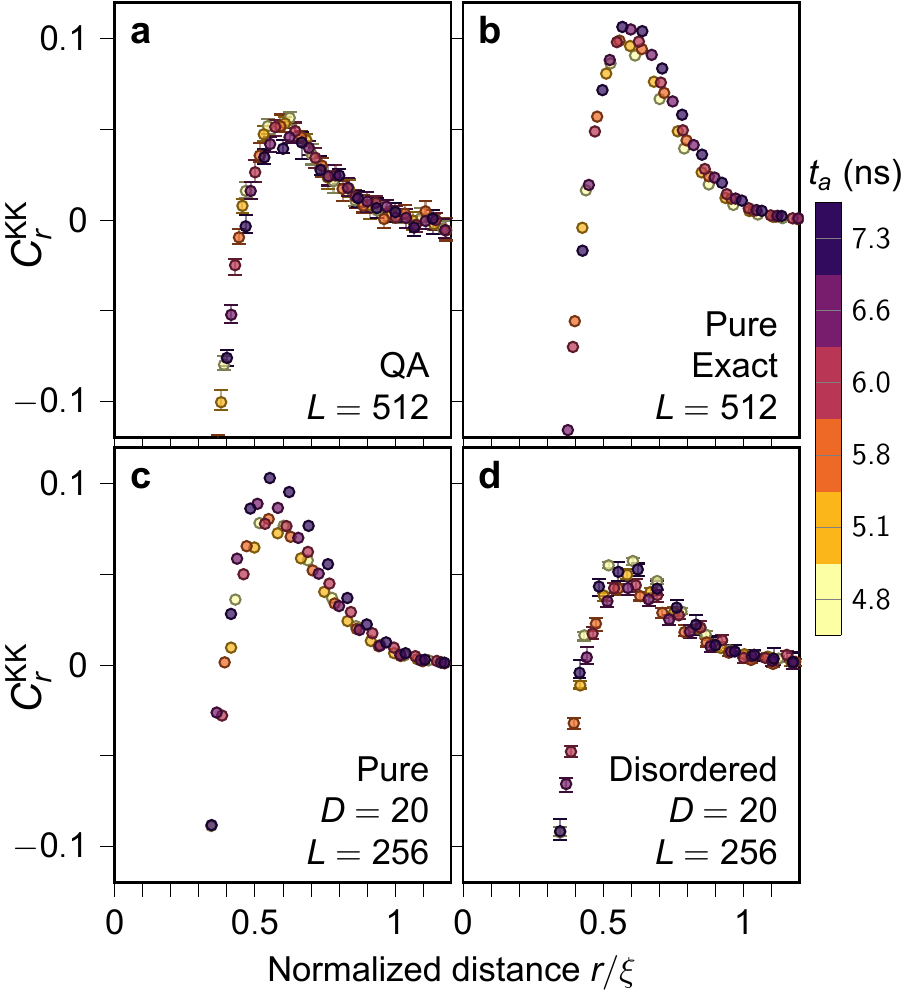}
  \caption{{\bf Normalized kink-kink correlations}.  {\bf a}, QA with $J=-1.4$ (left) has a positive peak in $C^{\textrm{KK}}_r$, which we compare to closed-system quantum models. {\bf b}, Exact time-evolution of the fermionized model.  {\bf c}, TEBD with limited bond dimension $D=20$.  {\bf d}, TEBD with $D=20$ and $\sigma=0.05$ Gaussian disorder added to longitudinal fields and couplings.  All models have $C^{\textrm{KK}}_r \rightarrow -1$ as $r/\xi\rightarrow 0$.  Error bars in {\bf a} and {\bf d} indicate 95\% statistical confidence intervals across experiments and disorder realizations, respectively.}\label{fig:3}
\end{figure}

\begin{equation}
  C^{\textrm{KK}}_r =\frac{1}{L}\sum_{i=1}^L\frac{ \langle K_i K_{i+r} \rangle - {\bar n^2}}{\bar n^2}.
  \label{eq:correlator}
\end{equation}
In Fig.~\ref{fig:3}a we plot $C_r^{\textrm{KK}}$ against the normalized lattice distance $r/\xi= \bar n r$. For multiple annealing times, the data collapse on a curve with a positive peak around $r/\xi \approx 0.6$, as predicted in \cite{Nowak2021}.  QA data are compared against the solution of the fermionized model (Fig.~\ref{fig:3}b), which exhibits a similar but higher peak.

The suppression of the peak in QA is expected from coarsening dynamics \cite{Roychowdhury2021} or other mechanisms such as dephasing \cite{Nowak2021} or kink diffusion outside the regime of validity of the adiabatic/impulse description of KZM.  Indeed, $C^{\textrm{KK}}_r$ does become purely negative for longer anneals (SM, Fig.~\ref{fig:kkc_longer}).  However, thermal effects do not appear to play a role (SM, Fig.~\ref{fig:kkc_backmatter}).  To probe potential effects of entanglement and disorder, we employ a tensor-network dynamics method known as time-evolution block decimation (TEBD) \cite{Oshiyama2020}.  Reducing TEBD bond dimension $D$ to $20$ provides a heuristic model of limited entanglement entropy $S$, given that $S\leq 2\log(D)$~\cite{Schuch:2008ve}; this lowers the peak slightly (Fig.~\ref{fig:3}c), but makes it $t_a$-dependent, inconsistent with the experimental data.  Further lowering $D$ worsens the agreement with QA (see Fig.~\ref{fig:tebd_disorder_varybd}), but combining $D=20$ with disorder in the QA Hamiltonian improves it (see SM).  Combining these effects gives a close match to QA results for $J=-1.4$ (Fig.~\ref{fig:3}d) and other coupling strengths (cf.~Fig.~\ref{fig:kkc_tebd_backmatter}).  Moreover, we find that $D=20$ is a lower bound on the bond dimension, in the sense that our QA data displays an  opposite trend with the anneal time $t_a$ (for short $t_a$) to that of TEBD for $D < 20$, but our QA data and TEBD agree for $D\ge 20$ (cf.~Fig.~\ref{fig:kkc_2rates}).

\begin{figure*}
  \includegraphics[width=0.9\linewidth]{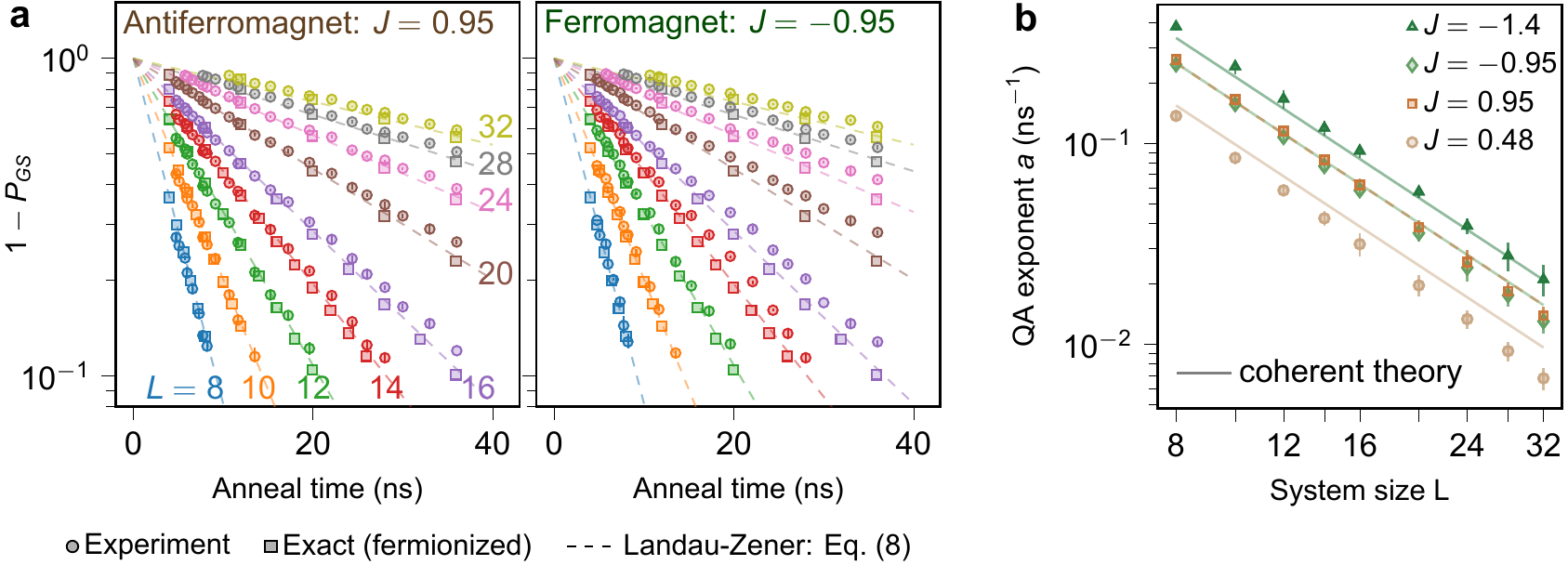}
  \caption{{\bf Crossover to adiabaticity.} {\bf a}, QA data (circles) for antiferromagnetic (left) and ferromagnetic (right) chains are compared against exact closed-system numerics (squares), using an independently extracted annealing schedule based on single-qubit measurements.  Both agree closely with a Landau-Zener model of diabatic transition occurring at a minimum gap (Eq.~(\ref{eq:a}), dashed lines).  {\bf b}, Exponent $a$ extracted from QA data in {\bf a} for varying $J$ and $L$.  Solid lines are analytical values from the closed-system model.}\label{fig:4}
\end{figure*}

Next, we investigate finite size effects.  When $t_a$ is sufficiently large as a function of $L$, the dynamics are dominated by a single Landau-Zener (LZ) transition \cite{Zener1932}, and the ground state probability $P_{GS}$  follows the adiabatic theorem \cite{Kato:50}.  The LZ transition probability is expected to decay exponentially in the annealing time, in contrast to the power-law dependence in the Kibble-Zurek regime. For 1D spin chains, it is possible obtain an analytical solution \cite{Dziarmaga2005} (see SM):
\begin{equation} \label{eq:a}
1-P_{GS} = e^{-at_a}, \qquad a = 2\pi^3b L^{-2}, 
\end{equation}
where $b$ is defined in Eq.~\eqref{eq:bconversion}.

Fig.~\ref{fig:4}a shows QA measurements for ferromagnetic and antiferromagnetic chains of equal coupling magnitude ($J=\pm 0.95$).  Since $L$ is even, the two Ising Hamiltonians are gauge-equivalent and we expect similar experimental outcomes.  We plot data in the range $\SI{5}{ns} \leq t_a \leq \SI{40}{ns}$ and $0.1 \leq P_{GS} \leq 0.9$ for values of $L$ ranging from $8$ to $32$. Fig.~\ref{fig:4}a also shows results of exact simulation of coherent Schr\"odinger dynamics for the fermionized system (squares; see SM) together with the analytical result of Eq.~\eqref{eq:a} (dashed lines), in remarkable agreement with the experimental data.
To test the agreement with closed-system theory for different $J$ values, Fig.~\ref{fig:4}b shows that $a$ as extracted from the empirical $P_{GS}$ data as per Eq.~\eqref{eq:a} remains consistent with the theoretical prediction $a\propto L^{-2}$ (solid lines).   

Although all of the above experimental results agree well with coherent quantum dynamics, an important question is whether they can also be explained by classical models. It is clearly impractical and even impossible to rule out every classical explanation; instead we consider the most plausible Monte Carlo methods that have been suggested as emulators for QA. In Appendix \ref{app:mc}, we consider simulated annealing, simulated quantum annealing based on path-integral Monte Carlo, and spin-vector Monte Carlo simulations.  We find that some of these models can reproduce some  aspects of the experimental data, but none of them can explain all experimental features.  We therefore conclude that only the coherent quantum model successfully explains all experimental results, and this view is strengthened considerably by the fact that we have not used any fitting parameters.

In conclusion, by tuning the parameters of a programmable quantum annealer, we have simulated quantum critical phenomena in 1D chains of up to 2000 spins. For fast anneals, we observe quantum Kibble-Zurek scaling in long chains and Landau-Zener scaling in short chains. In both regimes, kink densities are in quantitative agreement with coherent Schr\"odinger dynamics---remarkably, with no free parameters. In contrast, leading classical models can only reproduce some aspects of the experimental data---no single classical theory reproduces them all.  These results represent strong evidence for coherent evolution, with a significantly larger system and longer correlation length than previous quantum Kibble-Zurek demonstrations in a 1D system using Rydberg arrays \cite{Keesling2018}.  In addition, at longer anneal times we observe a crossover to the thermal regime, with anti-Kibble-Zurek behavior as theoretically predicted~\cite{Bando2020}.

We have used QA as a quantum simulator, producing results that are challenging to simulate classically, even in this widely-studied and simple model.  Path-integral Monte Carlo can simulate systems near thermal equilibrium \cite{Suzuki1976}, but cannot be used to describe or simulate quantum dynamics \cite{Liu2015, King2021, Bando2021}.  Likewise, open-system quantum simulations such as master equations \cite{Yip:2018aa} become computationally intractable beyond system sizes of around $40$ qubits.  Thus, our results pave the way to coherent quantum simulation on a previously unattainable scale.  Moreover, the ability to program both signs and magnitudes of Hamiltonian terms in a coherently evolved system is a key ingredient in the simulation of frustrated models such as quantum spin glasses, and ultimately in quantum optimization.  The results reported here represent an important step toward this goal.

\section*{Acknowledgments}

We thank H.~Oshiyama, N.~Shibata, A.~del Campo, L.~Addario-Berry, T.~Albash, and A.~W.~Sandvik for fruitful discussions, and acknowledge the contributions of both technical and non-technical staff at D-Wave. The authors acknowledge the Center for Advanced Research
Computing (CARC) at the University of Southern California for providing computing resources that have contributed to the research results reported within this publication. URL: \url{https://carc.usc.edu}. DAL acknowledges support by the National Science Foundation ``the Quantum Leap Big Idea'' under Grant No.~OMA-1936388, and by DARPA under the RQMLS program, Agreement No. HR00112190071.

\let\oldaddcontentsline\addcontentsline
\renewcommand{\addcontentsline}[3]{}
\bibliography{paper_kzm}

\begin{thebibliography}{62}%
\makeatletter
\providecommand \@ifxundefined [1]{%
 \@ifx{#1\undefined}
}%
\providecommand \@ifnum [1]{%
 \ifnum #1\expandafter \@firstoftwo
 \else \expandafter \@secondoftwo
 \fi
}%
\providecommand \@ifx [1]{%
 \ifx #1\expandafter \@firstoftwo
 \else \expandafter \@secondoftwo
 \fi
}%
\providecommand \natexlab [1]{#1}%
\providecommand \enquote  [1]{``#1''}%
\providecommand \bibnamefont  [1]{#1}%
\providecommand \bibfnamefont [1]{#1}%
\providecommand \citenamefont [1]{#1}%
\providecommand \href@noop [0]{\@secondoftwo}%
\providecommand \href [0]{\begingroup \@sanitize@url \@href}%
\providecommand \@href[1]{\@@startlink{#1}\@@href}%
\providecommand \@@href[1]{\endgroup#1\@@endlink}%
\providecommand \@sanitize@url [0]{\catcode `\\12\catcode `\$12\catcode
  `\&12\catcode `\#12\catcode `\^12\catcode `\_12\catcode `\%12\relax}%
\providecommand \@@startlink[1]{}%
\providecommand \@@endlink[0]{}%
\providecommand \url  [0]{\begingroup\@sanitize@url \@url }%
\providecommand \@url [1]{\endgroup\@href {#1}{\urlprefix }}%
\providecommand \urlprefix  [0]{URL }%
\providecommand \Eprint [0]{\href }%
\providecommand \doibase [0]{https://doi.org/}%
\providecommand \selectlanguage [0]{\@gobble}%
\providecommand \bibinfo  [0]{\@secondoftwo}%
\providecommand \bibfield  [0]{\@secondoftwo}%
\providecommand \translation [1]{[#1]}%
\providecommand \BibitemOpen [0]{}%
\providecommand \bibitemStop [0]{}%
\providecommand \bibitemNoStop [0]{.\EOS\space}%
\providecommand \EOS [0]{\spacefactor3000\relax}%
\providecommand \BibitemShut  [1]{\csname bibitem#1\endcsname}%
\let\auto@bib@innerbib\@empty
\bibitem [{\citenamefont {Kandala}\ \emph {et~al.}(2017)\citenamefont
  {Kandala}, \citenamefont {Mezzacapo}, \citenamefont {Temme}, \citenamefont
  {Takita}, \citenamefont {Brink}, \citenamefont {Chow},\ and\ \citenamefont
  {Gambetta}}]{Kandala2017}%
  \BibitemOpen
  \bibfield  {author} {\bibinfo {author} {\bibfnamefont {A.}~\bibnamefont
  {Kandala}}, \bibinfo {author} {\bibfnamefont {A.}~\bibnamefont {Mezzacapo}},
  \bibinfo {author} {\bibfnamefont {K.}~\bibnamefont {Temme}}, \bibinfo
  {author} {\bibfnamefont {M.}~\bibnamefont {Takita}}, \bibinfo {author}
  {\bibfnamefont {M.}~\bibnamefont {Brink}}, \bibinfo {author} {\bibfnamefont
  {J.~M.}\ \bibnamefont {Chow}},\ and\ \bibinfo {author} {\bibfnamefont
  {J.~M.}\ \bibnamefont {Gambetta}},\ }\bibfield  {title} {\bibinfo {title}
  {{Hardware-efficient variational quantum eigensolver for small molecules and
  quantum magnets}},\ }\href {https://doi.org/10.1038/nature23879} {\bibfield
  {journal} {\bibinfo  {journal} {Nature}\ }\textbf {\bibinfo {volume} {549}},\
  \bibinfo {pages} {242} (\bibinfo {year} {2017})}\BibitemShut {NoStop}%
\bibitem [{\citenamefont {Zhang}\ \emph {et~al.}(2017)\citenamefont {Zhang},
  \citenamefont {Pagano}, \citenamefont {Hess}, \citenamefont {Kyprianidis},
  \citenamefont {Becker}, \citenamefont {Kaplan}, \citenamefont {Gorshkov},
  \citenamefont {Gong},\ and\ \citenamefont {Monroe}}]{Zhang2017}%
  \BibitemOpen
  \bibfield  {author} {\bibinfo {author} {\bibfnamefont {J.}~\bibnamefont
  {Zhang}}, \bibinfo {author} {\bibfnamefont {G.}~\bibnamefont {Pagano}},
  \bibinfo {author} {\bibfnamefont {P.~W.}\ \bibnamefont {Hess}}, \bibinfo
  {author} {\bibfnamefont {A.}~\bibnamefont {Kyprianidis}}, \bibinfo {author}
  {\bibfnamefont {P.}~\bibnamefont {Becker}}, \bibinfo {author} {\bibfnamefont
  {H.}~\bibnamefont {Kaplan}}, \bibinfo {author} {\bibfnamefont {A.~V.}\
  \bibnamefont {Gorshkov}}, \bibinfo {author} {\bibfnamefont {Z.}~\bibnamefont
  {Gong}},\ and\ \bibinfo {author} {\bibfnamefont {C.}~\bibnamefont {Monroe}},\
  }\bibfield  {title} {\bibinfo {title} {{Observation of a many-body dynamical
  phase transition with a 53-qubit quantum simulator}},\ }\href
  {https://doi.org/10.1038/nature24654} {\bibfield  {journal} {\bibinfo
  {journal} {Nature}\ }\textbf {\bibinfo {volume} {551}},\ \bibinfo {pages}
  {601} (\bibinfo {year} {2017})}\BibitemShut {NoStop}%
\bibitem [{\citenamefont {Keesling}\ \emph {et~al.}(2019)\citenamefont
  {Keesling}, \citenamefont {Omran}, \citenamefont {Levine}, \citenamefont
  {Bernien}, \citenamefont {Pichler}, \citenamefont {Choi}, \citenamefont
  {Samajdar}, \citenamefont {Schwartz}, \citenamefont {Silvi}, \citenamefont
  {Sachdev}, \citenamefont {Zoller}, \citenamefont {Endres}, \citenamefont
  {Greiner}, \citenamefont {Vuleti{\'{c}}},\ and\ \citenamefont
  {Lukin}}]{Keesling2018}%
  \BibitemOpen
  \bibfield  {author} {\bibinfo {author} {\bibfnamefont {A.}~\bibnamefont
  {Keesling}}, \bibinfo {author} {\bibfnamefont {A.}~\bibnamefont {Omran}},
  \bibinfo {author} {\bibfnamefont {H.}~\bibnamefont {Levine}}, \bibinfo
  {author} {\bibfnamefont {H.}~\bibnamefont {Bernien}}, \bibinfo {author}
  {\bibfnamefont {H.}~\bibnamefont {Pichler}}, \bibinfo {author} {\bibfnamefont
  {S.}~\bibnamefont {Choi}}, \bibinfo {author} {\bibfnamefont {R.}~\bibnamefont
  {Samajdar}}, \bibinfo {author} {\bibfnamefont {S.}~\bibnamefont {Schwartz}},
  \bibinfo {author} {\bibfnamefont {P.}~\bibnamefont {Silvi}}, \bibinfo
  {author} {\bibfnamefont {S.}~\bibnamefont {Sachdev}}, \bibinfo {author}
  {\bibfnamefont {P.}~\bibnamefont {Zoller}}, \bibinfo {author} {\bibfnamefont
  {M.}~\bibnamefont {Endres}}, \bibinfo {author} {\bibfnamefont
  {M.}~\bibnamefont {Greiner}}, \bibinfo {author} {\bibfnamefont
  {V.}~\bibnamefont {Vuleti{\'{c}}}},\ and\ \bibinfo {author} {\bibfnamefont
  {M.~D.}\ \bibnamefont {Lukin}},\ }\bibfield  {title} {\bibinfo {title}
  {{Quantum Kibble–Zurek mechanism and critical dynamics on a programmable
  Rydberg simulator}},\ }\href {https://doi.org/10.1038/s41586-019-1070-1}
  {\bibfield  {journal} {\bibinfo  {journal} {Nature}\ }\textbf {\bibinfo
  {volume} {568}},\ \bibinfo {pages} {207} (\bibinfo {year}
  {2019})}\BibitemShut {NoStop}%
\bibitem [{\citenamefont {Kadowaki}\ and\ \citenamefont
  {Nishimori}(1998)}]{Kadowaki1998}%
  \BibitemOpen
  \bibfield  {author} {\bibinfo {author} {\bibfnamefont {T.}~\bibnamefont
  {Kadowaki}}\ and\ \bibinfo {author} {\bibfnamefont {H.}~\bibnamefont
  {Nishimori}},\ }\bibfield  {title} {\bibinfo {title} {{Quantum annealing in
  the transverse Ising model}},\ }\href
  {https://doi.org/10.1103/PhysRevE.58.5355} {\bibfield  {journal} {\bibinfo
  {journal} {Physical Review E}\ }\textbf {\bibinfo {volume} {58}},\ \bibinfo
  {pages} {5355} (\bibinfo {year} {1998})}\BibitemShut {NoStop}%
\bibitem [{\citenamefont {Johnson}\ \emph {et~al.}(2011)\citenamefont
  {Johnson}, \citenamefont {Amin}, \citenamefont {Gildert}, \citenamefont
  {Lanting}, \citenamefont {Hamze}, \citenamefont {Dickson}, \citenamefont
  {Harris}, \citenamefont {Berkley}, \citenamefont {Johansson}, \citenamefont
  {Bunyk}, \citenamefont {Chapple}, \citenamefont {Enderud}, \citenamefont
  {Hilton}, \citenamefont {Karimi}, \citenamefont {Ladizinsky}, \citenamefont
  {Ladizinsky}, \citenamefont {Oh}, \citenamefont {Perminov}, \citenamefont
  {Rich}, \citenamefont {Thom}, \citenamefont {Tolkacheva}, \citenamefont
  {Truncik}, \citenamefont {Uchaikin}, \citenamefont {Wang}, \citenamefont
  {Wilson},\ and\ \citenamefont {Rose}}]{Johnson2011}%
  \BibitemOpen
  \bibfield  {author} {\bibinfo {author} {\bibfnamefont {M.~W.}\ \bibnamefont
  {Johnson}}, \bibinfo {author} {\bibfnamefont {M.~H.}\ \bibnamefont {Amin}},
  \bibinfo {author} {\bibfnamefont {S.}~\bibnamefont {Gildert}}, \bibinfo
  {author} {\bibfnamefont {T.}~\bibnamefont {Lanting}}, \bibinfo {author}
  {\bibfnamefont {F.}~\bibnamefont {Hamze}}, \bibinfo {author} {\bibfnamefont
  {N.~G.}\ \bibnamefont {Dickson}}, \bibinfo {author} {\bibfnamefont
  {R.}~\bibnamefont {Harris}}, \bibinfo {author} {\bibfnamefont {A.~J.}\
  \bibnamefont {Berkley}}, \bibinfo {author} {\bibfnamefont {J.}~\bibnamefont
  {Johansson}}, \bibinfo {author} {\bibfnamefont {P.~I.}\ \bibnamefont
  {Bunyk}}, \bibinfo {author} {\bibfnamefont {E.~M.}\ \bibnamefont {Chapple}},
  \bibinfo {author} {\bibfnamefont {C.}~\bibnamefont {Enderud}}, \bibinfo
  {author} {\bibfnamefont {J.~P.}\ \bibnamefont {Hilton}}, \bibinfo {author}
  {\bibfnamefont {K.}~\bibnamefont {Karimi}}, \bibinfo {author} {\bibfnamefont
  {E.}~\bibnamefont {Ladizinsky}}, \bibinfo {author} {\bibfnamefont
  {N.}~\bibnamefont {Ladizinsky}}, \bibinfo {author} {\bibfnamefont
  {T.}~\bibnamefont {Oh}}, \bibinfo {author} {\bibfnamefont {I.}~\bibnamefont
  {Perminov}}, \bibinfo {author} {\bibfnamefont {C.}~\bibnamefont {Rich}},
  \bibinfo {author} {\bibfnamefont {M.~C.}\ \bibnamefont {Thom}}, \bibinfo
  {author} {\bibfnamefont {E.}~\bibnamefont {Tolkacheva}}, \bibinfo {author}
  {\bibfnamefont {C.~J.~S.}\ \bibnamefont {Truncik}}, \bibinfo {author}
  {\bibfnamefont {S.}~\bibnamefont {Uchaikin}}, \bibinfo {author}
  {\bibfnamefont {J.}~\bibnamefont {Wang}}, \bibinfo {author} {\bibfnamefont
  {B.}~\bibnamefont {Wilson}},\ and\ \bibinfo {author} {\bibfnamefont
  {G.}~\bibnamefont {Rose}},\ }\bibfield  {title} {\bibinfo {title} {{Quantum
  annealing with manufactured spins}},\ }\href
  {https://doi.org/10.1038/nature10012} {\bibfield  {journal} {\bibinfo
  {journal} {Nature}\ }\textbf {\bibinfo {volume} {473}},\ \bibinfo {pages}
  {194} (\bibinfo {year} {2011})}\BibitemShut {NoStop}%
\bibitem [{\citenamefont {Harris}\ \emph {et~al.}(2018)\citenamefont {Harris},
  \citenamefont {Sato}, \citenamefont {Berkley}, \citenamefont {Reis},
  \citenamefont {Altomare}, \citenamefont {Amin}, \citenamefont {Boothby},
  \citenamefont {Bunyk}, \citenamefont {Deng}, \citenamefont {Enderud},
  \citenamefont {Huang}, \citenamefont {Hoskinson}, \citenamefont {Johnson},
  \citenamefont {Ladizinsky}, \citenamefont {Ladizinsky}, \citenamefont
  {Lanting}, \citenamefont {Li}, \citenamefont {Medina}, \citenamefont
  {Molavi}, \citenamefont {Neufeld}, \citenamefont {Oh}, \citenamefont
  {Pavlov}, \citenamefont {Perminov}, \citenamefont {Poulin-Lamarre},
  \citenamefont {Rich}, \citenamefont {Smirnov}, \citenamefont {Swenson},
  \citenamefont {Tsai}, \citenamefont {Volkmann}, \citenamefont {Whittaker},\
  and\ \citenamefont {Yao}}]{Harris2018}%
  \BibitemOpen
  \bibfield  {author} {\bibinfo {author} {\bibfnamefont {R.}~\bibnamefont
  {Harris}}, \bibinfo {author} {\bibfnamefont {Y.}~\bibnamefont {Sato}},
  \bibinfo {author} {\bibfnamefont {A.~J.}\ \bibnamefont {Berkley}}, \bibinfo
  {author} {\bibfnamefont {M.}~\bibnamefont {Reis}}, \bibinfo {author}
  {\bibfnamefont {F.}~\bibnamefont {Altomare}}, \bibinfo {author}
  {\bibfnamefont {M.~H.}\ \bibnamefont {Amin}}, \bibinfo {author}
  {\bibfnamefont {K.}~\bibnamefont {Boothby}}, \bibinfo {author} {\bibfnamefont
  {P.}~\bibnamefont {Bunyk}}, \bibinfo {author} {\bibfnamefont
  {C.}~\bibnamefont {Deng}}, \bibinfo {author} {\bibfnamefont {C.}~\bibnamefont
  {Enderud}}, \bibinfo {author} {\bibfnamefont {S.}~\bibnamefont {Huang}},
  \bibinfo {author} {\bibfnamefont {E.}~\bibnamefont {Hoskinson}}, \bibinfo
  {author} {\bibfnamefont {M.~W.}\ \bibnamefont {Johnson}}, \bibinfo {author}
  {\bibfnamefont {E.}~\bibnamefont {Ladizinsky}}, \bibinfo {author}
  {\bibfnamefont {N.}~\bibnamefont {Ladizinsky}}, \bibinfo {author}
  {\bibfnamefont {T.}~\bibnamefont {Lanting}}, \bibinfo {author} {\bibfnamefont
  {R.}~\bibnamefont {Li}}, \bibinfo {author} {\bibfnamefont {T.}~\bibnamefont
  {Medina}}, \bibinfo {author} {\bibfnamefont {R.}~\bibnamefont {Molavi}},
  \bibinfo {author} {\bibfnamefont {R.}~\bibnamefont {Neufeld}}, \bibinfo
  {author} {\bibfnamefont {T.}~\bibnamefont {Oh}}, \bibinfo {author}
  {\bibfnamefont {I.}~\bibnamefont {Pavlov}}, \bibinfo {author} {\bibfnamefont
  {I.}~\bibnamefont {Perminov}}, \bibinfo {author} {\bibfnamefont
  {G.}~\bibnamefont {Poulin-Lamarre}}, \bibinfo {author} {\bibfnamefont
  {C.}~\bibnamefont {Rich}}, \bibinfo {author} {\bibfnamefont {A.}~\bibnamefont
  {Smirnov}}, \bibinfo {author} {\bibfnamefont {L.}~\bibnamefont {Swenson}},
  \bibinfo {author} {\bibfnamefont {N.}~\bibnamefont {Tsai}}, \bibinfo {author}
  {\bibfnamefont {M.}~\bibnamefont {Volkmann}}, \bibinfo {author}
  {\bibfnamefont {J.}~\bibnamefont {Whittaker}},\ and\ \bibinfo {author}
  {\bibfnamefont {J.}~\bibnamefont {Yao}},\ }\bibfield  {title} {\bibinfo
  {title} {{Phase transitions in a programmable quantum spin glass
  simulator}},\ }\href {https://doi.org/10.1126/science.aat2025} {\bibfield
  {journal} {\bibinfo  {journal} {Science}\ }\textbf {\bibinfo {volume}
  {361}},\ \bibinfo {pages} {162} (\bibinfo {year} {2018})}\BibitemShut
  {NoStop}%
\bibitem [{\citenamefont {King}\ \emph {et~al.}(2018)\citenamefont {King},
  \citenamefont {Carrasquilla}, \citenamefont {Raymond}, \citenamefont
  {Ozfidan}, \citenamefont {Andriyash}, \citenamefont {Berkley}, \citenamefont
  {Reis}, \citenamefont {Lanting}, \citenamefont {Harris}, \citenamefont
  {Altomare}, \citenamefont {Boothby}, \citenamefont {Bunyk}, \citenamefont
  {Enderud}, \citenamefont {Fr{\'{e}}chette}, \citenamefont {Hoskinson},
  \citenamefont {Ladizinsky}, \citenamefont {Oh}, \citenamefont
  {Poulin-Lamarre}, \citenamefont {Rich}, \citenamefont {Sato}, \citenamefont
  {Smirnov}, \citenamefont {Swenson}, \citenamefont {Volkmann}, \citenamefont
  {Whittaker}, \citenamefont {Yao}, \citenamefont {Ladizinsky}, \citenamefont
  {Mark}, \citenamefont {Hilton},\ and\ \citenamefont {Amin}}]{King2018}%
  \BibitemOpen
  \bibfield  {author} {\bibinfo {author} {\bibfnamefont {A.~D.}\ \bibnamefont
  {King}}, \bibinfo {author} {\bibfnamefont {J.}~\bibnamefont {Carrasquilla}},
  \bibinfo {author} {\bibfnamefont {J.}~\bibnamefont {Raymond}}, \bibinfo
  {author} {\bibfnamefont {I.}~\bibnamefont {Ozfidan}}, \bibinfo {author}
  {\bibfnamefont {E.}~\bibnamefont {Andriyash}}, \bibinfo {author}
  {\bibfnamefont {A.~J.}\ \bibnamefont {Berkley}}, \bibinfo {author}
  {\bibfnamefont {M.}~\bibnamefont {Reis}}, \bibinfo {author} {\bibfnamefont
  {T.}~\bibnamefont {Lanting}}, \bibinfo {author} {\bibfnamefont
  {R.}~\bibnamefont {Harris}}, \bibinfo {author} {\bibfnamefont
  {F.}~\bibnamefont {Altomare}}, \bibinfo {author} {\bibfnamefont
  {K.}~\bibnamefont {Boothby}}, \bibinfo {author} {\bibfnamefont {P.~I.}\
  \bibnamefont {Bunyk}}, \bibinfo {author} {\bibfnamefont {C.}~\bibnamefont
  {Enderud}}, \bibinfo {author} {\bibfnamefont {A.}~\bibnamefont
  {Fr{\'{e}}chette}}, \bibinfo {author} {\bibfnamefont {E.~M.}\ \bibnamefont
  {Hoskinson}}, \bibinfo {author} {\bibfnamefont {N.}~\bibnamefont
  {Ladizinsky}}, \bibinfo {author} {\bibfnamefont {T.}~\bibnamefont {Oh}},
  \bibinfo {author} {\bibfnamefont {G.}~\bibnamefont {Poulin-Lamarre}},
  \bibinfo {author} {\bibfnamefont {C.}~\bibnamefont {Rich}}, \bibinfo {author}
  {\bibfnamefont {Y.}~\bibnamefont {Sato}}, \bibinfo {author} {\bibfnamefont
  {A.~Y.}\ \bibnamefont {Smirnov}}, \bibinfo {author} {\bibfnamefont {L.~J.}\
  \bibnamefont {Swenson}}, \bibinfo {author} {\bibfnamefont {M.~H.}\
  \bibnamefont {Volkmann}}, \bibinfo {author} {\bibfnamefont {J.}~\bibnamefont
  {Whittaker}}, \bibinfo {author} {\bibfnamefont {J.}~\bibnamefont {Yao}},
  \bibinfo {author} {\bibfnamefont {E.}~\bibnamefont {Ladizinsky}}, \bibinfo
  {author} {\bibfnamefont {W.}~\bibnamefont {Mark}}, \bibinfo {author}
  {\bibfnamefont {J.~P.}\ \bibnamefont {Hilton}},\ and\ \bibinfo {author}
  {\bibfnamefont {M.~H.}\ \bibnamefont {Amin}},\ }\bibfield  {title} {\bibinfo
  {title} {{Observation of topological phenomena in a programmable lattice of
  1,800 qubits}},\ }\href {https://doi.org/10.1038/s41586-018-0410-x}
  {\bibfield  {journal} {\bibinfo  {journal} {Nature}\ }\textbf {\bibinfo
  {volume} {560}},\ \bibinfo {pages} {456} (\bibinfo {year}
  {2018})}\BibitemShut {NoStop}%
\bibitem [{\citenamefont {Nishimura}\ \emph {et~al.}(2020)\citenamefont
  {Nishimura}, \citenamefont {Nishimori},\ and\ \citenamefont
  {Katzgraber}}]{Nishimura2020}%
  \BibitemOpen
  \bibfield  {author} {\bibinfo {author} {\bibfnamefont {K.}~\bibnamefont
  {Nishimura}}, \bibinfo {author} {\bibfnamefont {H.}~\bibnamefont
  {Nishimori}},\ and\ \bibinfo {author} {\bibfnamefont {H.~G.}\ \bibnamefont
  {Katzgraber}},\ }\bibfield  {title} {\bibinfo {title} {{Griffiths-McCoy
  singularity on the diluted Chimera graph: Monte Carlo simulations and
  experiments on quantum hardware}},\ }\href
  {https://doi.org/10.1103/PhysRevA.102.042403} {\bibfield  {journal} {\bibinfo
   {journal} {Physical Review A}\ }\textbf {\bibinfo {volume} {102}},\ \bibinfo
  {pages} {042403} (\bibinfo {year} {2020})}\BibitemShut {NoStop}%
\bibitem [{\citenamefont {Gardas}\ \emph {et~al.}(2018)\citenamefont {Gardas},
  \citenamefont {Dziarmaga}, \citenamefont {Zurek},\ and\ \citenamefont
  {Zwolak}}]{Gardas2018}%
  \BibitemOpen
  \bibfield  {author} {\bibinfo {author} {\bibfnamefont {B.}~\bibnamefont
  {Gardas}}, \bibinfo {author} {\bibfnamefont {J.}~\bibnamefont {Dziarmaga}},
  \bibinfo {author} {\bibfnamefont {W.~H.}\ \bibnamefont {Zurek}},\ and\
  \bibinfo {author} {\bibfnamefont {M.}~\bibnamefont {Zwolak}},\ }\bibfield
  {title} {\bibinfo {title} {{Defects in Quantum Computers}},\ }\href
  {https://doi.org/10.1038/s41598-018-22763-2} {\bibfield  {journal} {\bibinfo
  {journal} {Scientific Reports}\ }\textbf {\bibinfo {volume} {8}},\ \bibinfo
  {pages} {2} (\bibinfo {year} {2018})}\BibitemShut {NoStop}%
\bibitem [{\citenamefont {Bando}\ \emph {et~al.}(2020)\citenamefont {Bando},
  \citenamefont {Susa}, \citenamefont {Oshiyama}, \citenamefont {Shibata},
  \citenamefont {Ohzeki}, \citenamefont {G{\'{o}}mez-Ruiz}, \citenamefont
  {Lidar}, \citenamefont {Suzuki}, \citenamefont {del Campo},\ and\
  \citenamefont {Nishimori}}]{Bando2020}%
  \BibitemOpen
  \bibfield  {author} {\bibinfo {author} {\bibfnamefont {Y.}~\bibnamefont
  {Bando}}, \bibinfo {author} {\bibfnamefont {Y.}~\bibnamefont {Susa}},
  \bibinfo {author} {\bibfnamefont {H.}~\bibnamefont {Oshiyama}}, \bibinfo
  {author} {\bibfnamefont {N.}~\bibnamefont {Shibata}}, \bibinfo {author}
  {\bibfnamefont {M.}~\bibnamefont {Ohzeki}}, \bibinfo {author} {\bibfnamefont
  {F.~J.}\ \bibnamefont {G{\'{o}}mez-Ruiz}}, \bibinfo {author} {\bibfnamefont
  {D.~A.}\ \bibnamefont {Lidar}}, \bibinfo {author} {\bibfnamefont
  {S.}~\bibnamefont {Suzuki}}, \bibinfo {author} {\bibfnamefont
  {A.}~\bibnamefont {del Campo}},\ and\ \bibinfo {author} {\bibfnamefont
  {H.}~\bibnamefont {Nishimori}},\ }\bibfield  {title} {\bibinfo {title}
  {{Probing the universality of topological defect formation in a quantum
  annealer: Kibble-Zurek mechanism and beyond}},\ }\href
  {https://doi.org/10.1103/PhysRevResearch.2.033369} {\bibfield  {journal}
  {\bibinfo  {journal} {Physical Review Research}\ }\textbf {\bibinfo {volume}
  {2}},\ \bibinfo {pages} {033369} (\bibinfo {year} {2020})}\BibitemShut
  {NoStop}%
\bibitem [{\citenamefont {Weinberg}\ \emph {et~al.}(2020)\citenamefont
  {Weinberg}, \citenamefont {Tylutki}, \citenamefont {R{\"{o}}nkk{\"{o}}},
  \citenamefont {Westerholm}, \citenamefont {{\AA}str{\"{o}}m}, \citenamefont
  {Manninen}, \citenamefont {T{\"{o}}rm{\"{a}}},\ and\ \citenamefont
  {Sandvik}}]{Weinberg2020}%
  \BibitemOpen
  \bibfield  {author} {\bibinfo {author} {\bibfnamefont {P.}~\bibnamefont
  {Weinberg}}, \bibinfo {author} {\bibfnamefont {M.}~\bibnamefont {Tylutki}},
  \bibinfo {author} {\bibfnamefont {J.~M.}\ \bibnamefont {R{\"{o}}nkk{\"{o}}}},
  \bibinfo {author} {\bibfnamefont {J.}~\bibnamefont {Westerholm}}, \bibinfo
  {author} {\bibfnamefont {J.~A.}\ \bibnamefont {{\AA}str{\"{o}}m}}, \bibinfo
  {author} {\bibfnamefont {P.}~\bibnamefont {Manninen}}, \bibinfo {author}
  {\bibfnamefont {P.}~\bibnamefont {T{\"{o}}rm{\"{a}}}},\ and\ \bibinfo
  {author} {\bibfnamefont {A.~W.}\ \bibnamefont {Sandvik}},\ }\bibfield
  {title} {\bibinfo {title} {{Scaling and Diabatic Effects in Quantum Annealing
  with a D-Wave Device}},\ }\href
  {https://doi.org/10.1103/PhysRevLett.124.090502} {\bibfield  {journal}
  {\bibinfo  {journal} {Physical Review Letters}\ }\textbf {\bibinfo {volume}
  {124}},\ \bibinfo {pages} {090502} (\bibinfo {year} {2020})}\BibitemShut
  {NoStop}%
\bibitem [{\citenamefont {Dutta}\ \emph {et~al.}(2015)\citenamefont {Dutta},
  \citenamefont {Aeppli}, \citenamefont {Chakrabarti}, \citenamefont
  {Divakaran}, \citenamefont {Rosenbaum},\ and\ \citenamefont
  {Sen}}]{Dutta2015}%
  \BibitemOpen
  \bibfield  {author} {\bibinfo {author} {\bibfnamefont {A.}~\bibnamefont
  {Dutta}}, \bibinfo {author} {\bibfnamefont {G.}~\bibnamefont {Aeppli}},
  \bibinfo {author} {\bibfnamefont {B.~K.}\ \bibnamefont {Chakrabarti}},
  \bibinfo {author} {\bibfnamefont {U.}~\bibnamefont {Divakaran}}, \bibinfo
  {author} {\bibfnamefont {T.~F.}\ \bibnamefont {Rosenbaum}},\ and\ \bibinfo
  {author} {\bibfnamefont {D.}~\bibnamefont {Sen}},\ }\href
  {https://doi.org/10.1017/CBO9781107706057} {\emph {\bibinfo {title} {{Quantum
  phase transitions in transverse field spin models: from statistical physics
  to quantum information}}}}\ (\bibinfo  {publisher} {Cambridge University
  Press},\ \bibinfo {year} {2015})\BibitemShut {NoStop}%
\bibitem [{\citenamefont {Albash}\ and\ \citenamefont
  {Lidar}(2018)}]{Albash2016}%
  \BibitemOpen
  \bibfield  {author} {\bibinfo {author} {\bibfnamefont {T.}~\bibnamefont
  {Albash}}\ and\ \bibinfo {author} {\bibfnamefont {D.~A.}\ \bibnamefont
  {Lidar}},\ }\bibfield  {title} {\bibinfo {title} {Adiabatic quantum
  computation},\ }\href {https://doi.org/10.1103/RevModPhys.90.015002}
  {\bibfield  {journal} {\bibinfo  {journal} {Rev. Mod. Phys.}\ }\textbf
  {\bibinfo {volume} {90}},\ \bibinfo {pages} {015002} (\bibinfo {year}
  {2018})}\BibitemShut {NoStop}%
\bibitem [{\citenamefont {Zurek}(1985)}]{Zurek1985}%
  \BibitemOpen
  \bibfield  {author} {\bibinfo {author} {\bibfnamefont {W.~H.}\ \bibnamefont
  {Zurek}},\ }\bibfield  {title} {\bibinfo {title} {{Cosmological experiments
  in superfluid helium?}},\ }\href {https://doi.org/10.1038/317505a0}
  {\bibfield  {journal} {\bibinfo  {journal} {Nature}\ }\textbf {\bibinfo
  {volume} {317}},\ \bibinfo {pages} {505} (\bibinfo {year}
  {1985})}\BibitemShut {NoStop}%
\bibitem [{\citenamefont {Anquez}\ \emph {et~al.}(2016)\citenamefont {Anquez},
  \citenamefont {Robbins}, \citenamefont {Bharath}, \citenamefont
  {Boguslawski}, \citenamefont {Hoang},\ and\ \citenamefont
  {Chapman}}]{Anquez2016}%
  \BibitemOpen
  \bibfield  {author} {\bibinfo {author} {\bibfnamefont {M.}~\bibnamefont
  {Anquez}}, \bibinfo {author} {\bibfnamefont {B.~A.}\ \bibnamefont {Robbins}},
  \bibinfo {author} {\bibfnamefont {H.~M.}\ \bibnamefont {Bharath}}, \bibinfo
  {author} {\bibfnamefont {M.}~\bibnamefont {Boguslawski}}, \bibinfo {author}
  {\bibfnamefont {T.~M.}\ \bibnamefont {Hoang}},\ and\ \bibinfo {author}
  {\bibfnamefont {M.~S.}\ \bibnamefont {Chapman}},\ }\bibfield  {title}
  {\bibinfo {title} {{Quantum Kibble-Zurek mechanism in a spin-1 Bose-Einstein
  condensate}},\ }\href {https://doi.org/10.1103/PhysRevLett.116.155301}
  {\bibfield  {journal} {\bibinfo  {journal} {Physical Review Letters}\
  }\textbf {\bibinfo {volume} {116}},\ \bibinfo {pages} {1} (\bibinfo {year}
  {2016})}\BibitemShut {NoStop}%
\bibitem [{\citenamefont {Browaeys}\ and\ \citenamefont
  {Lahaye}(2020)}]{Browaeys2020}%
  \BibitemOpen
  \bibfield  {author} {\bibinfo {author} {\bibfnamefont {A.}~\bibnamefont
  {Browaeys}}\ and\ \bibinfo {author} {\bibfnamefont {T.}~\bibnamefont
  {Lahaye}},\ }\bibfield  {title} {\bibinfo {title} {{Many-body physics with
  individually controlled Rydberg atoms}},\ }\href
  {https://doi.org/10.1038/s41567-019-0733-z} {\bibfield  {journal} {\bibinfo
  {journal} {Nature Physics}\ }\textbf {\bibinfo {volume} {16}},\ \bibinfo
  {pages} {132} (\bibinfo {year} {2020})}\BibitemShut {NoStop}%
\bibitem [{\citenamefont {Cui}\ \emph {et~al.}(2020)\citenamefont {Cui},
  \citenamefont {G{\'{o}}mez-Ruiz}, \citenamefont {Huang}, \citenamefont {Li},
  \citenamefont {Guo},\ and\ \citenamefont {del Campo}}]{Cui2020}%
  \BibitemOpen
  \bibfield  {author} {\bibinfo {author} {\bibfnamefont {J.~M.}\ \bibnamefont
  {Cui}}, \bibinfo {author} {\bibfnamefont {F.~J.}\ \bibnamefont
  {G{\'{o}}mez-Ruiz}}, \bibinfo {author} {\bibfnamefont {Y.~F.}\ \bibnamefont
  {Huang}}, \bibinfo {author} {\bibfnamefont {C.~F.}\ \bibnamefont {Li}},
  \bibinfo {author} {\bibfnamefont {G.~C.}\ \bibnamefont {Guo}},\ and\ \bibinfo
  {author} {\bibfnamefont {A.}~\bibnamefont {del Campo}},\ }\bibfield  {title}
  {\bibinfo {title} {{Experimentally testing quantum critical dynamics beyond
  the Kibble–Zurek mechanism}},\ }\href
  {https://doi.org/10.1038/s42005-020-0306-6} {\bibfield  {journal} {\bibinfo
  {journal} {Communications Physics}\ }\textbf {\bibinfo {volume} {3}},\
  \bibinfo {pages} {1} (\bibinfo {year} {2020})}\BibitemShut {NoStop}%
\bibitem [{\citenamefont {Zurek}\ \emph {et~al.}(2005)\citenamefont {Zurek},
  \citenamefont {Dorner},\ and\ \citenamefont {Zoller}}]{Zurek2005}%
  \BibitemOpen
  \bibfield  {author} {\bibinfo {author} {\bibfnamefont {W.~H.}\ \bibnamefont
  {Zurek}}, \bibinfo {author} {\bibfnamefont {U.}~\bibnamefont {Dorner}},\ and\
  \bibinfo {author} {\bibfnamefont {P.}~\bibnamefont {Zoller}},\ }\bibfield
  {title} {\bibinfo {title} {{Dynamics of a quantum phase transition}},\ }\href
  {https://doi.org/10.1103/PhysRevLett.95.105701} {\bibfield  {journal}
  {\bibinfo  {journal} {Physical Review Letters}\ }\textbf {\bibinfo {volume}
  {95}},\ \bibinfo {pages} {2} (\bibinfo {year} {2005})}\BibitemShut {NoStop}%
\bibitem [{\citenamefont {Dziarmaga}(2005)}]{Dziarmaga2005}%
  \BibitemOpen
  \bibfield  {author} {\bibinfo {author} {\bibfnamefont {J.}~\bibnamefont
  {Dziarmaga}},\ }\bibfield  {title} {\bibinfo {title} {{Dynamics of a quantum
  phase transition: Exact solution of the quantum Ising model}},\ }\href
  {https://doi.org/10.1103/PhysRevLett.95.245701} {\bibfield  {journal}
  {\bibinfo  {journal} {Physical Review Letters}\ }\textbf {\bibinfo {volume}
  {95}},\ \bibinfo {pages} {1} (\bibinfo {year} {2005})}\BibitemShut {NoStop}%
\bibitem [{\citenamefont {Cincio}\ \emph {et~al.}(2009)\citenamefont {Cincio},
  \citenamefont {Dziarmaga}, \citenamefont {Meisner},\ and\ \citenamefont
  {Rams}}]{Cincio2009}%
  \BibitemOpen
  \bibfield  {author} {\bibinfo {author} {\bibfnamefont {L.}~\bibnamefont
  {Cincio}}, \bibinfo {author} {\bibfnamefont {J.}~\bibnamefont {Dziarmaga}},
  \bibinfo {author} {\bibfnamefont {J.}~\bibnamefont {Meisner}},\ and\ \bibinfo
  {author} {\bibfnamefont {M.~M.}\ \bibnamefont {Rams}},\ }\bibfield  {title}
  {\bibinfo {title} {{Dynamics of a quantum phase transition with decoherence:
  Quantum Ising chain in a static spin environment}},\ }\href
  {https://doi.org/10.1103/PhysRevB.79.094421} {\bibfield  {journal} {\bibinfo
  {journal} {Physical Review B}\ }\textbf {\bibinfo {volume} {79}},\ \bibinfo
  {pages} {1} (\bibinfo {year} {2009})}\BibitemShut {NoStop}%
\bibitem [{\citenamefont {Arceci}\ \emph {et~al.}(2018)\citenamefont {Arceci},
  \citenamefont {Barbarino}, \citenamefont {Rossini},\ and\ \citenamefont
  {Santoro}}]{Arceci2018}%
  \BibitemOpen
  \bibfield  {author} {\bibinfo {author} {\bibfnamefont {L.}~\bibnamefont
  {Arceci}}, \bibinfo {author} {\bibfnamefont {S.}~\bibnamefont {Barbarino}},
  \bibinfo {author} {\bibfnamefont {D.}~\bibnamefont {Rossini}},\ and\ \bibinfo
  {author} {\bibfnamefont {G.~E.}\ \bibnamefont {Santoro}},\ }\bibfield
  {title} {\bibinfo {title} {{Optimal working point in dissipative quantum
  annealing}},\ }\href {https://doi.org/10.1103/PhysRevB.98.064307} {\bibfield
  {journal} {\bibinfo  {journal} {Physical Review B}\ }\textbf {\bibinfo
  {volume} {98}},\ \bibinfo {pages} {064307} (\bibinfo {year}
  {2018})}\BibitemShut {NoStop}%
\bibitem [{\citenamefont {{del Campo}}(2018)}]{DelCampo2018}%
  \BibitemOpen
  \bibfield  {author} {\bibinfo {author} {\bibfnamefont {A.}~\bibnamefont {{del
  Campo}}},\ }\bibfield  {title} {\bibinfo {title} {{Universal Statistics of
  Topological Defects Formed in a Quantum Phase Transition}},\ }\href
  {https://doi.org/10.1103/PhysRevLett.121.200601} {\bibfield  {journal}
  {\bibinfo  {journal} {Physical Review Letters}\ }\textbf {\bibinfo {volume}
  {121}},\ \bibinfo {pages} {200601} (\bibinfo {year} {2018})}\BibitemShut
  {NoStop}%
\bibitem [{Note1()}]{Note1}%
  \BibitemOpen
  \bibinfo {note} {For detailed modeling, we determine the schedule terms
  $\Gamma (s)$ and $\protect \mathcal J(s)$ for each programmed value of $J$,
  based on a radio-frequency SQUID flux qubit model (see Supplementary
  Materials).}\BibitemShut {Stop}%
\bibitem [{\citenamefont {Mayo}\ \emph {et~al.}(2021)\citenamefont {Mayo},
  \citenamefont {Fan}, \citenamefont {Chern},\ and\ \citenamefont {del
  Campo}}]{Mayo2021}%
  \BibitemOpen
  \bibfield  {author} {\bibinfo {author} {\bibfnamefont {J.~J.}\ \bibnamefont
  {Mayo}}, \bibinfo {author} {\bibfnamefont {Z.}~\bibnamefont {Fan}}, \bibinfo
  {author} {\bibfnamefont {G.-W.}\ \bibnamefont {Chern}},\ and\ \bibinfo
  {author} {\bibfnamefont {A.}~\bibnamefont {del Campo}},\ }\bibfield  {title}
  {\bibinfo {title} {{Distribution of kinks in an Ising ferromagnet after
  annealing and the generalized Kibble-Zurek mechanism}},\ }\href
  {https://doi.org/10.1103/PhysRevResearch.3.033150} {\bibfield  {journal}
  {\bibinfo  {journal} {Physical Review Research}\ }\textbf {\bibinfo {volume}
  {3}},\ \bibinfo {pages} {033150} (\bibinfo {year} {2021})}\BibitemShut
  {NoStop}%
\bibitem [{\citenamefont {Krebs}\ \emph {et~al.}(1995)\citenamefont {Krebs},
  \citenamefont {Pfannm{\"{u}}ller}, \citenamefont {Wehefritz},\ and\
  \citenamefont {Hinrichsen}}]{Krebs1995}%
  \BibitemOpen
  \bibfield  {author} {\bibinfo {author} {\bibfnamefont {K.}~\bibnamefont
  {Krebs}}, \bibinfo {author} {\bibfnamefont {M.~P.}\ \bibnamefont
  {Pfannm{\"{u}}ller}}, \bibinfo {author} {\bibfnamefont {B.}~\bibnamefont
  {Wehefritz}},\ and\ \bibinfo {author} {\bibfnamefont {H.}~\bibnamefont
  {Hinrichsen}},\ }\bibfield  {title} {\bibinfo {title} {{Finite-size scaling
  studies of one-dimensional reaction-diffusion systems. Part I. Analytical
  results}},\ }\href {https://doi.org/10.1007/BF02180138} {\bibfield  {journal}
  {\bibinfo  {journal} {Journal of Statistical Physics}\ }\textbf {\bibinfo
  {volume} {78}},\ \bibinfo {pages} {1429} (\bibinfo {year}
  {1995})}\BibitemShut {NoStop}%
\bibitem [{\citenamefont {Roychowdhury}\ \emph {et~al.}(2021)\citenamefont
  {Roychowdhury}, \citenamefont {Moessner},\ and\ \citenamefont
  {Das}}]{Roychowdhury2021}%
  \BibitemOpen
  \bibfield  {author} {\bibinfo {author} {\bibfnamefont {K.}~\bibnamefont
  {Roychowdhury}}, \bibinfo {author} {\bibfnamefont {R.}~\bibnamefont
  {Moessner}},\ and\ \bibinfo {author} {\bibfnamefont {A.}~\bibnamefont
  {Das}},\ }\bibfield  {title} {\bibinfo {title} {{Dynamics and correlations at
  a quantum phase transition beyond Kibble-Zurek}},\ }\href
  {https://doi.org/10.1103/PhysRevB.104.014406} {\bibfield  {journal} {\bibinfo
   {journal} {Physical Review B}\ }\textbf {\bibinfo {volume} {104}},\ \bibinfo
  {pages} {014406} (\bibinfo {year} {2021})}\BibitemShut {NoStop}%
\bibitem [{\citenamefont {Nowak}\ and\ \citenamefont
  {Dziarmaga}(2021)}]{Nowak2021}%
  \BibitemOpen
  \bibfield  {author} {\bibinfo {author} {\bibfnamefont {R.~J.}\ \bibnamefont
  {Nowak}}\ and\ \bibinfo {author} {\bibfnamefont {J.}~\bibnamefont
  {Dziarmaga}},\ }\bibfield  {title} {\bibinfo {title} {{Quantum Kibble-Zurek
  mechanism: Kink correlations after a quench in the quantum Ising chain}},\
  }\href {https://doi.org/10.1103/PhysRevB.104.075448} {\bibfield  {journal}
  {\bibinfo  {journal} {Physical Review B}\ }\textbf {\bibinfo {volume}
  {104}},\ \bibinfo {pages} {075448} (\bibinfo {year} {2021})}\BibitemShut
  {NoStop}%
\bibitem [{\citenamefont {Oshiyama}\ \emph {et~al.}(2020)\citenamefont
  {Oshiyama}, \citenamefont {Shibata},\ and\ \citenamefont
  {Suzuki}}]{Oshiyama2020}%
  \BibitemOpen
  \bibfield  {author} {\bibinfo {author} {\bibfnamefont {H.}~\bibnamefont
  {Oshiyama}}, \bibinfo {author} {\bibfnamefont {N.}~\bibnamefont {Shibata}},\
  and\ \bibinfo {author} {\bibfnamefont {S.}~\bibnamefont {Suzuki}},\
  }\bibfield  {title} {\bibinfo {title} {{Kibble–Zurek mechanism in a
  dissipative transverse Ising chain}},\ }\href
  {https://doi.org/10.7566/JPSJ.89.104002} {\bibfield  {journal} {\bibinfo
  {journal} {Journal of the Physical Society of Japan}\ }\textbf {\bibinfo
  {volume} {89}},\ \bibinfo {pages} {1} (\bibinfo {year} {2020})}\BibitemShut
  {NoStop}%
\bibitem [{\citenamefont {Schuch}\ \emph {et~al.}(2008)\citenamefont {Schuch},
  \citenamefont {Wolf}, \citenamefont {Verstraete},\ and\ \citenamefont
  {Cirac}}]{Schuch:2008ve}%
  \BibitemOpen
  \bibfield  {author} {\bibinfo {author} {\bibfnamefont {N.}~\bibnamefont
  {Schuch}}, \bibinfo {author} {\bibfnamefont {M.~M.}\ \bibnamefont {Wolf}},
  \bibinfo {author} {\bibfnamefont {F.}~\bibnamefont {Verstraete}},\ and\
  \bibinfo {author} {\bibfnamefont {J.~I.}\ \bibnamefont {Cirac}},\ }\bibfield
  {title} {\bibinfo {title} {Entropy scaling and simulability by matrix product
  states},\ }\href {https://doi.org/10.1103/PhysRevLett.100.030504} {\bibfield
  {journal} {\bibinfo  {journal} {Physical Review Letters}\ }\textbf {\bibinfo
  {volume} {100}},\ \bibinfo {pages} {030504} (\bibinfo {year}
  {2008})}\BibitemShut {NoStop}%
\bibitem [{\citenamefont {Zener}(1932)}]{Zener1932}%
  \BibitemOpen
  \bibfield  {author} {\bibinfo {author} {\bibfnamefont {C.}~\bibnamefont
  {Zener}},\ }\bibfield  {title} {\bibinfo {title} {{Non-Adiabatic Crossing of
  Energy Levels}},\ }\href {https://doi.org/10.1098/rspa.1932.0165} {\bibfield
  {journal} {\bibinfo  {journal} {Proceedings of the Royal Society of London A:
  Mathematical, Physical and Engineering Sciences}\ }\textbf {\bibinfo {volume}
  {137}},\ \bibinfo {pages} {696} (\bibinfo {year} {1932})}\BibitemShut
  {NoStop}%
\bibitem [{\citenamefont {Kato}(1950)}]{Kato:50}%
  \BibitemOpen
  \bibfield  {author} {\bibinfo {author} {\bibfnamefont {T.}~\bibnamefont
  {Kato}},\ }\bibfield  {title} {\bibinfo {title} {On the adiabatic theorem of
  quantum mechanics},\ }\href {https://doi.org/10.1143/JPSJ.5.435} {\bibfield
  {journal} {\bibinfo  {journal} {J. Phys. Soc. Jpn.}\ }\textbf {\bibinfo
  {volume} {5}},\ \bibinfo {pages} {435} (\bibinfo {year} {1950})}\BibitemShut
  {NoStop}%
\bibitem [{\citenamefont {Suzuki}(1976)}]{Suzuki1976}%
  \BibitemOpen
  \bibfield  {author} {\bibinfo {author} {\bibfnamefont {M.}~\bibnamefont
  {Suzuki}},\ }\bibfield  {title} {\bibinfo {title} {{Relationship between
  d-Dimensional Quantal Spin Systems and (d+1)-Dimensional Ising Systems:
  Equivalence, Critical Exponents and Systematic Approximants of the Partition
  Function and Spin Correlations}},\ }\href
  {https://doi.org/10.1143/PTP.56.1454} {\bibfield  {journal} {\bibinfo
  {journal} {Progress of Theoretical Physics}\ }\textbf {\bibinfo {volume}
  {56}},\ \bibinfo {pages} {1454} (\bibinfo {year} {1976})}\BibitemShut
  {NoStop}%
\bibitem [{\citenamefont {Liu}\ \emph {et~al.}(2015)\citenamefont {Liu},
  \citenamefont {Polkovnikov},\ and\ \citenamefont {Sandvik}}]{Liu2015}%
  \BibitemOpen
  \bibfield  {author} {\bibinfo {author} {\bibfnamefont {C.~W.}\ \bibnamefont
  {Liu}}, \bibinfo {author} {\bibfnamefont {A.}~\bibnamefont {Polkovnikov}},\
  and\ \bibinfo {author} {\bibfnamefont {A.~W.}\ \bibnamefont {Sandvik}},\
  }\bibfield  {title} {\bibinfo {title} {{Quantum versus classical annealing:
  Insights from scaling theory and results for spin glasses on 3-regular
  graphs}},\ }\href {https://doi.org/10.1103/PhysRevLett.114.147203} {\bibfield
   {journal} {\bibinfo  {journal} {Physical Review Letters}\ }\textbf {\bibinfo
  {volume} {114}},\ \bibinfo {pages} {1} (\bibinfo {year} {2015})}\BibitemShut
  {NoStop}%
\bibitem [{\citenamefont {King}\ \emph
  {et~al.}(2021{\natexlab{a}})\citenamefont {King}, \citenamefont {Raymond},
  \citenamefont {Lanting}, \citenamefont {Isakov}, \citenamefont {Mohseni},
  \citenamefont {Poulin-Lamarre}, \citenamefont {Ejtemaee}, \citenamefont
  {Bernoudy}, \citenamefont {Ozfidan}, \citenamefont {Smirnov}, \citenamefont
  {Reis}, \citenamefont {Altomare}, \citenamefont {Babcock}, \citenamefont
  {Baron}, \citenamefont {Berkley}, \citenamefont {Boothby}, \citenamefont
  {Bunyk}, \citenamefont {Christiani}, \citenamefont {Enderud}, \citenamefont
  {Evert}, \citenamefont {Harris}, \citenamefont {Hoskinson}, \citenamefont
  {Huang}, \citenamefont {Jooya}, \citenamefont {Khodabandelou}, \citenamefont
  {Ladizinsky}, \citenamefont {Li}, \citenamefont {Lott}, \citenamefont
  {MacDonald}, \citenamefont {Marsden}, \citenamefont {Marsden}, \citenamefont
  {Medina}, \citenamefont {Molavi}, \citenamefont {Neufeld}, \citenamefont
  {Norouzpour}, \citenamefont {Oh}, \citenamefont {Pavlov}, \citenamefont
  {Perminov}, \citenamefont {Prescott}, \citenamefont {Rich}, \citenamefont
  {Sato}, \citenamefont {Sheldan}, \citenamefont {Sterling}, \citenamefont
  {Swenson}, \citenamefont {Tsai}, \citenamefont {Volkmann}, \citenamefont
  {Whittaker}, \citenamefont {Wilkinson}, \citenamefont {Yao}, \citenamefont
  {Neven}, \citenamefont {Hilton}, \citenamefont {Ladizinsky}, \citenamefont
  {Johnson},\ and\ \citenamefont {Amin}}]{King2021}%
  \BibitemOpen
  \bibfield  {author} {\bibinfo {author} {\bibfnamefont {A.~D.}\ \bibnamefont
  {King}}, \bibinfo {author} {\bibfnamefont {J.}~\bibnamefont {Raymond}},
  \bibinfo {author} {\bibfnamefont {T.}~\bibnamefont {Lanting}}, \bibinfo
  {author} {\bibfnamefont {S.~V.}\ \bibnamefont {Isakov}}, \bibinfo {author}
  {\bibfnamefont {M.}~\bibnamefont {Mohseni}}, \bibinfo {author} {\bibfnamefont
  {G.}~\bibnamefont {Poulin-Lamarre}}, \bibinfo {author} {\bibfnamefont
  {S.}~\bibnamefont {Ejtemaee}}, \bibinfo {author} {\bibfnamefont
  {W.}~\bibnamefont {Bernoudy}}, \bibinfo {author} {\bibfnamefont
  {I.}~\bibnamefont {Ozfidan}}, \bibinfo {author} {\bibfnamefont {A.~Y.}\
  \bibnamefont {Smirnov}}, \bibinfo {author} {\bibfnamefont {M.}~\bibnamefont
  {Reis}}, \bibinfo {author} {\bibfnamefont {F.}~\bibnamefont {Altomare}},
  \bibinfo {author} {\bibfnamefont {M.}~\bibnamefont {Babcock}}, \bibinfo
  {author} {\bibfnamefont {C.}~\bibnamefont {Baron}}, \bibinfo {author}
  {\bibfnamefont {A.~J.}\ \bibnamefont {Berkley}}, \bibinfo {author}
  {\bibfnamefont {K.}~\bibnamefont {Boothby}}, \bibinfo {author} {\bibfnamefont
  {P.~I.}\ \bibnamefont {Bunyk}}, \bibinfo {author} {\bibfnamefont
  {H.}~\bibnamefont {Christiani}}, \bibinfo {author} {\bibfnamefont
  {C.}~\bibnamefont {Enderud}}, \bibinfo {author} {\bibfnamefont
  {B.}~\bibnamefont {Evert}}, \bibinfo {author} {\bibfnamefont
  {R.}~\bibnamefont {Harris}}, \bibinfo {author} {\bibfnamefont
  {E.}~\bibnamefont {Hoskinson}}, \bibinfo {author} {\bibfnamefont
  {S.}~\bibnamefont {Huang}}, \bibinfo {author} {\bibfnamefont
  {K.}~\bibnamefont {Jooya}}, \bibinfo {author} {\bibfnamefont
  {A.}~\bibnamefont {Khodabandelou}}, \bibinfo {author} {\bibfnamefont
  {N.}~\bibnamefont {Ladizinsky}}, \bibinfo {author} {\bibfnamefont
  {R.}~\bibnamefont {Li}}, \bibinfo {author} {\bibfnamefont {P.~A.}\
  \bibnamefont {Lott}}, \bibinfo {author} {\bibfnamefont {A.~J.~R.}\
  \bibnamefont {MacDonald}}, \bibinfo {author} {\bibfnamefont {D.}~\bibnamefont
  {Marsden}}, \bibinfo {author} {\bibfnamefont {G.}~\bibnamefont {Marsden}},
  \bibinfo {author} {\bibfnamefont {T.}~\bibnamefont {Medina}}, \bibinfo
  {author} {\bibfnamefont {R.}~\bibnamefont {Molavi}}, \bibinfo {author}
  {\bibfnamefont {R.}~\bibnamefont {Neufeld}}, \bibinfo {author} {\bibfnamefont
  {M.}~\bibnamefont {Norouzpour}}, \bibinfo {author} {\bibfnamefont
  {T.}~\bibnamefont {Oh}}, \bibinfo {author} {\bibfnamefont {I.}~\bibnamefont
  {Pavlov}}, \bibinfo {author} {\bibfnamefont {I.}~\bibnamefont {Perminov}},
  \bibinfo {author} {\bibfnamefont {T.}~\bibnamefont {Prescott}}, \bibinfo
  {author} {\bibfnamefont {C.}~\bibnamefont {Rich}}, \bibinfo {author}
  {\bibfnamefont {Y.}~\bibnamefont {Sato}}, \bibinfo {author} {\bibfnamefont
  {B.}~\bibnamefont {Sheldan}}, \bibinfo {author} {\bibfnamefont
  {G.}~\bibnamefont {Sterling}}, \bibinfo {author} {\bibfnamefont {L.~J.}\
  \bibnamefont {Swenson}}, \bibinfo {author} {\bibfnamefont {N.}~\bibnamefont
  {Tsai}}, \bibinfo {author} {\bibfnamefont {M.~H.}\ \bibnamefont {Volkmann}},
  \bibinfo {author} {\bibfnamefont {J.~D.}\ \bibnamefont {Whittaker}}, \bibinfo
  {author} {\bibfnamefont {W.}~\bibnamefont {Wilkinson}}, \bibinfo {author}
  {\bibfnamefont {J.}~\bibnamefont {Yao}}, \bibinfo {author} {\bibfnamefont
  {H.}~\bibnamefont {Neven}}, \bibinfo {author} {\bibfnamefont {J.~P.}\
  \bibnamefont {Hilton}}, \bibinfo {author} {\bibfnamefont {E.}~\bibnamefont
  {Ladizinsky}}, \bibinfo {author} {\bibfnamefont {M.~W.}\ \bibnamefont
  {Johnson}},\ and\ \bibinfo {author} {\bibfnamefont {M.~H.}\ \bibnamefont
  {Amin}},\ }\bibfield  {title} {\bibinfo {title} {{Scaling advantage over
  path-integral Monte Carlo in quantum simulation of geometrically frustrated
  magnets}},\ }\href {https://doi.org/10.1038/s41467-021-20901-5} {\bibfield
  {journal} {\bibinfo  {journal} {Nature Communications}\ }\textbf {\bibinfo
  {volume} {12}},\ \bibinfo {pages} {1113} (\bibinfo {year}
  {2021}{\natexlab{a}})}\BibitemShut {NoStop}%
\bibitem [{\citenamefont {Bando}\ and\ \citenamefont
  {Nishimori}(2021)}]{Bando2021}%
  \BibitemOpen
  \bibfield  {author} {\bibinfo {author} {\bibfnamefont {Y.}~\bibnamefont
  {Bando}}\ and\ \bibinfo {author} {\bibfnamefont {H.}~\bibnamefont
  {Nishimori}},\ }\bibfield  {title} {\bibinfo {title} {{Simulated quantum
  annealing as a simulator of nonequilibrium quantum dynamics}},\ }\href
  {https://doi.org/10.1103/PhysRevA.104.022607} {\bibfield  {journal} {\bibinfo
   {journal} {Physical Review A}\ }\textbf {\bibinfo {volume} {104}},\ \bibinfo
  {pages} {022607} (\bibinfo {year} {2021})}\BibitemShut {NoStop}%
\bibitem [{\citenamefont {Yip}\ \emph {et~al.}(2018)\citenamefont {Yip},
  \citenamefont {Albash},\ and\ \citenamefont {Lidar}}]{Yip:2018aa}%
  \BibitemOpen
  \bibfield  {author} {\bibinfo {author} {\bibfnamefont {K.~W.}\ \bibnamefont
  {Yip}}, \bibinfo {author} {\bibfnamefont {T.}~\bibnamefont {Albash}},\ and\
  \bibinfo {author} {\bibfnamefont {D.~A.}\ \bibnamefont {Lidar}},\ }\bibfield
  {title} {\bibinfo {title} {Quantum trajectories for time-dependent adiabatic
  master equations},\ }\href {https://doi.org/10.1103/PhysRevA.97.022116}
  {\bibfield  {journal} {\bibinfo  {journal} {Physical Review A}\ }\textbf
  {\bibinfo {volume} {97}},\ \bibinfo {pages} {022116} (\bibinfo {year}
  {2018})}\BibitemShut {NoStop}%
\bibitem [{\citenamefont {King}\ \emph
  {et~al.}(2021{\natexlab{b}})\citenamefont {King}, \citenamefont {Nisoli},
  \citenamefont {Dahl}, \citenamefont {Poulin-Lamarre},\ and\ \citenamefont
  {Lopez-Bezanilla}}]{King2021a}%
  \BibitemOpen
  \bibfield  {author} {\bibinfo {author} {\bibfnamefont {A.~D.}\ \bibnamefont
  {King}}, \bibinfo {author} {\bibfnamefont {C.}~\bibnamefont {Nisoli}},
  \bibinfo {author} {\bibfnamefont {E.~D.}\ \bibnamefont {Dahl}}, \bibinfo
  {author} {\bibfnamefont {G.}~\bibnamefont {Poulin-Lamarre}},\ and\ \bibinfo
  {author} {\bibfnamefont {A.}~\bibnamefont {Lopez-Bezanilla}},\ }\bibfield
  {title} {\bibinfo {title} {{Qubit spin ice}},\ }\href
  {https://doi.org/10.1126/science.abe2824} {\bibfield  {journal} {\bibinfo
  {journal} {Science}\ }\textbf {\bibinfo {volume} {373}},\ \bibinfo {pages}
  {576} (\bibinfo {year} {2021}{\natexlab{b}})}\BibitemShut {NoStop}%
\bibitem [{\citenamefont {King}\ \emph
  {et~al.}(2021{\natexlab{c}})\citenamefont {King}, \citenamefont {Batista},
  \citenamefont {Raymond}, \citenamefont {Lanting}, \citenamefont {Ozfidan},
  \citenamefont {Poulin-Lamarre}, \citenamefont {Zhang},\ and\ \citenamefont
  {Amin}}]{King2021b}%
  \BibitemOpen
  \bibfield  {author} {\bibinfo {author} {\bibfnamefont {A.~D.}\ \bibnamefont
  {King}}, \bibinfo {author} {\bibfnamefont {C.~D.}\ \bibnamefont {Batista}},
  \bibinfo {author} {\bibfnamefont {J.}~\bibnamefont {Raymond}}, \bibinfo
  {author} {\bibfnamefont {T.}~\bibnamefont {Lanting}}, \bibinfo {author}
  {\bibfnamefont {I.}~\bibnamefont {Ozfidan}}, \bibinfo {author} {\bibfnamefont
  {G.}~\bibnamefont {Poulin-Lamarre}}, \bibinfo {author} {\bibfnamefont
  {H.}~\bibnamefont {Zhang}},\ and\ \bibinfo {author} {\bibfnamefont {M.~H.}\
  \bibnamefont {Amin}},\ }\bibfield  {title} {\bibinfo {title} {{Quantum
  Annealing Simulation of Out-of-Equilibrium Magnetization in a Spin-Chain
  Compound}},\ }\href {https://doi.org/10.1103/PRXQuantum.2.030317} {\bibfield
  {journal} {\bibinfo  {journal} {PRX Quantum}\ }\textbf {\bibinfo {volume}
  {2}},\ \bibinfo {pages} {030317} (\bibinfo {year}
  {2021}{\natexlab{c}})}\BibitemShut {NoStop}%
\bibitem [{\citenamefont {Lanting}\ \emph {et~al.}(2017)\citenamefont
  {Lanting}, \citenamefont {King}, \citenamefont {Evert},\ and\ \citenamefont
  {Hoskinson}}]{Lanting2017}%
  \BibitemOpen
  \bibfield  {author} {\bibinfo {author} {\bibfnamefont {T.}~\bibnamefont
  {Lanting}}, \bibinfo {author} {\bibfnamefont {A.~D.}\ \bibnamefont {King}},
  \bibinfo {author} {\bibfnamefont {B.}~\bibnamefont {Evert}},\ and\ \bibinfo
  {author} {\bibfnamefont {E.~M.}\ \bibnamefont {Hoskinson}},\ }\bibfield
  {title} {\bibinfo {title} {{Experimental demonstration of perturbative
  anticrossing mitigation using nonuniform driver Hamiltonians}},\ }\href
  {https://doi.org/10.1103/PhysRevA.96.042322} {\bibfield  {journal} {\bibinfo
  {journal} {Physical Review A}\ }\textbf {\bibinfo {volume} {96}},\ \bibinfo
  {pages} {042322} (\bibinfo {year} {2017})}\BibitemShut {NoStop}%
\bibitem [{\citenamefont {Kirkpatrick}\ \emph {et~al.}(1983)\citenamefont
  {Kirkpatrick}, \citenamefont {Gelatt},\ and\ \citenamefont
  {Vecchi}}]{Kirkpatrick1983}%
  \BibitemOpen
  \bibfield  {author} {\bibinfo {author} {\bibfnamefont {S.}~\bibnamefont
  {Kirkpatrick}}, \bibinfo {author} {\bibfnamefont {C.~D.}\ \bibnamefont
  {Gelatt}},\ and\ \bibinfo {author} {\bibfnamefont {M.~P.}\ \bibnamefont
  {Vecchi}},\ }\bibfield  {title} {\bibinfo {title} {{Optimization by Simulated
  Annealing}},\ }\href {https://doi.org/10.1126/science.220.4598.671}
  {\bibfield  {journal} {\bibinfo  {journal} {Science}\ }\textbf {\bibinfo
  {volume} {220}},\ \bibinfo {pages} {671} (\bibinfo {year}
  {1983})}\BibitemShut {NoStop}%
\bibitem [{\citenamefont {Rieger}\ and\ \citenamefont
  {Kawashima}(1999)}]{Rieger1999}%
  \BibitemOpen
  \bibfield  {author} {\bibinfo {author} {\bibfnamefont {H.}~\bibnamefont
  {Rieger}}\ and\ \bibinfo {author} {\bibfnamefont {N.}~\bibnamefont
  {Kawashima}},\ }\bibfield  {title} {\bibinfo {title} {{Application of a
  continuous time cluster algorithm to the two-dimensional random quantum Ising
  ferromagnet}},\ }\href {https://doi.org/10.1007/s100510050761} {\bibfield
  {journal} {\bibinfo  {journal} {The European Physical Journal B}\ }\textbf
  {\bibinfo {volume} {9}},\ \bibinfo {pages} {233} (\bibinfo {year}
  {1999})}\BibitemShut {NoStop}%
\bibitem [{\citenamefont {Boixo}\ \emph {et~al.}(2014)\citenamefont {Boixo},
  \citenamefont {R{\o}nnow}, \citenamefont {Isakov}, \citenamefont {Wang},
  \citenamefont {Wecker}, \citenamefont {Lidar}, \citenamefont {Martinis},\
  and\ \citenamefont {Troyer}}]{Boixo2014a}%
  \BibitemOpen
  \bibfield  {author} {\bibinfo {author} {\bibfnamefont {S.}~\bibnamefont
  {Boixo}}, \bibinfo {author} {\bibfnamefont {T.~F.}\ \bibnamefont
  {R{\o}nnow}}, \bibinfo {author} {\bibfnamefont {S.~V.}\ \bibnamefont
  {Isakov}}, \bibinfo {author} {\bibfnamefont {Z.}~\bibnamefont {Wang}},
  \bibinfo {author} {\bibfnamefont {D.}~\bibnamefont {Wecker}}, \bibinfo
  {author} {\bibfnamefont {D.~A.}\ \bibnamefont {Lidar}}, \bibinfo {author}
  {\bibfnamefont {J.~M.}\ \bibnamefont {Martinis}},\ and\ \bibinfo {author}
  {\bibfnamefont {M.}~\bibnamefont {Troyer}},\ }\bibfield  {title} {\bibinfo
  {title} {{Evidence for quantum annealing with more than one hundred
  qubits}},\ }\href {https://doi.org/10.1038/nphys2900} {\bibfield  {journal}
  {\bibinfo  {journal} {Nature Physics}\ }\textbf {\bibinfo {volume} {10}},\
  \bibinfo {pages} {218} (\bibinfo {year} {2014})}\BibitemShut {NoStop}%
\bibitem [{\citenamefont {Heim}\ \emph {et~al.}(2015)\citenamefont {Heim},
  \citenamefont {R{\o}nnow}, \citenamefont {Isakov},\ and\ \citenamefont
  {Troyer}}]{Heim2015}%
  \BibitemOpen
  \bibfield  {author} {\bibinfo {author} {\bibfnamefont {B.}~\bibnamefont
  {Heim}}, \bibinfo {author} {\bibfnamefont {T.~F.}\ \bibnamefont {R{\o}nnow}},
  \bibinfo {author} {\bibfnamefont {S.~V.}\ \bibnamefont {Isakov}},\ and\
  \bibinfo {author} {\bibfnamefont {M.}~\bibnamefont {Troyer}},\ }\bibfield
  {title} {\bibinfo {title} {{Quantum versus classical annealing of Ising spin
  glasses}},\ }\href {https://doi.org/10.1126/science.aaa4170} {\bibfield
  {journal} {\bibinfo  {journal} {Science}\ }\textbf {\bibinfo {volume}
  {348}},\ \bibinfo {pages} {215} (\bibinfo {year} {2015})}\BibitemShut
  {NoStop}%
\bibitem [{\citenamefont {Shin}\ \emph {et~al.}(2014)\citenamefont {Shin},
  \citenamefont {Smith}, \citenamefont {Smolin},\ and\ \citenamefont
  {Vazirani}}]{Shin2014}%
  \BibitemOpen
  \bibfield  {author} {\bibinfo {author} {\bibfnamefont {S.~W.}\ \bibnamefont
  {Shin}}, \bibinfo {author} {\bibfnamefont {G.}~\bibnamefont {Smith}},
  \bibinfo {author} {\bibfnamefont {J.~A.}\ \bibnamefont {Smolin}},\ and\
  \bibinfo {author} {\bibfnamefont {U.}~\bibnamefont {Vazirani}},\ }\href
  {http://arxiv.org/abs/arXiv:1401.7087v2 http://arxiv.org/abs/1401.7087}
  {\bibinfo {title} {{How ``Quantum'' is the D-Wave Machine?}}} (\bibinfo
  {year} {2014}),\ \Eprint {https://arxiv.org/abs/1401.7087} {arXiv:1401.7087}
  \BibitemShut {NoStop}%
\bibitem [{\citenamefont {Albash}\ \emph {et~al.}(2015)\citenamefont {Albash},
  \citenamefont {Vinci}, \citenamefont {Mishra}, \citenamefont {Warburton},\
  and\ \citenamefont {Lidar}}]{q-sig2}%
  \BibitemOpen
  \bibfield  {author} {\bibinfo {author} {\bibfnamefont {T.}~\bibnamefont
  {Albash}}, \bibinfo {author} {\bibfnamefont {W.}~\bibnamefont {Vinci}},
  \bibinfo {author} {\bibfnamefont {A.}~\bibnamefont {Mishra}}, \bibinfo
  {author} {\bibfnamefont {P.~A.}\ \bibnamefont {Warburton}},\ and\ \bibinfo
  {author} {\bibfnamefont {D.~A.}\ \bibnamefont {Lidar}},\ }\bibfield  {title}
  {\bibinfo {title} {Consistency tests of classical and quantum models for a
  quantum annealer},\ }\href
  {http://link.aps.org/doi/10.1103/PhysRevA.91.042314} {\bibfield  {journal}
  {\bibinfo  {journal} {Phys. Rev. A}\ }\textbf {\bibinfo {volume} {91}},\
  \bibinfo {pages} {042314} (\bibinfo {year} {2015})}\BibitemShut {NoStop}%
\bibitem [{\citenamefont {Yamashiro}\ \emph {et~al.}(2019)\citenamefont
  {Yamashiro}, \citenamefont {Ohkuwa}, \citenamefont {Nishimori},\ and\
  \citenamefont {Lidar}}]{Yamashiro:2019aa}%
  \BibitemOpen
  \bibfield  {author} {\bibinfo {author} {\bibfnamefont {Y.}~\bibnamefont
  {Yamashiro}}, \bibinfo {author} {\bibfnamefont {M.}~\bibnamefont {Ohkuwa}},
  \bibinfo {author} {\bibfnamefont {H.}~\bibnamefont {Nishimori}},\ and\
  \bibinfo {author} {\bibfnamefont {D.~A.}\ \bibnamefont {Lidar}},\ }\bibfield
  {title} {\bibinfo {title} {Dynamics of reverse annealing for the fully
  connected $p$-spin model},\ }\href
  {https://link.aps.org/doi/10.1103/PhysRevA.100.052321} {\bibfield  {journal}
  {\bibinfo  {journal} {Physical Review A}\ }\textbf {\bibinfo {volume}
  {100}},\ \bibinfo {pages} {052321} (\bibinfo {year} {2019})}\BibitemShut
  {NoStop}%
\bibitem [{\citenamefont {Albash}\ and\ \citenamefont
  {Marshall}(2021)}]{Albash2021}%
  \BibitemOpen
  \bibfield  {author} {\bibinfo {author} {\bibfnamefont {T.}~\bibnamefont
  {Albash}}\ and\ \bibinfo {author} {\bibfnamefont {J.}~\bibnamefont
  {Marshall}},\ }\bibfield  {title} {\bibinfo {title} {{Comparing Relaxation
  Mechanisms in Quantum and Classical Transverse-Field Annealing}},\ }\href
  {https://doi.org/10.1103/PhysRevApplied.15.014029} {\bibfield  {journal}
  {\bibinfo  {journal} {Physical Review Applied}\ }\textbf {\bibinfo {volume}
  {15}},\ \bibinfo {pages} {014029} (\bibinfo {year} {2021})}\BibitemShut
  {NoStop}%
\bibitem [{\citenamefont {Dziarmaga}(2006)}]{Dziarmaga2006}%
  \BibitemOpen
  \bibfield  {author} {\bibinfo {author} {\bibfnamefont {J.}~\bibnamefont
  {Dziarmaga}},\ }\bibfield  {title} {\bibinfo {title} {{Dynamics of a quantum
  phase transition in the random Ising model: Logarithmic dependence of the
  defect density on the transition rate}},\ }\href
  {https://doi.org/10.1103/physrevb.74.064416} {\bibfield  {journal} {\bibinfo
  {journal} {Physical Review B}\ }\textbf {\bibinfo {volume} {74}},\ \bibinfo
  {pages} {064416} (\bibinfo {year} {2006})}\BibitemShut {NoStop}%
\bibitem [{\citenamefont {Caneva}\ \emph {et~al.}(2007)\citenamefont {Caneva},
  \citenamefont {Fazio},\ and\ \citenamefont {Santoro}}]{Caneva2007}%
  \BibitemOpen
  \bibfield  {author} {\bibinfo {author} {\bibfnamefont {T.}~\bibnamefont
  {Caneva}}, \bibinfo {author} {\bibfnamefont {R.}~\bibnamefont {Fazio}},\ and\
  \bibinfo {author} {\bibfnamefont {G.~E.}\ \bibnamefont {Santoro}},\
  }\bibfield  {title} {\bibinfo {title} {{Adiabatic quantum dynamics of a
  random Ising chain across its quantum critical point}},\ }\href
  {https://doi.org/10.1103/PhysRevB.76.144427} {\bibfield  {journal} {\bibinfo
  {journal} {Physical Review B}\ }\textbf {\bibinfo {volume} {76}},\ \bibinfo
  {pages} {144427} (\bibinfo {year} {2007})}\BibitemShut {NoStop}%
\bibitem [{\citenamefont {Jordan}\ and\ \citenamefont
  {Wigner}(1928)}]{Jordan1928}%
  \BibitemOpen
  \bibfield  {author} {\bibinfo {author} {\bibfnamefont {P.}~\bibnamefont
  {Jordan}}\ and\ \bibinfo {author} {\bibfnamefont {E.}~\bibnamefont
  {Wigner}},\ }\bibfield  {title} {\bibinfo {title} {{{\"{U}}ber das Paulische
  {\"{A}}quivalenzverbot}},\ }\href {https://doi.org/10.1007/BF01331938}
  {\bibfield  {journal} {\bibinfo  {journal} {Zeitschrift f{\"{u}}r Physik}\
  }\textbf {\bibinfo {volume} {47}},\ \bibinfo {pages} {631} (\bibinfo {year}
  {1928})}\BibitemShut {NoStop}%
\bibitem [{\citenamefont {Schollw{\"{o}}ck}(2011)}]{Schollwock2011}%
  \BibitemOpen
  \bibfield  {author} {\bibinfo {author} {\bibfnamefont {U.}~\bibnamefont
  {Schollw{\"{o}}ck}},\ }\bibfield  {title} {\bibinfo {title} {{The
  density-matrix renormalization group in the age of matrix product states}},\
  }\href {https://doi.org/10.1016/j.aop.2010.09.012} {\bibfield  {journal}
  {\bibinfo  {journal} {Annals of Physics}\ }\textbf {\bibinfo {volume}
  {326}},\ \bibinfo {pages} {96} (\bibinfo {year} {2011})}\BibitemShut
  {NoStop}%
\bibitem [{\citenamefont {White}(1992)}]{White1992}%
  \BibitemOpen
  \bibfield  {author} {\bibinfo {author} {\bibfnamefont {S.~R.}\ \bibnamefont
  {White}},\ }\bibfield  {title} {\bibinfo {title} {Density matrix formulation
  for quantum renormalization groups},\ }\href
  {https://doi.org/10.1103/PhysRevLett.69.2863} {\bibfield  {journal} {\bibinfo
   {journal} {Phys. Rev. Lett.}\ }\textbf {\bibinfo {volume} {69}},\ \bibinfo
  {pages} {2863} (\bibinfo {year} {1992})}\BibitemShut {NoStop}%
\bibitem [{\citenamefont {White}(1993)}]{White1993}%
  \BibitemOpen
  \bibfield  {author} {\bibinfo {author} {\bibfnamefont {S.~R.}\ \bibnamefont
  {White}},\ }\bibfield  {title} {\bibinfo {title} {Density-matrix algorithms
  for quantum renormalization groups},\ }\href
  {https://doi.org/10.1103/PhysRevB.48.10345} {\bibfield  {journal} {\bibinfo
  {journal} {Phys. Rev. B}\ }\textbf {\bibinfo {volume} {48}},\ \bibinfo
  {pages} {10345} (\bibinfo {year} {1993})}\BibitemShut {NoStop}%
\bibitem [{\citenamefont {Vidal}(2004)}]{Vidal2004}%
  \BibitemOpen
  \bibfield  {author} {\bibinfo {author} {\bibfnamefont {G.}~\bibnamefont
  {Vidal}},\ }\bibfield  {title} {\bibinfo {title} {Efficient simulation of
  one-dimensional quantum many-body systems},\ }\href
  {https://doi.org/10.1103/PhysRevLett.93.040502} {\bibfield  {journal}
  {\bibinfo  {journal} {Phys. Rev. Lett.}\ }\textbf {\bibinfo {volume} {93}},\
  \bibinfo {pages} {040502} (\bibinfo {year} {2004})}\BibitemShut {NoStop}%
\bibitem [{\citenamefont {White}\ and\ \citenamefont
  {Feiguin}(2004)}]{White2004}%
  \BibitemOpen
  \bibfield  {author} {\bibinfo {author} {\bibfnamefont {S.~R.}\ \bibnamefont
  {White}}\ and\ \bibinfo {author} {\bibfnamefont {A.~E.}\ \bibnamefont
  {Feiguin}},\ }\bibfield  {title} {\bibinfo {title} {{Real-Time Evolution
  Using the Density Matrix Renormalization Group}},\ }\href
  {https://doi.org/10.1103/PhysRevLett.93.076401} {\bibfield  {journal}
  {\bibinfo  {journal} {Physical Review Letters}\ }\textbf {\bibinfo {volume}
  {93}},\ \bibinfo {pages} {076401} (\bibinfo {year} {2004})}\BibitemShut
  {NoStop}%
\bibitem [{\citenamefont {Suzuki}\ and\ \citenamefont
  {Okada}(2007)}]{Suzuki2007}%
  \BibitemOpen
  \bibfield  {author} {\bibinfo {author} {\bibfnamefont {S.}~\bibnamefont
  {Suzuki}}\ and\ \bibinfo {author} {\bibfnamefont {M.}~\bibnamefont {Okada}},\
  }\bibfield  {title} {\bibinfo {title} {{Study on Quantum Annealing Using the
  Density Matrix Renormalization Group}},\ }\href
  {https://doi.org/10.4036/iis.2007.49} {\bibfield  {journal} {\bibinfo
  {journal} {Interdisciplinary Information Sciences}\ }\textbf {\bibinfo
  {volume} {13}},\ \bibinfo {pages} {49} (\bibinfo {year} {2007})}\BibitemShut
  {NoStop}%
\bibitem [{\citenamefont {Suzuki}\ \emph {et~al.}(2019)\citenamefont {Suzuki},
  \citenamefont {Oshiyama},\ and\ \citenamefont {Shibata}}]{Suzuki2019}%
  \BibitemOpen
  \bibfield  {author} {\bibinfo {author} {\bibfnamefont {S.}~\bibnamefont
  {Suzuki}}, \bibinfo {author} {\bibfnamefont {H.}~\bibnamefont {Oshiyama}},\
  and\ \bibinfo {author} {\bibfnamefont {N.}~\bibnamefont {Shibata}},\
  }\bibfield  {title} {\bibinfo {title} {{Quantum Annealing of Pure and Random
  Ising Chains Coupled to a Bosonic Environment}},\ }\href
  {https://doi.org/10.7566/JPSJ.88.061003} {\bibfield  {journal} {\bibinfo
  {journal} {Journal of the Physical Society of Japan}\ }\textbf {\bibinfo
  {volume} {88}},\ \bibinfo {pages} {061003} (\bibinfo {year}
  {2019})}\BibitemShut {NoStop}%
\bibitem [{\citenamefont {Fitzpatrick}\ \emph {et~al.}(2021)\citenamefont
  {Fitzpatrick}, \citenamefont {Raymond},\ and\ \citenamefont
  {Kennett}}]{Fitzpatrick2021}%
  \BibitemOpen
  \bibfield  {author} {\bibinfo {author} {\bibfnamefont {M.~R.~C.}\
  \bibnamefont {Fitzpatrick}}, \bibinfo {author} {\bibfnamefont
  {J.}~\bibnamefont {Raymond}},\ and\ \bibinfo {author} {\bibfnamefont {M.~P.}\
  \bibnamefont {Kennett}},\ }\href@noop {} {\bibinfo {title} {{Essentially
  exact numerical modelling of flux qubit chains subject to charge and flux
  noise}}} (\bibinfo {year} {2021}),\ \Eprint
  {https://arxiv.org/abs/2110.01647} {arXiv:2110.01647} \BibitemShut {NoStop}%
\bibitem [{Spi(2021)}]{Spinbosonchain}%
  \BibitemOpen
  \href@noop {} {\bibinfo {title} {{Spin-Boson Chain}}},\ \bibinfo
  {howpublished} {\url{https://github.com/dwavesystems/spin-boson-chain}}
  (\bibinfo {year} {2021})\BibitemShut {NoStop}%
\bibitem [{\citenamefont {Makri}(1992)}]{Makri:1992wt}%
  \BibitemOpen
  \bibfield  {author} {\bibinfo {author} {\bibfnamefont {N.}~\bibnamefont
  {Makri}},\ }\bibfield  {title} {\bibinfo {title} {{Improved Feynman
  propagators on a grid and non-adiabatic corrections within the path integral
  framework}},\ }\href
  {https://doi.org/https://doi.org/10.1016/0009-2614(92)85654-S} {\bibfield
  {journal} {\bibinfo  {journal} {Chemical Physics Letters}\ }\textbf {\bibinfo
  {volume} {193}},\ \bibinfo {pages} {435} (\bibinfo {year}
  {1992})}\BibitemShut {NoStop}%
\bibitem [{\citenamefont {Strathearn}\ \emph {et~al.}(2018)\citenamefont
  {Strathearn}, \citenamefont {Kirton}, \citenamefont {Kilda}, \citenamefont
  {Keeling},\ and\ \citenamefont {Lovett}}]{Strathearn2018}%
  \BibitemOpen
  \bibfield  {author} {\bibinfo {author} {\bibfnamefont {A.}~\bibnamefont
  {Strathearn}}, \bibinfo {author} {\bibfnamefont {P.}~\bibnamefont {Kirton}},
  \bibinfo {author} {\bibfnamefont {D.}~\bibnamefont {Kilda}}, \bibinfo
  {author} {\bibfnamefont {J.}~\bibnamefont {Keeling}},\ and\ \bibinfo {author}
  {\bibfnamefont {B.~W.}\ \bibnamefont {Lovett}},\ }\bibfield  {title}
  {\bibinfo {title} {{Efficient non-Markovian quantum dynamics using
  time-evolving matrix product operators}},\ }\href
  {https://doi.org/10.1038/s41467-018-05617-3} {\bibfield  {journal} {\bibinfo
  {journal} {Nature Communications}\ }\textbf {\bibinfo {volume} {9}},\
  \bibinfo {pages} {3322} (\bibinfo {year} {2018})}\BibitemShut {NoStop}%
\bibitem [{\citenamefont {Sanz}\ \emph {et~al.}(2016)\citenamefont {Sanz},
  \citenamefont {Egusquiza}, \citenamefont {Di~Candia}, \citenamefont {Saberi},
  \citenamefont {Lamata},\ and\ \citenamefont {Solano}}]{Sanz:2016tb}%
  \BibitemOpen
  \bibfield  {author} {\bibinfo {author} {\bibfnamefont {M.}~\bibnamefont
  {Sanz}}, \bibinfo {author} {\bibfnamefont {I.~L.}\ \bibnamefont {Egusquiza}},
  \bibinfo {author} {\bibfnamefont {R.}~\bibnamefont {Di~Candia}}, \bibinfo
  {author} {\bibfnamefont {H.}~\bibnamefont {Saberi}}, \bibinfo {author}
  {\bibfnamefont {L.}~\bibnamefont {Lamata}},\ and\ \bibinfo {author}
  {\bibfnamefont {E.}~\bibnamefont {Solano}},\ }\bibfield  {title} {\bibinfo
  {title} {Entanglement classification with matrix product states},\ }\href
  {https://doi.org/10.1038/srep30188} {\bibfield  {journal} {\bibinfo
  {journal} {Scientific Reports}\ }\textbf {\bibinfo {volume} {6}},\ \bibinfo
  {pages} {30188} (\bibinfo {year} {2016})}\BibitemShut {NoStop}%
\end{thebibliography}%
\let\addcontentsline\oldaddcontentsline

\appendix

\section*{Methods}

\subsection*{Quantum annealing experiments}

Quantum annealing is performed on a D-Wave 2000Q lower noise (LN) processor using multiple randomly-generated embeddings (one for $L=2000$, three for $L=512$, up to 100 for $L=8$) in parallel.  Each data point represents data taken over $300$ iterations for $L=2000$ and $L=512$, and $50$ iterations for smaller values of $L$.  In each iteration the qubits are annealed $100$ times, providing $100$ spin states.  Each spin state consists of values $\{s_i\}_{i=1}^L$, where $s_i=\pm 1$ is the qubit readout state in the computational basis.

For each data point in the plots, we refine the general-purpose calibration by fine-tuning individual Hamiltonian terms based on trivial symmetries of the chain:  We tune per-qubit linear flux biases to bring qubits to degeneracy $(\braket{s_i}\approx 0)$, and tune two-qubit couplers to homogenize average correlations across chain bonds ($\braket{s_is_j} \approx (\sum_{\braket{k,\ell}}\braket{s_ks_\ell}/L$), as in previous studies of degenerate systems \cite{King2018,Nishimura2020}.  To mitigate desynchronization of annealing schedules between different qubits for the fastest anneals, we additionally refine anneal offsets based on annealing lines, although in this case there is little effect.  We describe these methods in the SM.

To generate error bars, a statistical bootstrap is performed.  For individual data points, the method treats each QPU call as an individual trial and resamples with replacement.  In particular, estimates of $C^{\textrm{KK}}_r$ are computed for each QPU call, then bootstrapped, so each estimate of $\bar n$ represents a QPU call, not an overall average.  To compute QA exponents $a$ for Fig.~\ref{fig:4}b), we treat every $t_a$ as a trial, and generate a distribution of fit slopes based on bootstrapped sets of annealing times.  Data markers and error bars represent the median and 95\% confidence interval of the resampling median.

\subsection*{Annealing schedule}

The annealing schedule depicted in Fig.~\ref{fig:1} is based on qubit parameters extracted through averaged single-qubit measurements.  Since qubits are actually multi-level objects rather than perfect two-level Ising spins, we convert the qubit Hamiltonian to an effective Ising Hamiltonian following the method laid out in recent studies of geometrically frustrated lattices \cite{King2021,King2021a}.  We perform approximate diagonalization of the $s$-dependent eigenspectrum of a 12-qubit periodic chain Hamiltonian.  We simplify the computation by dividing the qubits into four chains of three qubits each, and retaining only the 12 lowest energy levels of each three-qubit chain.  Once this eigenspectrum is computed for a given coupling strength $J$, we perform a two-parameter fit on $\Gamma(s)$ and $\mathcal J(s)|J|$ in equation (\ref{eq:ham}), minimizing a weighted average of the differences in the first eight eigengaps between the qubit Hamiltonian and the transverse-field Ising Hamiltonian.  Effective qubit temperatures were measured using single-qubit susceptibility measurements, as described in Section II.D of the Supplementary Information of Ref.~\cite{Johnson2011}.

\subsection*{Fermionized models and TEBD}

Calculations using the fermionized system were performed on the same number of spins as in QA, i.e., $L=512$ in Fig.~\ref{fig:3} and a range of $L$ for Fig.~\ref{fig:4}.   TEBD data in Fig.~\ref{fig:3} were produced using $L=256$ to reduce computation time. This has a negligible effect on results since this is much larger than the correlation length at the values of $t_a$ investigated, as we confirmed by solving the fermionized model at both $L=256$ and $L=512$.  The average and error bars representing 95\% statistical confidence in TEBD data were obtained for 300 realizations of disorder.

\widetext
\clearpage

\begin{center}
\textbf{\large Supplementary Material:\\ \mytitle}
\end{center}

\tableofcontents
\setcounter{equation}{0}
\setcounter{figure}{0}
\renewcommand{\thefigure}{S\arabic{figure}}
\renewcommand{\theHfigure}{S\arabic{figure}}
\makeatother

\section{Quantum annealing methods}

\subsection{Extracting annealing schedules}

We follow the method used in recent experiments \cite{King2021,King2021a} to extract an effective transverse-field Ising Hamiltonian from a qubit model based on single-qubit measurements.  The first task is to generate an eigenspectrum capturing the energy of fundamental local excitations.  This was previously done by making a small representative ``gadget'' that shares the general local structure of the larger system, while being small enough to diagonalize approximately.  In this case, our gadget is a periodic 12-qubit chain.  We account for more than two energy levels per rf-SQUID flux qubit, so even at this scale the computation is nontrivial.  To make the computation simpler we divide the chain into four subchains of three qubits each, compute the spectrum of each subchain as a four-level object, and then compute the spectrum of the full 12-qubit chain. 

\begin{figure}
  \includegraphics[width=\linewidth]{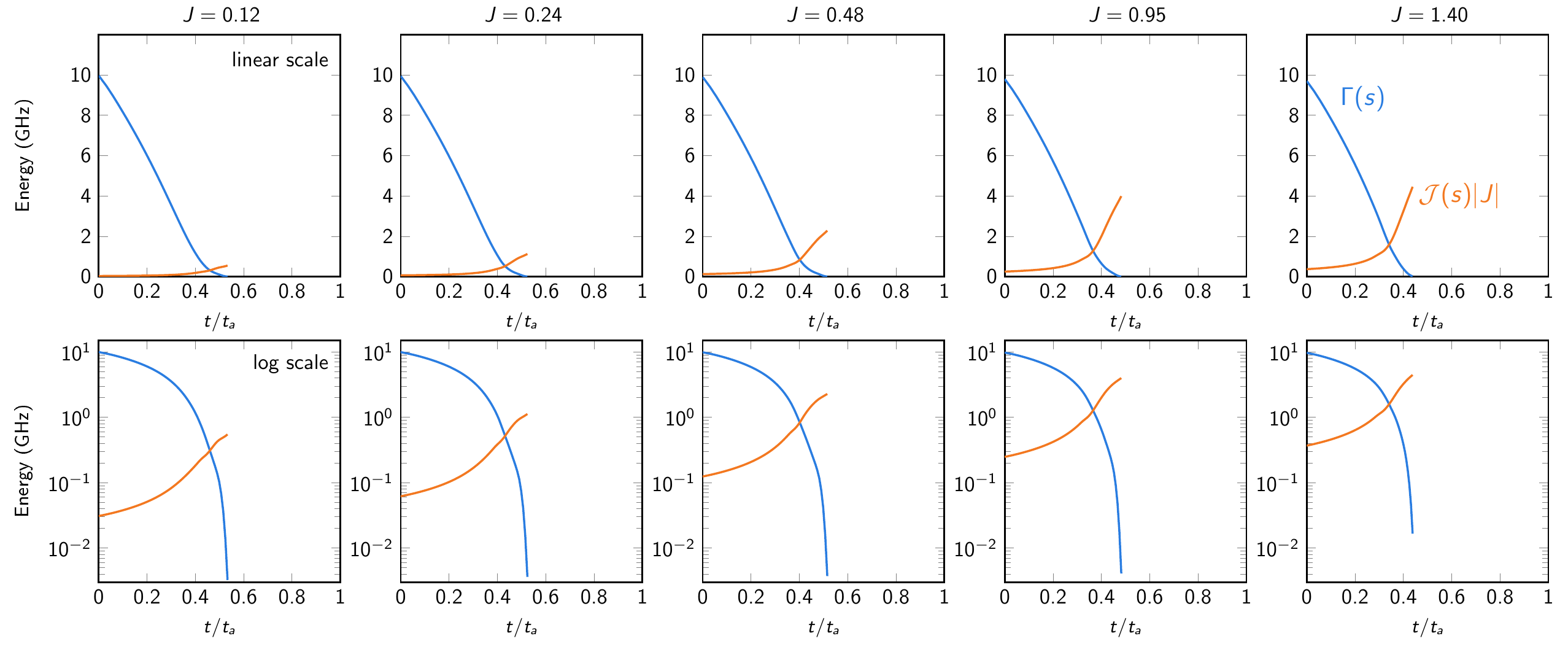}
  \caption{{\bf Extracted annealing schedules.} For each coupling magnitude $J$, we extract an effective transverse-field Ising model using a qubit model and single-qubit measurements.  These are shown in linear (top) and log (bottom) energy scales.  }\label{fig:schedules}
\end{figure}

After computing the spectrum of the 12-qubit chain, we determine a best-fit transverse-field Ising Hamiltonian using $\Gamma(s)$ and $\mathcal J(s)|J|$ as fitting parameters at each value of $s$.  The objective function for the fit is a weighted sum of the first eight eigengaps.  The resulting schedules, which we use in our software simulations, are shown in Fig.~\ref{fig:schedules}.  Note that the schedules are not computed all the way to $s=t/t_a=1$, because when $\mathcal J(s)|J| \gg \Gamma(s)$ we can safely presume that qubit dynamics have ceased, and the numerical methods used to extract the schedule become unstable.

\subsection{Calibration refinement shim}

We can refine the QA calibration, suppressing disorder that may arise from calibration imperfections and crosstalk, by exploiting two trivial symmetries of the Ising chain that hold everywhere in the phase diagram:

\begin{enumerate}
    \item All qubits have average magnetization zero, i.e.,
    \begin{equation}
        \forall i,\qquad\langle \sigma_i^z\rangle=0.
    \end{equation}
    \item All couplers have the same average correlation, i.e.,
    \begin{equation}
        \forall i,\qquad\langle \sigma_i^z\sigma_{i+1}^z \rangle = \frac{1}{L}\sum_{j=1}^L \langle \sigma_j^z\sigma_{j+1}^z \rangle.
    \end{equation}
\end{enumerate}
Any behavior that systematically violates these symmetries indicates a bias, which we suppress using the iterative methods described here.  

\begin{figure}
    \centering
    \includegraphics[height=20cm]{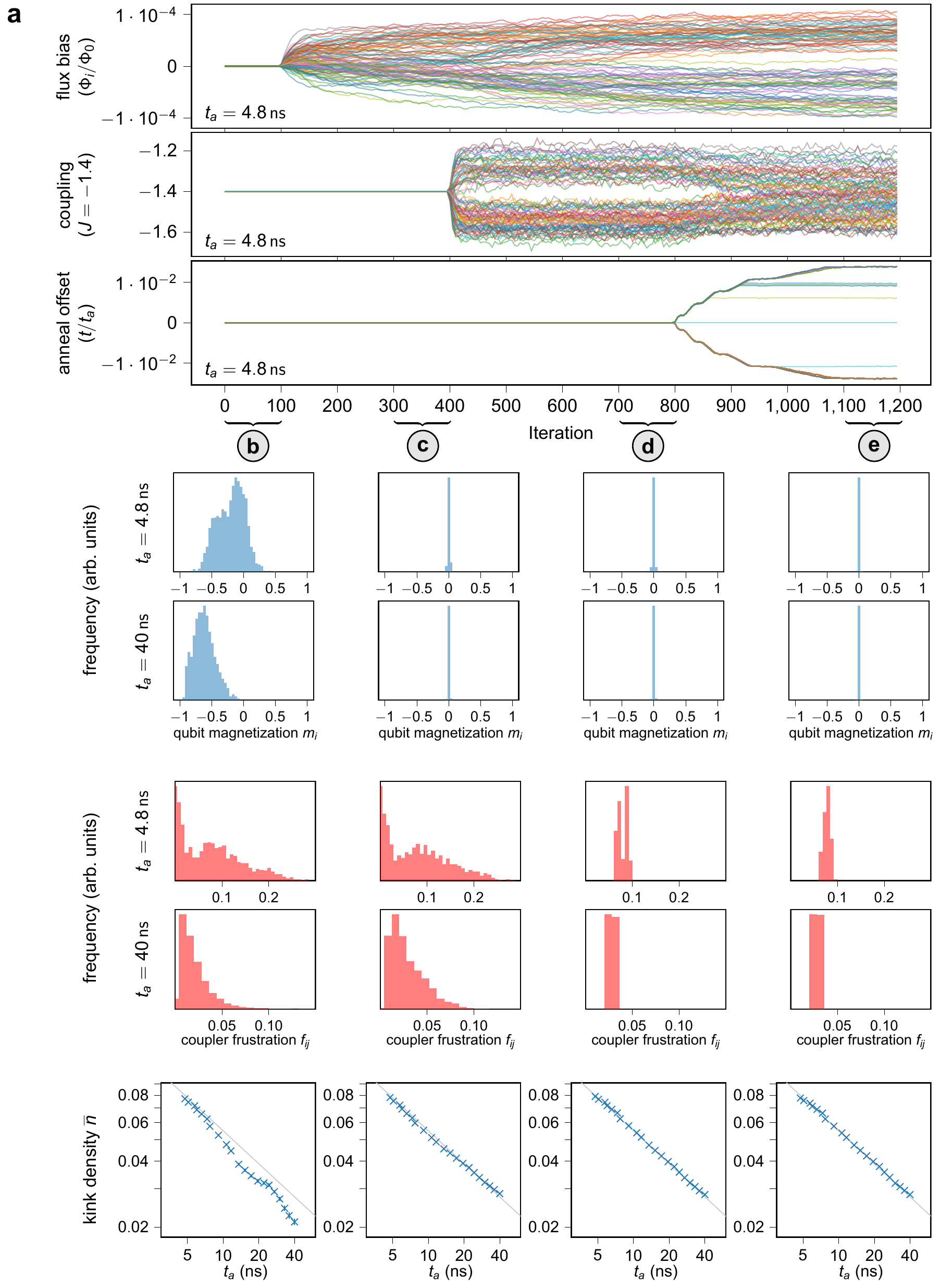}
    \caption{{\bf Calibration refinement shim demonstration.} {\bf a}, We run 1,200 iterations with $L=2000$, $J=-1.4$, sequentially turning on different parts of the shim one at a time, to illustrate their effects.  Shown are shim values for $t_a=\SI{4.8}{ns}$, where flux biases (top), individual coupling terms (middle) and anneal offsets (bottom) are adjusted.  100 of 2000 statistic lines are shown.  {\bf b--e} We measure 2000 average qubit magnetizations and 2000 coupler frustration probabilities for 100 iterations of 100 samples each, with different adjustments activated: {\bf b}, No shim.  {\bf c}, Flux-bias shim added.  {\bf d}, Coupler shim added.  {\bf e}, Anneal offset shim added.  Gray lines in kink density plots are a guide to the eye with $\bar n \propto t_a^{-1/2}$.}
    \label{fig:shim}
\end{figure}

Let $s_i$ indicate an output state for qubit $i$, i.e., a measurement of $\sigma_i^z$ at $s=1$.  If we measure $\langle s_i \rangle$ as being systematically nonzero, the qubit is biased.  In this case we compensate by adding a per-qubit flux bias to qubit $i$, which we denote by $\Phi_i$.  Similarly, if we measure correlation between two coupled qubits $\langle s_is_{i+1} \rangle$ that is systematically different from the average over all $i$, we compensate by making the coupler $J_{i,i+1}$ either slightly stronger or slightly weaker.  These two approaches, shown in Fig.~\ref{fig:shim}a, have become an important and standard ingredient of simulations of degenerate systems \cite{King2018, Nishimura2020, King2021, King2021a, King2021b}.  Here we give a more detailed description of these methods than in previous works.  Furthermore, we introduce an additional element of per-qubit anneal offsets, which are important for synchronizing qubits during fast anneals.

The qubits in the D-Wave 2000Q LN processor are controlled by four annealing lines, such that each eight-qubit unit cell has four qubits on each of two lines, and no two coupled qubits are annealed by the same line.  When annealing fast ($t_a \ll \SI{20}{ns}$), desynchronization between these lines can become significant.  In principle, if a qubit is delayed in its annealing schedule relative to other qubits, the couplers incident to that qubit will be expressed relatively weakly, leading to the couplers being frustrated more often than other couplers, i.e., hosting more kinks than on average.  This desynchronization can be compensated using the ``anneal offset'' feature \cite{Lanting2017}, which allows the definition of $s$ to be shifted slightly forwards or backwards for qubit $i$, by $\mathcal O_i$.  To compute appropriate anneal offsets, we compute an average frustration term for each annealing line.  For $\ell$ from $1$ to $4$, let $F_\ell$ denote the empirical probability that a coupler between two qubits, one of which is on the annealing line indexed $\ell$, is frustrated.  Note that the average over all lines, i.e., $\frac 14\sum_{j=1}^4F_\ell$ is equal to $\bar n$.  Similar to the flux-bias and coupler-tuning refinements, if $F_\ell < \bar n$, we advance the qubits on line $\ell$ using anneal offsets, by increasing $\mathcal O_i$ for any qubit $i$ on line $\ell$.

We now formalize these ideas.  We define three constants: $\alpha_\Phi$, $\alpha_J$, and $\alpha_a$ which define step sizes for the flux offset, coupler tuning, and anneal offset adjustments respectively.  We further define two damping constants $\delta_J$ and $\delta_a$ for the latter two adjustments, to prevent erratic iterative behavior, since these two adjustments can affect one another significantly.

For fixed $L$, $J$, and $t_a$, we run multiple iterations; each iteration consists of a call to the QA processor that draws 100 samples. 

Upon receiving the output samples, we compute the following statistics for the $100$ samples:
\begin{itemize}
    \item magnetization $m_i = \langle s_i\rangle$ for each qubit $i$
    \item correlation $c_{ij} = \langle s_is_j\rangle$ for each coupled pair of qubits $i,j$
    \item frustration $f_{ij} = (\text{sign}(J) c_{ij}+1)/2$ for each coupled pair of qubits $i,j$
    \item average line frustration $F_\ell = \sum_{\{i,j\} \in V_\ell}f_{ij}/|V_\ell|$ where $V_\ell$ is the set of coupled pairs $\{i,j\}$ such that either qubit $i$ or qubit $j$ is on annealing line $\ell$. 
    \item overall average frustration $\bar n$.
\end{itemize}
We then adjust QA Hamiltonian terms as follows:
\begin{itemize}
    \item for each qubit $i$, $\Phi_i \leftarrow \Phi_i - \alpha_\Phi m_i$
\item for each coupler $ij$, $J_{ij} \leftarrow J_{ij} + \alpha_J(f_{ij}-\bar n)$
\item for each annealing line $\ell$, $\mathcal O_\ell \leftarrow \mathcal O_\ell + \alpha_a(F_\ell - \bar n)$
\end{itemize}
Finally we damp the two latter adjustments:
\begin{itemize}
\item for each coupler $ij$, $J_{ij} \leftarrow (1-\delta_J)J_{ij} + \delta_JJ$
\item for each annealing line $\ell$, $\mathcal O_\ell \leftarrow (1-\delta_a)\mathcal O_\ell$.
\end{itemize}

We demonstrate these methods in Fig.~\ref{fig:shim}, showing an example of this calibration refinement shim using $L=2000$, $J=-1.4$, and $t_a$ ranging from $\SI{4.8}{ns}$ to $\SI{40}{ns}$.  The first $100$ iterations are run with $\alpha_\Phi=\alpha_J=\alpha_a=0$.  We then turn on $\alpha_\Phi =5e-6$ for $300$ iterations.  We then turn on $\alpha_J=0.2$, $\delta_J=0.02$ for $400$ iterations.  Finally we turn on $\alpha_a=0.02$, $\delta_a=0.002$ for $400$ iterations.

With no calibration refinement (Fig.~\ref{fig:shim}b), crosstalk and other biases---which can arise from calibration imperfections and time-dependent noise---lead to qubit magnetizations that are far from the desired value of zero.  Similarly, when flux biases have balanced the qubits  (Fig.~\ref{fig:shim}c), coupling inhomogeneity can lead to a broad distribution of frustration among the couplers in the chain.  When couplings are additionally tuned (Fig.~\ref{fig:shim}d), kink distributions look much more uniform, but at $t_a=\SI{4.8}{ns}$ there still remains a visibly bimodal distribution of coupler frustrations; this is mitigated with the additional tuning of anneal offsets (Fig.~\ref{fig:shim}e).  This experimental protocol, in which shim parts are activated in sequence, is shown for illustrative purposes only, and does not reflect actual experimental methods.  In the main experiments we activate all parts of the shim immediately.  Furthermore, since the compensations vary smoothly as functions of $T$, $J$, and $t_a$, we do not need to compute them from scratch for each data point.

\begin{figure}
    \centering
    \includegraphics[width=14cm]{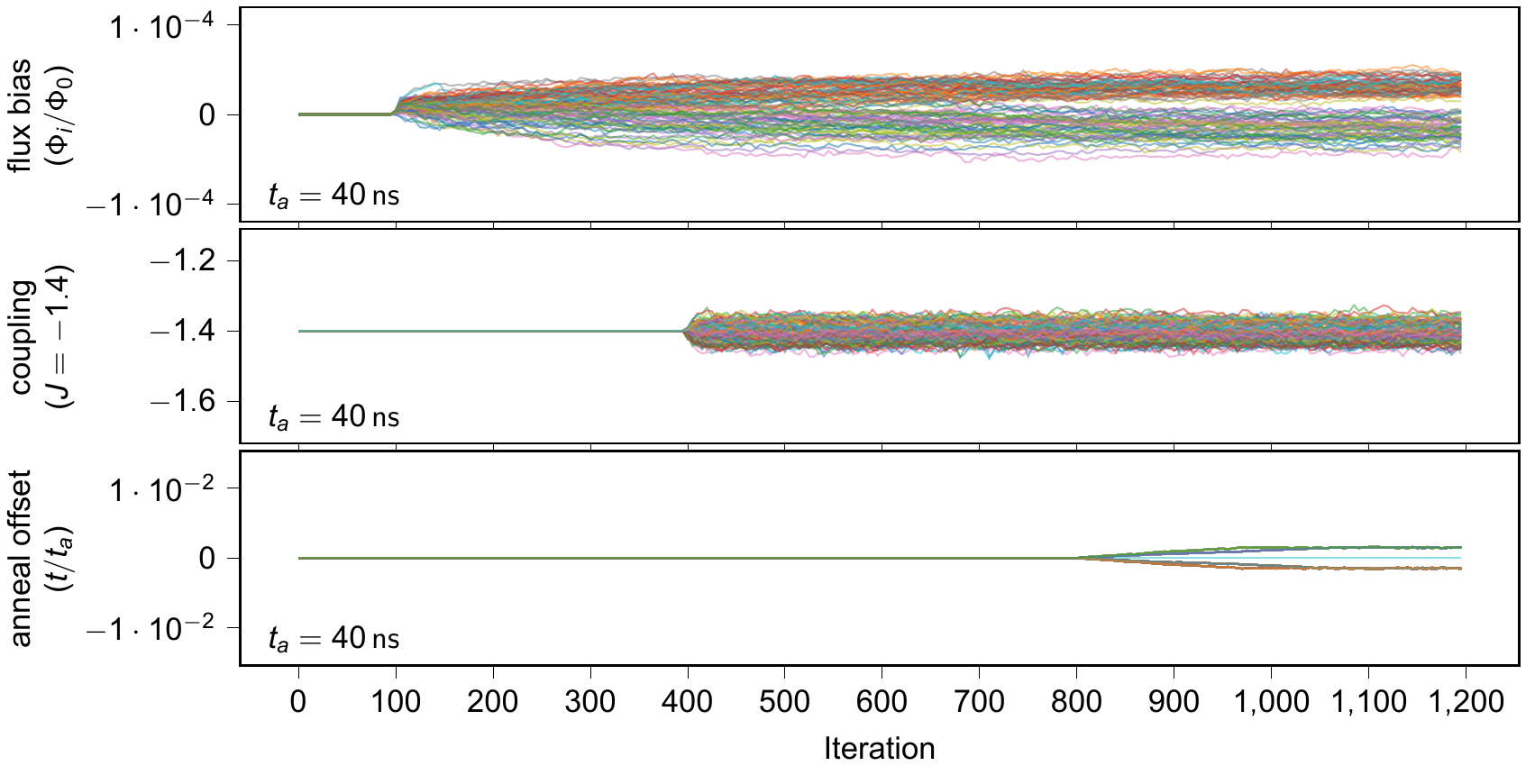}
    \caption{{\bf Calibration refinement shim for slower anneal.} Data analogous to Fig.~\ref{fig:shim}a are presented for $t_a=\SI{40}{ns}$, indicating that the required calibration tuning decreases significantly as anneals become slower.}
    \label{fig:shimslow}
\end{figure}

One can see in (Fig.~\ref{fig:shim}a) that the anneal offsets saturate for some qubits; this is due to limited per-qubit range in the anneal offset parameter.  The compensations required at $t_a=\SI{4.8}{ns}$ are large compared to required compensations at slower anneals; we show data for $t_a=\SI{40}{ns}$ in Fig.~\ref{fig:shimslow} for comparison.

In this work the anneal offsets have only a small impact on observables.  However, we expect them to be an important aspect in future work on fast anneals.

\section{Monte Carlo methods}\label{app:mc}

As mentioned in the main text, certain behaviors of the quantum Ising chain can be reproduced in classical models.  While we have shown quantitative agreement between QA and the quantum model for the kink density distribution in the Kibble-Zurek regime and success probability in the Landau-Zener regime, it is also useful to determine which behavior, if any, of the quantum system can be emulated with classical Monte Carlo approaches.  Here we consider three classical Monte Carlo approaches, and compare their behavior to that of QA.

In this section, as with QA data for $L=512$, error bars represent 95\% bootstrap confidence intervals, using resampling on a set of 300 experiments, each experiment consisting of 300 samples.

\subsection{Simulated annealing}

Simulated annealing (SA) \cite{Kirkpatrick1983} involves gradually lowering the temperature of a spin system in the Metropolis-Hastings algorithm, and is a purely thermal method.  In the setting of a 1D chain, we expect the scaling of kink densities to resemble diffusive scaling, i.e., $\bar n  \propto 1/\sqrt{t_a}$, since at low temperatures the kinks follow random walks along a locally degenerate energy landscape until they meet and annihilate pairwise with another kink.

The SA we study generates a new random ordering of variables to update for each Monte Carlo sweep, and we follow a geometric schedule from the high-temperature limit $\beta=0.1$ to the low-temperature limit $\beta=100$.

\begin{figure}
  \includegraphics[width=0.7\linewidth]{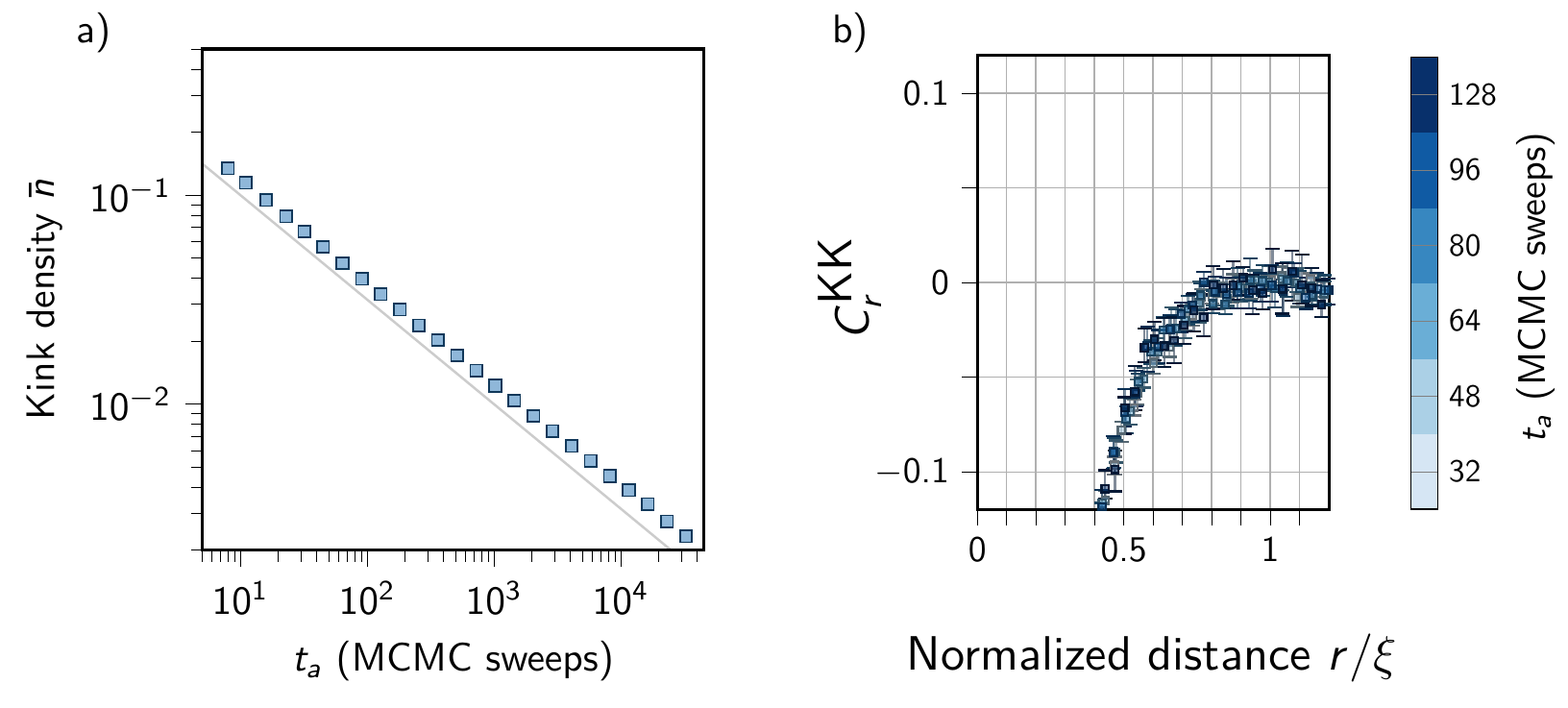}
  \caption{{\bf Kink densities and kink-kink correlators in SA.}  {\bf a}, Kink scaling is shown for a range of anneal times, with diffusive scaling $\bar n  \propto 1/\sqrt{t_a}$ indicated with a gray line.  {\bf b}, Kink-kink correlators are shown, with no positive peak.}\label{fig:sa_meta}
\end{figure}

In Fig.~\ref{fig:sa_meta} we show kink density scaling and kink-kink correlators for SA.  While SA does closely follow the $\bar n \propto 1/\sqrt{t_a}$ scaling that is characteristic of both the quantum KZM and classical diffusion, there is no positive peak in the correlator.  This is consistent with a diffusive picture.

\subsection{SQA-PIMC}

Path-integral Monte Carlo (PIMC) is a method of simulating finite-temperature quantum systems at thermal equilibrium using the Suzuki-Trotter decomposition \cite{Suzuki1976}.  We use a variant that approaches the continuous imaginary time limit and employs Swendsen-Wang cluster updates \cite{Rieger1999}.  By running PIMC along an annealing schedule we can simulate equilibrium aspects of quantum annealing; we call this method SQA-PIMC \cite{Boixo2014a}.  We use the QA annealing schedule for $J=1.4$.  To make the energy scales dimensionless we divide by the intersection point $\Gamma(s_c)=\mathcal J(s_c)|J|$.  Monte Carlo sweeps are performed at equally spaced points in the schedule from $s=0$ to $s=0.44$, at which point $\Gamma/J = 0.004$ and we assume the dynamics to be frozen.

\begin{figure}
  \includegraphics[width=\linewidth]{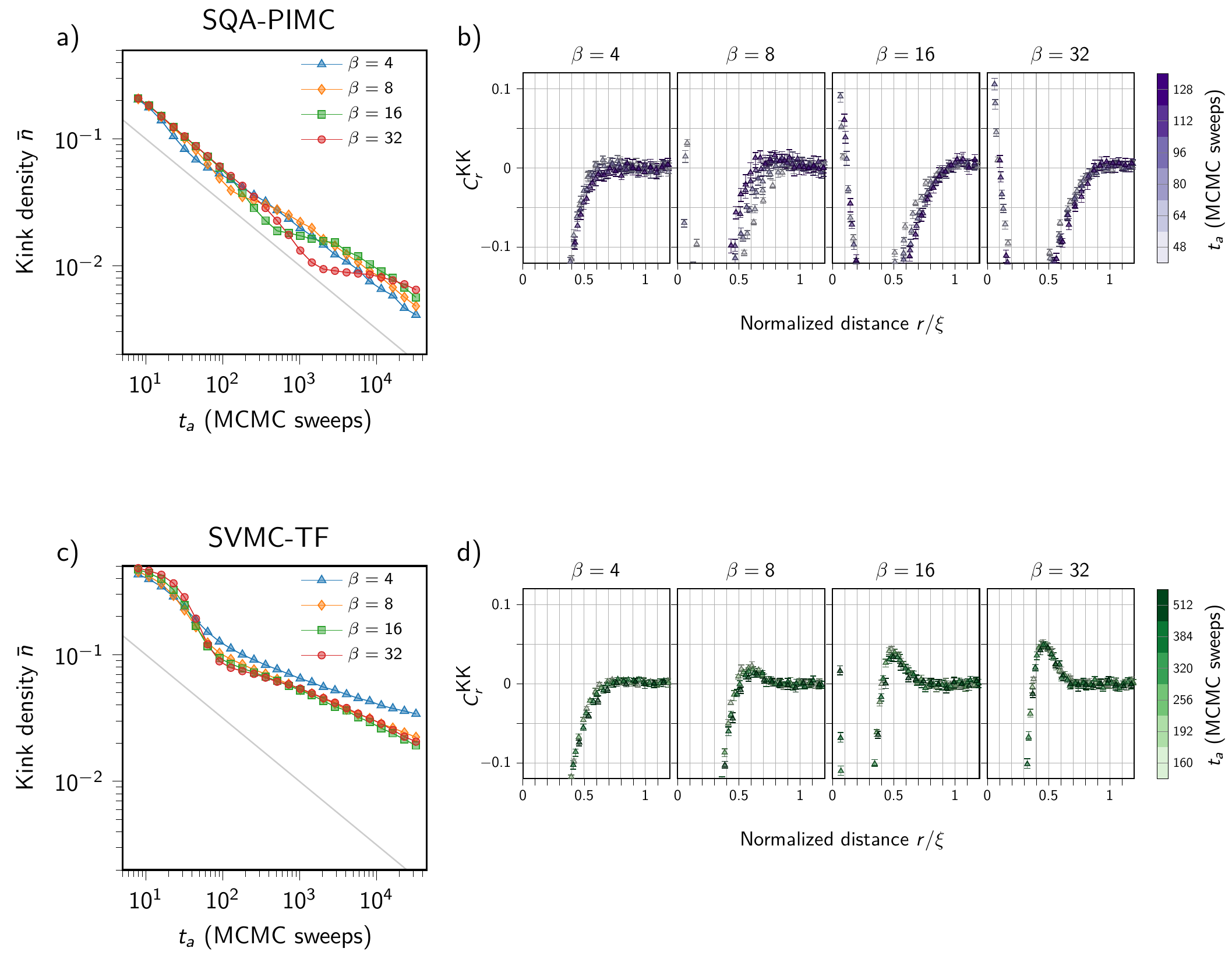}
  \caption{{\bf Effect of temperature on SQA-PIMC and SVMC-TF.}  Inverse temperatures $\beta=4$, $8$, $16$, and $32$ are used in SQA-PIMC (a--b) and SVMC-TF (c--d).}\label{fig:mcmc_temperatures}
\end{figure}

In Fig.~\ref{fig:mcmc_temperatures}a--b we show kink density scaling and kink-kink correlators for a range of inverse temperatures in SQA-PIMC.  Unlike SVMC-TF, which we describe in the next section, there are significant differences between $\beta=16$ and $\beta=32$ in terms of kink density scaling.  Generally speaking, kink density deviates {\em downwards} from $\bar n \propto 1/\sqrt{t_a}$, before plateauing at a kink density roughly proportional to $1/\beta$, then again trending downward towards $\bar n \propto 1/\sqrt{t_a}$.  This is different from our QA data, and similar to the situation seen in Ref.~\cite{Heim2015} Fig.~3A, with the main distinctions being that their results are in the setting of Gaussian 2D spin glasses and for $M=64$ Trotter slices, away from the continuous-time limit. The kink-kink correlator results are also different from our QA data (Fig.~\ref{fig:3} of the main text): there is no positive correlator peak except very weakly at $\beta=8$, and at high $\beta$ values the correlator develops a sharply rising positive tail with decreasing normalized distance, which is not seen in our QA data either.

\subsection{SVMC-TF}

Spin-vector Monte Carlo (SVMC) \cite{Shin2014} is a semiclassical model of quantum annealing wherein qubits are modeled by two-dimensional rotors and dynamics proceed via Metropolis-Hastings updates. The model has had mixed success in reproducing results from previous QA experiments \cite{q-sig2,Yamashiro:2019aa}. SVMC-TF \cite{Albash2021} is an improved variant of this model with ``transverse field'' updates, in which proposed update angles are chosen from a region around a rotor's current angle, and the region's width depends on $\Gamma(s)/(\mathcal J(s)|J|)$.  To test the ability of SVMC-TF to mimic the behavior of coherent quantum annealing, we ran SVMC-TF for a range of anneal lengths using the QA schedule.  We set the inverse temperature to $\beta=32$, meaning that at the quantum critical point (QCP) we have $\Gamma(s_c)=\mathcal J(s_c)|J|=32$ in terms of dimensionless temperature.

In Fig.~\ref{fig:mcmc_temperatures}c--d we show kink density scaling and kink-kink correlators for a range of inverse temperatures in SVMC-TF, using the QA schedule for $J=1.4$ normalized at the QCP, as in SQA-PIMC.  The results confirm that we are running SVMC-TF in the low-temperature limit, and do not expect significant changes to scaling for even lower temperatures.  For significantly smaller $\beta$, the positive peak in the normalized kink-kink correlator $C^{\textrm{KK}}_r$ disappears.

\begin{figure}
  \includegraphics[width=0.7\linewidth]{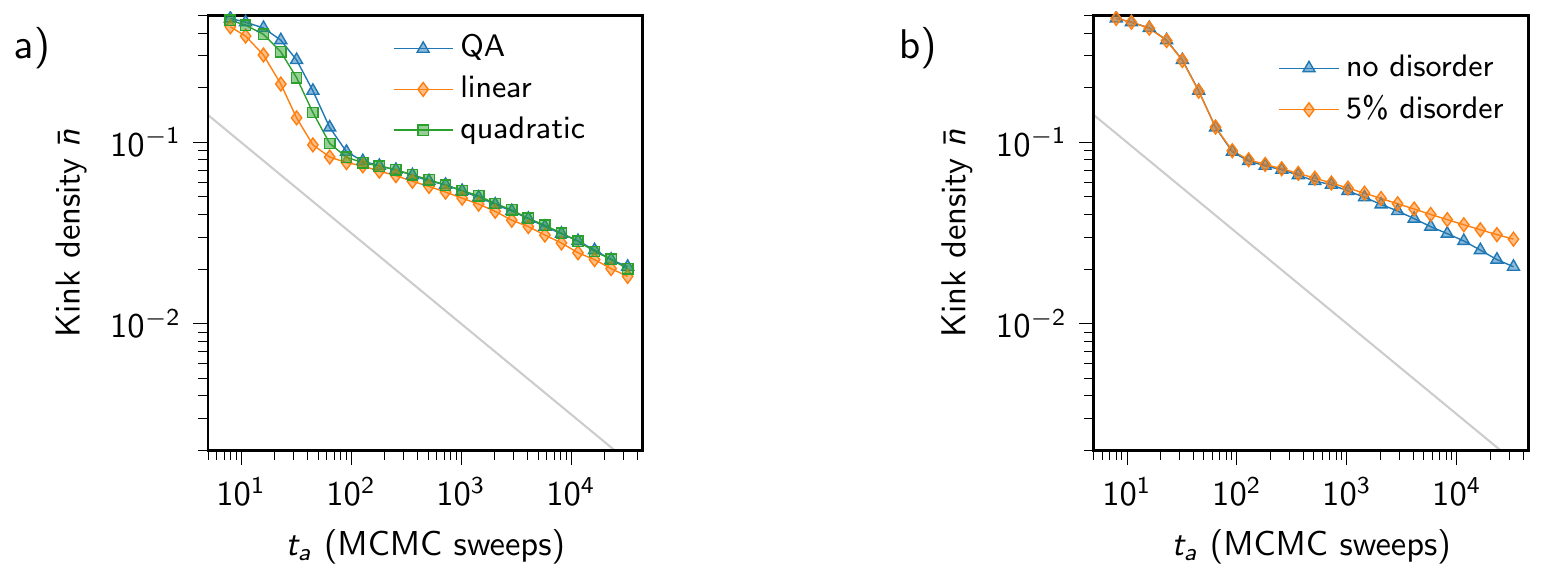}
  \caption{{\bf Effect of annealing schedule shape and disorder on SVMC-TF kink density scaling.}  {\bf a}, We probe kink densities for linear and quadratic annealing schedules, in addition to the $J=1.4$ QA schedule, at $\beta=32$. {\bf b}, We probe the addition of disorder to the QA schedule and its effect on kink density scaling.}\label{fig:svmctf_density_comparison}
\end{figure}

SVMC-TF is the only one of the three MCMC methods investigated that shows a positive peak in kink-kink correlation.  We are therefore interested in probing the robustness of the kink density scaling, which deviates significantly from the quantum KZM.  We test adjustments to the annealing schedule and the addition of disorder to individual $h$ and $J$ terms, with results shown in Fig.~\ref{fig:svmctf_density_comparison}.  In Fig.~\ref{fig:svmctf_density_comparison}a we test three schedules.  In addition to the QA schedule for $J=1.4$ (see Fig.~\ref{fig:schedules}), we test two others: a linear schedule, for which $\Gamma(s)=\beta(1-s)$ and $\mathcal J(s)|J|=\beta \cdot s$ and a quadratic schedule, for which $\Gamma(s)=4\beta(1-s)^2$ and $\mathcal J(s)|J| = 4\beta s^2$.  Note that all schedules are normalized so the crossing of $\Gamma(s)$ and $\mathcal J(s)|J|$ occurs at $1$. In Fig.~\ref{fig:svmctf_density_comparison}b we test the effect of adding disorder. Neither modifications to the schedule nor Hamiltonian disorder improve the agreement in kink density scaling between SVMC-TF and the quantum model.

\section{Analytical solutions}\label{app:conversion}

Previous theoretical papers \cite{Nowak2021,Zurek2005,Dziarmaga2005,DelCampo2018} have studied quantum spin chains using the fermionization technique and obtained analytical solutions for kink density statistics and the Landau-Zener probability.
To use those results we need to convert our Hamiltonian into the form used in those papers.

Hamiltonian (\ref{eq:ham}) can be written as
\begin{equation}
  H(s) = -\sum_i \bigg( \Gamma(s)\sigma_i^x  - \mathcal J(s)  J \sigma_i^z\sigma_{i+1}^z \bigg),
\end{equation}
where $s=t/t_a$, with $t\in [0,t_a]$ being time and $t_a$ being the annealing time. For simplicity, we assume  $\mathcal J(s)  J<0$, i.e., ferromagnetic coupling. 
This is to be compared to the dimensionless Hamiltonian (e.g., Eq.~(5) in Ref.~\cite{Nowak2021}):
\begin{equation}\label{eq:Nowak}
\tilde  H = -\sum_i (g \sigma_i^x  +  \sigma_i^z\sigma_{i{+}1}^z).
\end{equation}
The parameter $g$ is given as
\begin{equation} \label{eq:19}
g = -{\frac{\tilde t}{\tau_Q}},
\end{equation}
where $\tilde t \in (-\infty, 0]$ is the dimensionless time and $\tau_Q$ characterizes the time it takes to traverse the critical point.  Our goal is to find $\tau_Q$ in terms of the Hamiltonian parameters in Eq.~\eqref{eq:ham}, which we rewrite as
\begin{equation} \label{e1}
  H(s) =   \mathcal J(s) J \sum_i \bigg( \big[-\Gamma(s)/\mathcal J(s) J \big]\sigma_i^x + \sigma_i^z\sigma_{i{+}1}^z \bigg).
\end{equation}
Comparing with \eqref{eq:Nowak}, we define
\begin{equation}
g(s) = {\Gamma(s) \over |\mathcal J(s) J|},
\end{equation}
and \eqref{e1} becomes
\begin{equation} \label{e2}
  H(s) =   -|\mathcal J(s) J| \sum_i \big[g(s) \sigma_i^x  +  \sigma_i^z\sigma_{i{+}1}^z \big].
\end{equation}
Hamiltonian \eqref{eq:Nowak} is dimensionless, with Schr\"odinger equation
\begin{equation} \label{e3}
i {d \over d\tilde t} \ket{\psi(\tilde t)} =  -\sum_i (g \sigma_i^x  +  \sigma_i^z\sigma_{i{+}1}^z) \ket{\psi(\tilde t)}.
\end{equation}
Hamiltonian \eqref{e2}, on the other hand, is dimensionful with an overall energy scale $|\mathcal J(s) J|$ that is time dependent. The corresponding Schr\"odinger equation reads
\begin{equation} \label{seq}
i \hbar {d \over d t} \ket{\psi(t)} =  -|\mathcal J(s) J| \sum_i \big[g(s) \sigma_i^x  +  \sigma_i^z\sigma_{i{+}1}^z \big]\ket{\psi(t)}.
\end{equation}
 
We would like the two Hamiltonians to lead to the same dynamics near the critical point $s = s_c$ defined by
 \begin{equation} \label{CPoint}
 \Gamma(s_c) = \mathcal J(s_c) |J|.
 \end{equation}
We therefore expand the schedule near the critical point:
\begin{eqnarray}
\Gamma(s) \!\!&=&\!\! \Gamma(s_c) +  \Gamma'(s_c) (s - s_c) 
\label{expn1}\\
\mathcal J(s) \!\!&=&\!\! \mathcal J(s_c) +  \mathcal J'(s_c) (s - s_c).
\label{expn2}
\end{eqnarray}
We also linearly expand $g(s)$ as
\begin{equation} \label{e5}
g(s) = {\Gamma(s_c) \over |\mathcal J(s_c) J|} +  {\Gamma'(s_c)\mathcal J(s_c) -  \Gamma(s_c)\mathcal J'(s_c) \over \mathcal J^2(s_c) |J|} (s - s_c).
\end{equation}
Let us write \eqref{seq} as
\begin{equation}
{i\hbar \over |\mathcal J(t) J|} {d \over d t} \ket{\psi(t)} =  - \sum_i \big[g(t) \sigma_i^x  +  \sigma_i^z\sigma_{i{+}1}^z \big]\ket{\psi(t)}.
\end{equation}
For this to agree with \eqref{e3}, we need
\begin{equation}
\tilde t = {1\over \hbar}\int^t |\mathcal J(t') J| dt' = {t_a |J|  \over \hbar}\int^s \mathcal J(s') ds'. \label{ttlde}
\end{equation}
Substituting \eqref{expn2} into \eqref{ttlde}, near the critical point $\tilde t_c$ we obtain the linear expansion
\begin{equation} \label{e4}
\tilde t = \tilde t_c + {t_a |J|  \over \hbar}\mathcal J(s_c)(s - s_c).
\end{equation}
Substituting \eqref{e4} into \eqref{eq:19} and equating the coefficient of $s$ with that in \eqref{e5}, we get
\begin{equation}
-{t_a |J| \over \hbar \tau_Q}\mathcal J(s_c)  = {\Gamma'(s_c)\mathcal J(s_c) -  \Gamma(s_c)\mathcal J'(s_c) \over \mathcal J^2(s_c) |J|}.
\end{equation}
Solving for $\tau_Q$ and using \eqref{CPoint}, we obtain
\begin{equation}\label{eq:tauq}
\tau_Q = {\Gamma(s_c) t_a/\hbar \over  \mathcal J'(s_c)/\mathcal J(s_c)  - \Gamma'(s_c)/\Gamma(s_c)} = b \, t_a.
\end{equation}
where
\begin{equation}\label{eq:bsolution}
b = {\Gamma(s_c) /\hbar \over  \mathcal J'(s_c)/\mathcal J(s_c)  - \Gamma'(s_c)/\Gamma(s_c)}.
\end{equation}
Equation \eqref{eq:tauq} is used to generate the data given as ``coherent theory'' in Fig.~\ref{fig:2}, with \cite{Dziarmaga2005}
\begin{equation}\label{eq:tauq2}
    \bar n  = \frac{1}{2\pi\sqrt{2\tau_Q}}  = \frac{t_a^{-1/2}}{2\pi\sqrt{2b}} .
\end{equation}

For small chains the minimum gap is large and excitation is dominated by a single Landau-Zener transition, with probability \cite{Dziarmaga2005}
\begin{equation}
1- P_{GS} \approx e^{-2\pi^3\tau_Q/L^{2}} = e^{-a t_a},
\end{equation} 
where
\begin{equation}\label{eq:lzmodel}
a = 2\pi^3b/L^{2}. 
\end{equation}

\section{Efficient simulation by fermionization}\label{fermionization}

At zero temperature and in the absence of coupling to an environment, the uniform and disordered transverse-field Ising models (\ref{eq:ham}) are efficiently diagonalizable, and annealing dynamics can be analyzed \cite{Dziarmaga2005,Dziarmaga2006,Caneva2007}. These three papers cover the numerical methods employed.  Aside from some details concerning the specific statistics evaluated, we only give a brief overview of these methods in this appendix. For the general case we might write our Hamiltonian of interest as
\begin{equation}
  H = - \sum_i \gamma_i \sigma_i^x + \sum_{i} J_i \sigma_i^z\sigma_{i+1}^z\;.
\end{equation}
When translated to a fermionic model via the Jordan-Wigner transform $\sigma^z_i=-(a_i + a_i^\dagger) \prod_{l<i}[1-2a_l^\dagger a_l] $ and $\sigma^x_i=2a_i^\dagger a_i - 1$ \cite{Jordan1928}, and restricting the basis for consistency with the initial condition (so called anti-periodic boundary conditions) in quantum annealing, one obtains
\begin{equation}
  H = - \sum_i \gamma_i (2 a_i^\dagger a_i - 1) + \sum_{i} J_i (a_{i}^\dagger - a_{i}) (a_{i+1}^\dagger + a_{i+1}),
\end{equation}
where $a$ and $a^\dagger$ are the standard annihilation and creation fermionic operators respectively.

For the disordered case we can simulate dynamics as evolution of complex coefficients, exploiting the Bogoliubov transformation of $a_i$ in the Heisenberg representation:
\begin{equation}
  a_i(s) = \sum_{m} u_{im}(s) b_m + v_{im}^*(s) b_m^\dagger,
\end{equation}
where $b$ and $b^\dagger$ are time-independent fermionic operators that diagonalize the Hamiltonian at the beginning of the anneal ($J_i=0$); $s$ is the normalized anneal duration.
The Heisenberg equation for $a_i(s)$ leads to the Bogoliubov-de Gennes equation of
the complex coefficients $u_{im}$ and $v_{im}$:
\begin{eqnarray}
 \mathrm{i}\hbar\frac{d}{ds}u_{im}(s) &=& 
  t_a\sum_{j}\left(A_{ij}u_{jm}(s) + B_{ij}v_{jm}(s)\right) ,\\
 \mathrm{i}\hbar\frac{d}{ds}v_{im}^*(s) &=&
  t_a\sum_{j}\left(A_{ij}v_{jm}^*(s) + B_{ij}u_{jm}^*(s)\right) ,
\end{eqnarray}
where $A_{ij} = -2\Gamma_i\delta_{i,j} + J_i\delta_{j,i+1} + J_{i-1}\delta_{j,i-1}$, $B_{ij} = J_i\delta_{j,i+1} - J_{i-1}\delta_{j,i-1}$ and $A_{L 1} = A_{1 L} = B_{L 1} = - B_{1 L} = -J_L$. The initial condition for
$u_{im}(s)$ and $v_{im}(s)$ is given so that $H(s=0)$ is diagonalized by $b_{m}$
and $b_{m}^{\dagger}$.

Our main interest in this paper is the evaluation of kink statistics for a smooth evolution of the Hamiltonian with $\gamma_i = \Gamma(s)$ and $J_i = +\mathcal{J}(s) J$ per (\ref{eq:ham}). We also consider the impact of quenched disorder in $J$ as shown in Fig.~\ref{fig:3}; disorder in the transverse field is not thought to explain a dominant portion of the deviation between experiment and theory.  The kink operator at site $i$ is $K_i=(1+\mathrm{sign}(J_i)\sigma^z_{i}\sigma^z_{i+1})/2$. For a chain of $L$ couplers, with periodic boundary conditions as studied in this paper, the operator yielding the density of kinks is $n = \frac{1}{L}\sum_i K_i$. The full kink-kink correlation operator at distance $r$ is given by the operator $\chi_r = \frac{1}{L}\sum_{i} K_i K_{i+r}$. All of these can be efficiently represented via the Jordan-Wigner transformation.
In practice, assuming the ground state of $H(0)$ as the initial state, the expectation values of
$\sigma^z_i\sigma^z_{i+1}$ and $\sigma^z_i\sigma^z_{i+1}\sigma^z_{i+r}\sigma^z_{i+r+1}$ 
at time $s$ are given by
\begin{equation}
 \langle\psi(s)|\sigma^z_i\sigma^z_{i+1}|\psi(s)\rangle = \eta_i
  \left[Q(s)P^{\dagger}(s)\right]_{i\, i+1} ,
\end{equation}
\begin{eqnarray}
 \langle\psi(s)|\sigma^z_i\sigma^z_{i+1}\sigma^z_{i+r}\sigma^z_{i+r+1}|\psi(s)\rangle &=& 
 \eta_i\eta_{i+r}\left[ - \left[Q(s)P^{\dagger}(s)\right]_{i\, i+r+1}
  \left[P(s)Q^{\dagger}(s)\right]_{i+1\, i+r} \right.\nonumber\\
 &&+ \left[Q(s)Q^{\dagger}(s)\right]_{i\, i+r}
  \left[P(s)P^{\dagger}(s)\right]_{i+1\, i+r+1} \nonumber\\
 &&+ \left.\left[Q(s)P^{\dagger}(s)\right]_{i\, i+1} 
  \left[Q(s)P^{\dagger}(s)\right]_{i+r\, i+r+1} \right],
\end{eqnarray}
where we define $P(s) = u(s) + v(s)$, $Q(s) = v(s) - u(s)$,
$\eta_i = 1$ for $i = 1,2,\cdots, L-1$, and $\eta_L = -1$. 
We assume the periodicity of the site index, i.e., $i+L = i$.

For the pure model, or per realization of disorder, we determine the kink rate as ${\bar n} = \langle \psi(s=1)| n |\psi(s=1)\rangle$. ${\bar n}$ can be compared to experimentally evaluated quantity  (\ref{eq:kinkdensity}). Similarly we define the kink-kink correlator as
\begin{equation}
C_r^{\textrm{KK}} = \frac{\langle \psi(s=1)| \chi_r |\psi(s=1)\rangle - {\bar n}^2}{{\bar n}^2}\;, \label{eq:Cd}
\end{equation}
which is to be compared to the experimentally determined quantity (\ref{eq:correlator}). It should be noted that it is sufficient to approximate the $s=1$ result by annealing to the final calibrated schedule point (at $s<1$), as shown in Fig.~\ref{fig:schedules}; dynamics at larger $s$ do not lead to evolution of the statistics. Disorder averages of each quantity, after inclusion of any additional normalization in (\ref{eq:Cd}), are taken by simulating many instances drawn from the disordered distribution (see Appendix \ref{app:disorder}). Disorder in either the transverse field or couplings can be accommodated by the fermionized model, whereas disorder in the longitudinal field can only be modeled using tensor network methods as described in Appendix \ref{TEBD}. 

We can efficiently simulate transverse-field Ising model kink statistics and ground state probability throughout the annealing dynamics, including cases of quenched transverse-field and coupler disorder. These results are found to be consistent with TEBD (Appendix \ref{TEBD}), a method relying on a controlled bond-dimension approximation, which is typically slower but allows generalization for inclusion of longitudinal field perturbations either in the form of quenched noise or a bosonic bath at finite or zero temperature.

\section{Tensor network methods: Time-evolution block decimation}
\label{TEBD}
Matrix product states are an efficient means to represent and evolve quantum Ising spin states subject to local interactions, and are closely tied to density matrix renormalization group methods, which extend insights to the large system limit and phase transition phenomena~\cite{Schollwock2011,White1992,White1993}. In these frameworks quantum states are represented efficiently by a sequence of tensors, and system symmetries are exploited when possible.

The evolution of quantum states from a prepared initial condition can involve the growth of quantum correlations.  Using, for example, time-evolution block decimation (TEBD) \cite{Vidal2004,White2004}, the tensor size required for exact dynamics can grow to a scale exponential in the system size. However, it is understood that quantum correlations can be captured by finite bond dimension in gapped one-dimensional systems, and as such it is possible to represent quantum states efficiently by bounding the size of the tensors \cite{Schollwock2011}. The parameter bounding the size of the tensors is called the bond dimension $D$, and the complexity of a TEBD algorithm scales with this bond dimension. We simulate finite systems where the gap is finite throughout the anneal, but potentially very small. Nevertheless, we can observe a convergence of statistics at practically accessible bond dimensions.

Matrix product state and density matrix renormalization group methods have recently been employed for the modeling of annealing dynamics in one dimensional systems consisting of a transverse-field Ising model coupled to a bosonic bath at both zero and finite temperatures, for both finite systems and the infinite system limit \cite{Oshiyama2020,Suzuki2007,Suzuki2019,Fitzpatrick2021,Spinbosonchain}. Coupling to the bosonic bath is handled by use of a quasi-adiabatic path integral (QUAPI) method \cite{Makri:1992wt,Strathearn2018}. Bosonic degrees of freedom are integrated out and have the effect of inducing a memory in the system, which must itself be controlled with a second time parameter, as with bond dimension (which can also be bounded at practically accessible scales).  This, in combination with an extension of TEBD called infinite-TEBD (iTEBD), was recently used to simulate the quantum Ising chain \cite{Oshiyama2020, Bando2020}. Statistics may be evaluated at the end of the anneal per the discussion of Appendix \ref{fermionization}, where the disorder average is taken where necessary by an average over sampled instances.

\subsection{Simulation without disorder}\label{app:nodisorder}

\begin{figure}[t]
 \includegraphics[width=12cm, bb = 0 0 559.705 109.674]{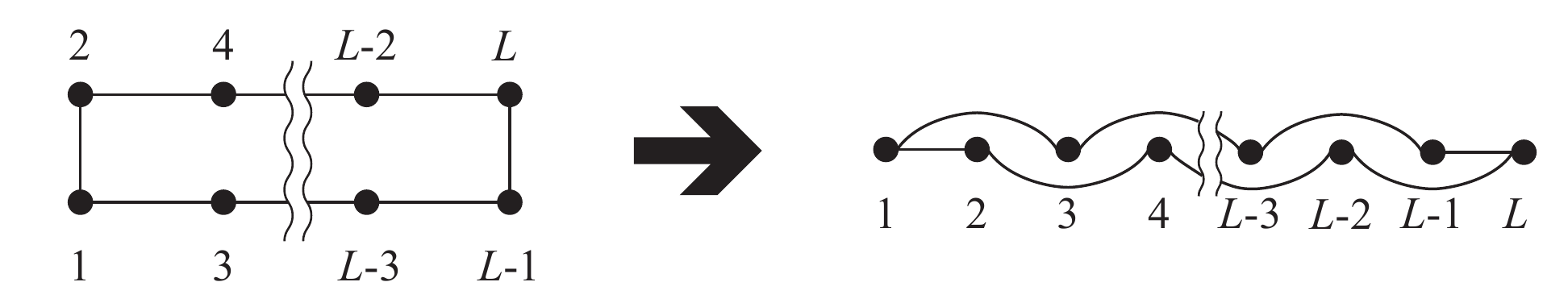}
\caption{Mapping of a periodic chain to a linear chain. Dots and solid lines show spins and bonds between spins, respectively. A periodic chain with nearest neighbor interactions is mapped to a linear chain with next nearest neighbor interactions and nearest neighbor interactions at both edges.}
\label{fig:periodic_chain}
\end{figure}

Let us consider the periodic quantum Ising chain. To apply TEBD, we map the periodic chain to a linear chain with next nearest neighbor interactions and nearest neighbor interactions at both edges, as shown in Fig.~\ref{fig:periodic_chain}. The corresponding Hamiltonian for the linear chain is given by
\begin{equation}
 H(s) = - \Gamma(s)\left(\sum_{i=1}^L \sigma^x_i\right) + \mathcal{J}(s) J \left(\sigma^z_1\sigma^z_2 + \sum_{i=1}^{L-2}\sigma^z_{i}\sigma^z_{i+2} + \sigma^z_{L-1}\sigma^z_L\right) 
  = h_{12}(s) + \sum_{i=1}^{L-2}h_{i i+2}(s) + h_{L-1 L}(s),
\end{equation}
where $h_{ij}$ is defined, for convenience, as
\begin{equation}
 h_{ij}(s) = - \Gamma(s) \frac{1}{2}\left(\sigma^x_i + \sigma^x_j\right) + \mathcal{J}(s)J\sigma^z_i\sigma^z_j .
\end{equation}

In order to simulate the unitary time evolution of a quantum Ising chain within the framework of matrix product states, we discretize the time as $t = l\mathit{\Delta}t$ with a small time width 
$\mathit{\Delta}t$ and decompose the time-evolution operator from $t=0$ to $t=t_a=M\mathit{\Delta}t$
into a product of Trotter slices:
\begin{equation}
 \mathcal{U}(t_a,0) \approx e^{-i H((M-1)\mathit{\Delta}t/t_a)\mathit{\Delta}t}\cdots
e^{-i H(\mathit{\Delta}t/t_a)\mathit{\Delta}t}e^{-i H(0)\mathit{\Delta}t} .
\end{equation}
Each Trotter slice is decomposed into a product of local operators:
\begin{equation}
 e^{-i H(l\mathit{\Delta}t/t_a)\mathit{\Delta}t} \approx
\mathcal{U}_{12}\mathcal{U}_{13}\cdots\mathcal{U}_{L-2 L}
\mathcal{U}_{L-1 L}^2\mathcal{U}_{L-2 L}
\cdots\mathcal{U}_{13}\mathcal{U}_{12} ,
\label{eq:symm_decomp}
\end{equation}
where $\mathcal{U}_{ij} \equiv e^{-ih_{ij}(l\mathit{\Delta}t/t_a)\frac{\mathit{\Delta}t}{2}}$.
Note that the $(l\mathit{\Delta}t/t_a)$-dependence in $\mathcal{U}_{ij}$ is 
omitted to simplify the notation.
We consider the time evolution of the state from $|\psi\rangle$ to
$|\psi'\rangle =  e^{-i H(l\mathit{\Delta}t/t_a)\mathit{\Delta}t}|\psi\rangle$ by 
using Eq.~(\ref{eq:symm_decomp}). We apply the local operators $\mathcal{U}_{ij}$ one by one from the rightmost $\mathcal{U}_{12}$ to the leftmost $\mathcal{U}_{12}$. The right and left half parts of Eq.~(\ref{eq:symm_decomp}) lead the right- and left-ward moves, respectively. For the right half
part, let
\begin{equation}
 |\psi^{(1)}_{\to}\rangle = \mathcal{U}_{13}\mathcal{U}_{12}|\psi\rangle, ~~
  |\psi^{(k)}_{\to}\rangle = \mathcal{U}_{k k+2}|\psi^{(k-1)}_{\to}\rangle ~~ \mbox{for} ~~ 
  k=2,3,\cdots,L-2, ~~ \mbox{and} ~~
  |\psi^{(L-1)}_{\to}\rangle = \mathcal{U}_{L-1 L}|\psi^{(L-2)}_{\to}\rangle .
\end{equation}
and for the left half part, 
\begin{equation}
 |\psi^{(L)}_{\leftarrow}\rangle = \mathcal{U}_{L-1 L}|\psi^{(L-1)}_{\to}\rangle, ~~
  |\psi^{(k)}_{\leftarrow}\rangle = \mathcal{U}_{k-1 k+1}|\psi^{(k+1)}_{\leftarrow}\rangle ~~ \mbox{for} ~~~
  k = 2,3,\cdots,L-1, ~~ \mbox{and} ~~ 
  |\psi'\rangle = \mathcal{U}_{12}|\psi^{(2)}_{\leftarrow}\rangle ,
\end{equation}
where the arrows indicate a right or left move.

We now introduce a matrix product representation of $|\psi\rangle$. 
Using the computational basis
$|\sigma_1\sigma_2\cdots\sigma_L\rangle$ diagonalizing
$\sigma^z_i$ $\forall i$, we write 
$|\psi\rangle = \sum_{\sigma_1,\sigma_2,\cdots,\sigma_L} 
\psi_{\sigma_1\sigma_2\cdots\sigma_L}
|\sigma_1\sigma_2\cdots\sigma_L\rangle$ with
\begin{equation}
 \psi_{\sigma_1\sigma_2\cdots\sigma_L}
  = \sum_{q_2,\cdots,q_{L-1}} 
  \psi_{\sigma_1 q_2}v^{(2)}_{q_2;\sigma_2q_3}v^{(3)}_{q_3;\sigma_3q_4}\cdots 
v^{(L-1)}_{q_{L-1};\sigma_{L-1}\sigma_L} .
\label{eq:psi_MPS}
\end{equation}
As mentioned above, the bond dimension $D$ corresponds to the size of matrices
$v^{(l)}_{q;\sigma q'}$ for a fixed $\sigma$ and can be kept finite 
even for $L\to\infty$ 
in the ground state of a one-dimensional gapped spin chain. 
On the basis of this expression, 
we apply the slice of time-evolution operator in Eq.~(\ref{eq:symm_decomp})
to $|\psi\rangle$. 
First, we focus on 
$|\psi^{(2)}_{\to}\rangle = \mathcal{U}_{24}|\psi^{(1)}_{\to}\rangle
= \mathcal{U}_{24}\mathcal{U}_{13}\mathcal{U}_{12}|\psi\rangle$. This can be wrtten as
\begin{equation}
 |\psi^{(2)}_{\to}\rangle =\sum_{\sigma_1,\sigma_2,\cdots,\sigma_L}\sum_{q_5,\cdots,q_{L-1}}
 \Psi^{(2)}_{\sigma_1\sigma_2\sigma_3\sigma_4q_5}
 v^{(5)}_{q_5;\sigma_5q_6}\cdots 
 v^{(L-1)}_{q_{L-1};\sigma_{L-1}\sigma_L}
 |\sigma_1\sigma_2\cdots\sigma_L\rangle ,
\label{eq:psi1_raw}
\end{equation}
where $\Psi^{(2)}$ is defined as
\begin{equation}
 \Psi^{(2)}_{\sigma_1\sigma_2\sigma_3\sigma_4q_5} :=
 \sum_{\sigma_1',\sigma_2',\sigma_3',\sigma_4'}\sum_{q_2,q_3,q_4}\left[
  \mathcal{U}_{24}\mathcal{U}_{13}\mathcal{U}_{12}
       \right]_{\sigma_1\sigma_2\sigma_3\sigma_4;\sigma_1'\sigma_2'\sigma_3'\sigma_4'}
 \psi_{\sigma_1'q_2}v^{(2)}_{q_2;\sigma_2'q_3}v^{(3)}_{q_3;\sigma_3'q_4}
 v^{(4)}_{q_4;\sigma_4'q_5} 
\end{equation}
with $[\mathcal{O}]_{\sigma_1\sigma_2\sigma_3\sigma_4;\sigma_1'\sigma_2'\sigma_3'\sigma_4'}$
denoting a matrix element of an operator $\mathcal{O}$.
Regarding $\Psi^{(2)}$ as a rectangular matrix with the column index $(\sigma_1,\sigma_2)$
and the row index $(\sigma_3,\sigma_4,q_5)$, the singular value decomposition of $\Psi^{(2)}$ 
yields
\begin{equation}
 \Psi^{(2)}_{\sigma_1\sigma_2\sigma_3\sigma_4q_5}
  = \sum_{p}u^{(2)}_{\sigma_1\sigma_2;p}\lambda^{(2)}_{p}v^{(2)}_{p;\sigma_3\sigma_4q_5},
\end{equation}
where $u^{(2)}$ and $v^{(2)}$ are unitary matrices, and $\lambda^{(2)}_p$ stands for the singular value. 
Letting $\psi^{(2)}_{p\sigma_3\sigma_4q_5} := \lambda^{(2)}_pv^{(2)}_{p;\sigma_3\sigma_4q_5}$, 
one obtains a matrix product representation of Eq.~(\ref{eq:psi1_raw}):
\begin{equation}
|\psi^{(2)}_{\to}\rangle =\sum_{\sigma_1,\sigma_2,\cdots,\sigma_L}\sum_{p_2,q_4,\cdots,q_{L-1}}
 u^{(2)}_{\sigma_1\sigma_2;p_2}\psi^{(2)}_{p_2\sigma_3\sigma_4q_5}
 v^{(5)}_{q_5;\sigma_5q_6}\cdots 
 v^{(L-1)}_{q_{L-1};\sigma_{L-1}\sigma_L}
 |\sigma_1\sigma_2\cdots\sigma_L\rangle .
\end{equation}
Repeating the application of $\mathcal{U}_{i i+2}$ and the singular
value decomposition, one obtains
a matrix product state of the form:
\begin{equation}
 |\psi^{(L-1)}_{\to}\rangle = \sum_{\sigma_1,\sigma_2,\cdots,\sigma_L}
  \sum_{p_2,p_3,\cdots,p_{L-2}}
  u^{(2)}_{\sigma_1\sigma_2;p_2}u^{(3)}_{p_2\sigma_3;p_3}
  \cdots u^{(L-2)}_{p_{L-3}\sigma_{L-2};p_{L-2}}
  \psi^{(L-2)}_{p_{L-2}\sigma_{L-1}\sigma_L} 
   |\sigma_1\sigma_2\cdots\sigma_L\rangle .
\end{equation}
The leftward move can be calculated using the same procedure. Repeating the application of $\mathcal{U}_{i-2\, i}$ and
the singular value decomposition, one arrives at
\begin{equation}
 |\psi'\rangle
  = \sum_{\sigma_1,\sigma_2,\cdots,\sigma_L}
  \sum_{q_3,q_4\cdots,q_{L-1}}\psi'_{\sigma_1\sigma_2q_3}
  v^{(3)}_{q_3;\sigma_3q_4}v^{(4)}_{q_4;\sigma_4q_5}\cdots
  v^{(L-1)}_{q_{L-1};\sigma_{L-1}\sigma_L}
  |\sigma_1\sigma_2\cdots\sigma_L\rangle ,
\end{equation}
where we note that 
$v^{(k)}_{q;\sigma q'}$ is a unitary matrix yielded by the singular value decomposition
and is different from the one in Eq.~(\ref{eq:psi_MPS}) in general.
We remark here that the range of $p$ and $q$ in the matrices
$u^{(k)}_{p\sigma;p'}$ and $v^{(k)}_{q;\sigma q'}$ is bounded by the bond dimension $D$ we provide in the simulation.
This means that, assuming a descending order of the singular values,
we keep at most the largest $D$ singular values to express the
wavefunction.

In the present work, we studied $L=256$ chains by the present TEBD method with the bond dimension
ranging from $D = 8$ to $32$. The width of the Trotter slice is fixed at
$\mathit{\Delta}t = 0.01$.

\subsection{Simulation with disorder}\label{app:disorder}

To add disorder to the Hamiltonian (\ref{eq:ham}) we first reformulate the Hamiltonian more explicitly as 
\begin{equation}\label{eq:hamh}
  H(s) = -\Gamma(s)\bigg(\sum_i \sigma_i^x \bigg) + \mathcal J(s) \bigg( \sum_{i,j} J_{ij}\sigma_i^z\sigma_j^z  + \sum_i h_i\sigma_i^z\bigg),
\end{equation}
where nominally (i.e., with no disorder) the coupling terms $J_{ij}$ are all identical and the local longitudinal field terms $h_i$ are all zero.  We apply a naive model of disorder in the classical part of the Hamiltonian, in which the terms are perturbed:
\begin{eqnarray}\label{eq:disorder}
    J_{ij} &\leftarrow& J + \mathcal N(0,\sigma)\\\label{eq:disorder2}
    h_{i} &\leftarrow& \mathcal N(0,\sigma)
\end{eqnarray}
where $\mathcal N(\mu,\sigma)$ is a Gaussian variable with mean $\mu$ and standard deviation $\sigma$.  In Fig.~\ref{fig:3} we used $J=-1.4$ and $\sigma=0.05$.  To model the same absolute level of noise with smaller coupling strength, for other values of $J$ we use $\sigma=0.05*1.4/J$. We thus expect disorder to increase with $J$ decreasing below $1.4$.

\begin{figure}
  \includegraphics[width=\linewidth]{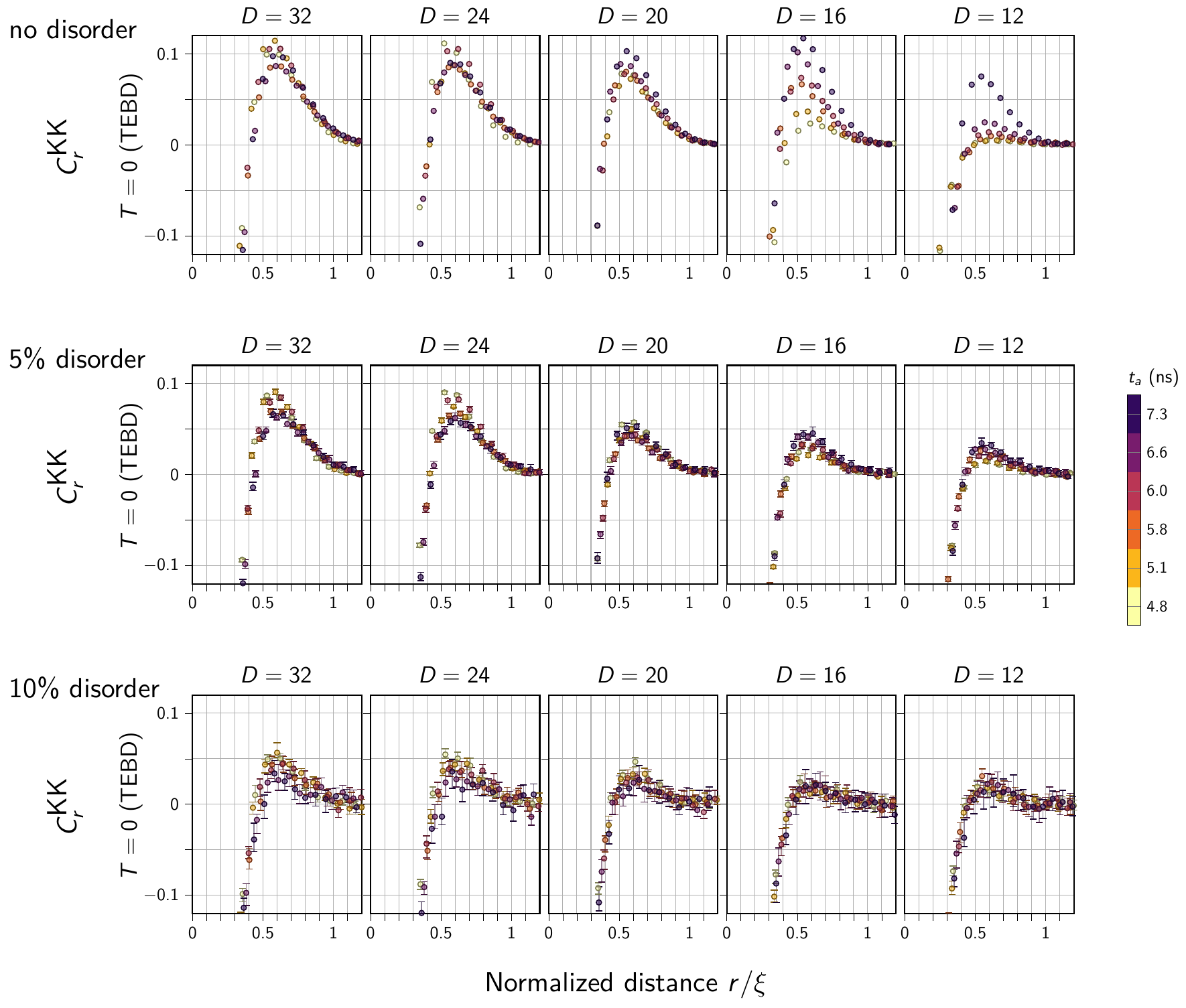}
  \caption{{\bf Kink-kink correlators for varying bond dimension.}  TEBD is run on systems with no disorder (top) and with $\sigma=0.05$ (middle) and $0.10$ (bottom) disorder added to $h$ and $J$ with $J=-1.4$; see Eqs.~(\ref{eq:disorder}) and (\ref{eq:disorder2}).  Bond dimension varies from $32$ to $12$.  As in QA experiments, error bars represent $95\%$ confidence intervals from a bootstrap across $300$ disorder realizations per experiment.}\label{fig:tebd_disorder_varybd}
\end{figure}

Fig.~\ref{fig:tebd_disorder_varybd} shows kink-kink correlators from TEBD with $5\%$ and $10\%$ disorder at $J=-1.4$, with varying bond dimension.

\begin{figure}
  \includegraphics[width=\linewidth]{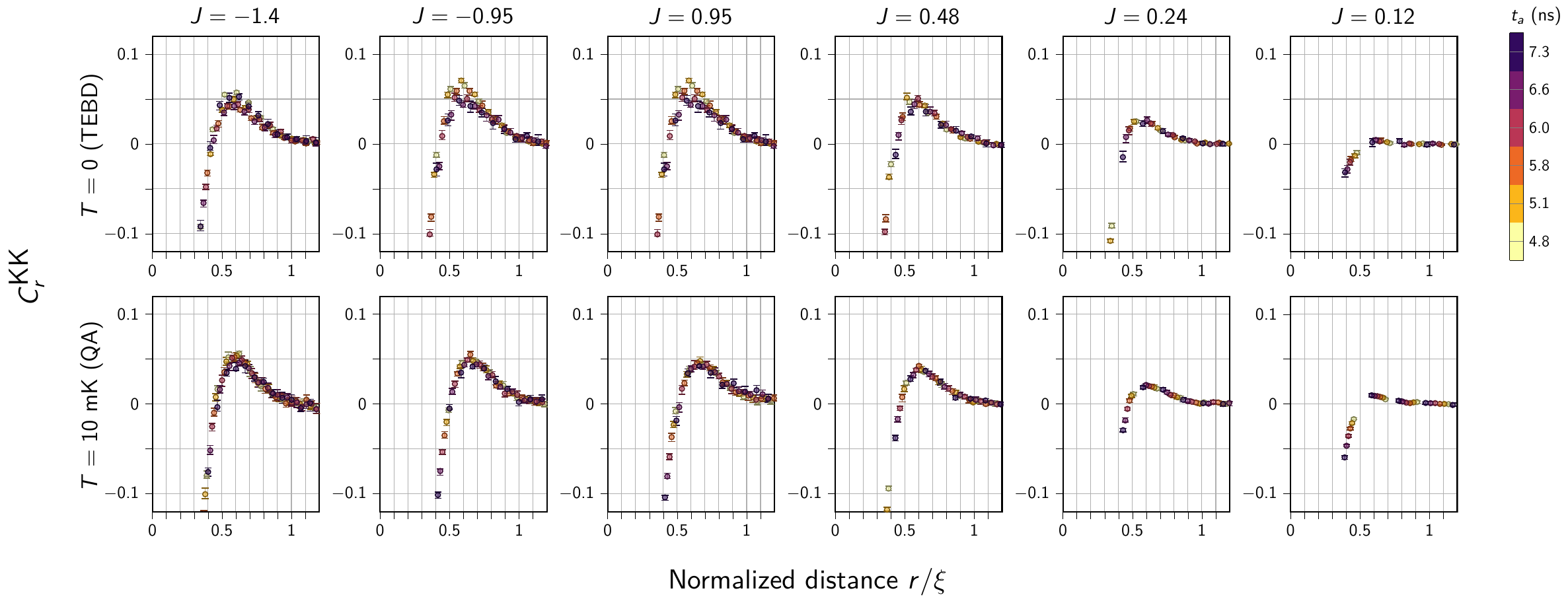}
  \caption{{\bf Kink-kink correlators for TEBD with disorder and for QA, with varying $J$.}  TEBD (top) is run on systems with disorder added to $h$ and $J$ ($\sigma=0.05\cdot 1.4/J$; see Eqs.~(\ref{eq:disorder}) and (\ref{eq:disorder2})).  Bond dimension is fixed at $D=20$.  QA (bottom) is run for varying $J$; compare with Fig.~\ref{fig:kkc_backmatter}.\label{fig:kkc_tebd_backmatter}}
\end{figure}

Fig.~\ref{fig:kkc_tebd_backmatter} shows kink-kink correlators generated using TEBD with these levels of disorder, with bond dimension $D=20$.  Although this is only a naive model of the effects of noise in the QA processor, it shows good agreement with QA experiments shown in Appendix \ref{sec:additionaldata}.

\section{Additional QA data}\label{sec:additionaldata}

All data shown in this section are for $L=512$.

\begin{figure}
  \includegraphics[width=\linewidth]{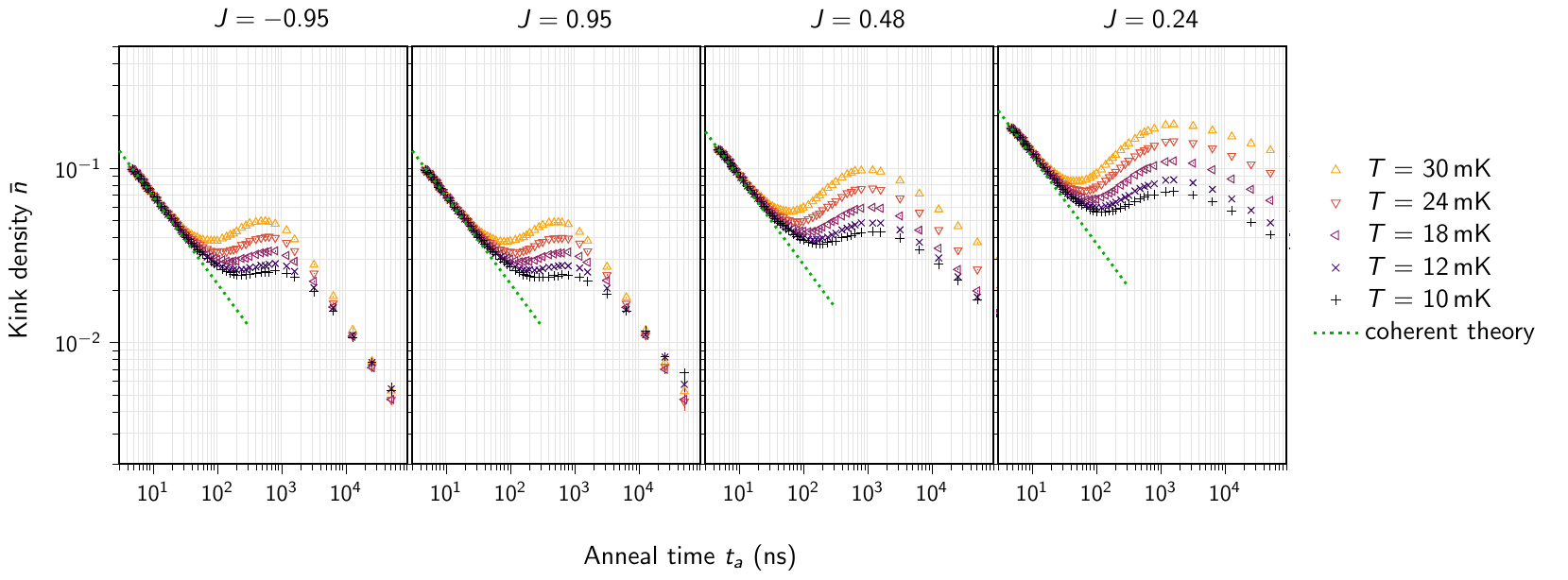}
  \caption{{\bf Kink densities for varying coupling strength.}  Extended data as in Fig.~\ref{fig:2}a are shown for additional values of $J$.}\label{fig:density_backmatter}
\end{figure}

\begin{figure}
  \includegraphics[width=.7\linewidth]{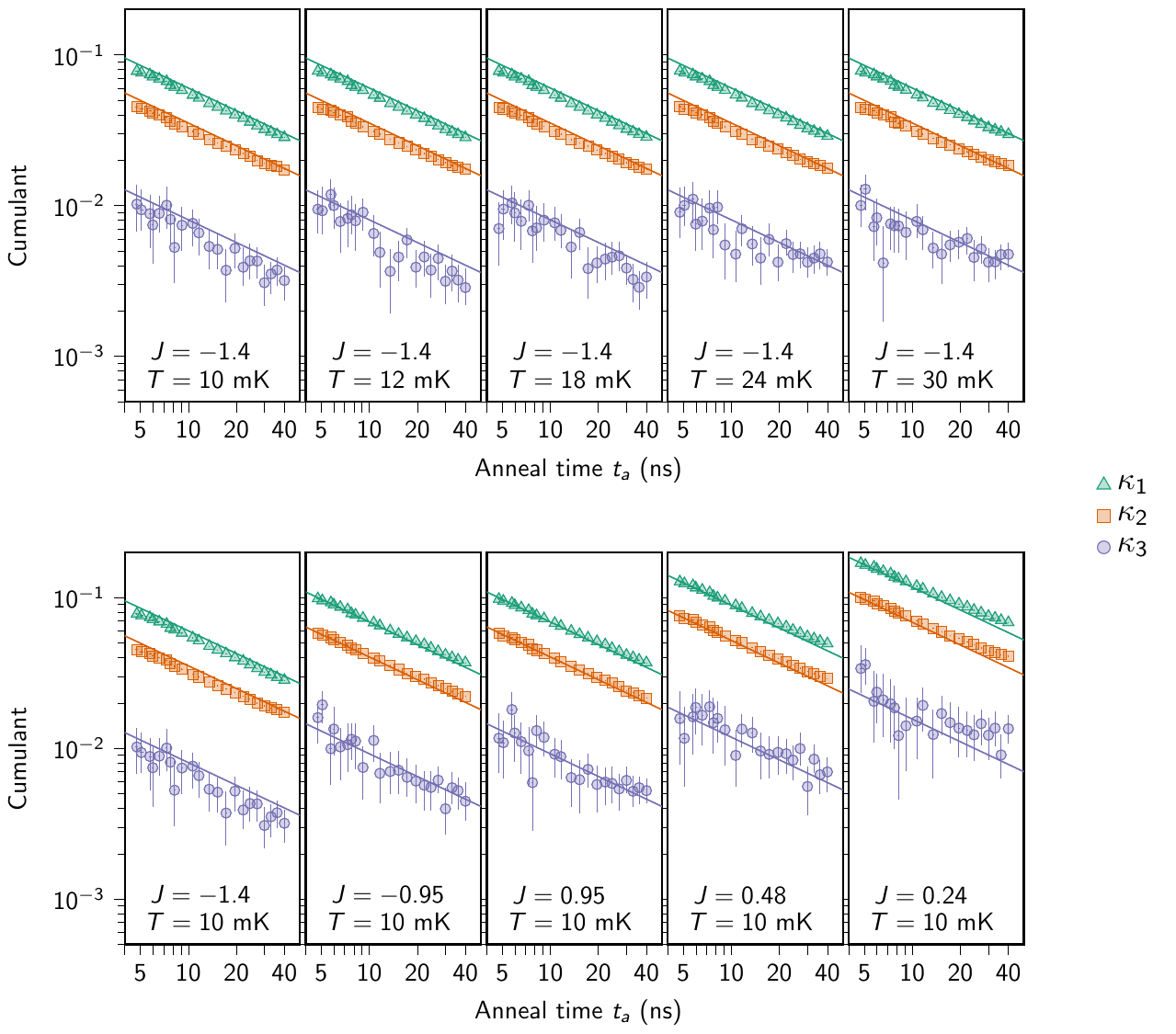}
  \caption{{\bf Cumulants of kink distribution.}  Data analogous to Fig.~\ref{fig:2}c are given for a sweep of $T$ for strong coupling ($J=-1.4$, top) and $J$ for low temperature ($T=10$, bottom). As in Fig.~\ref{fig:2}, lines are calculated from coherent theory.}\label{fig:cumulants_backmatter}
\end{figure}

\begin{figure}
  \includegraphics[width=\linewidth]{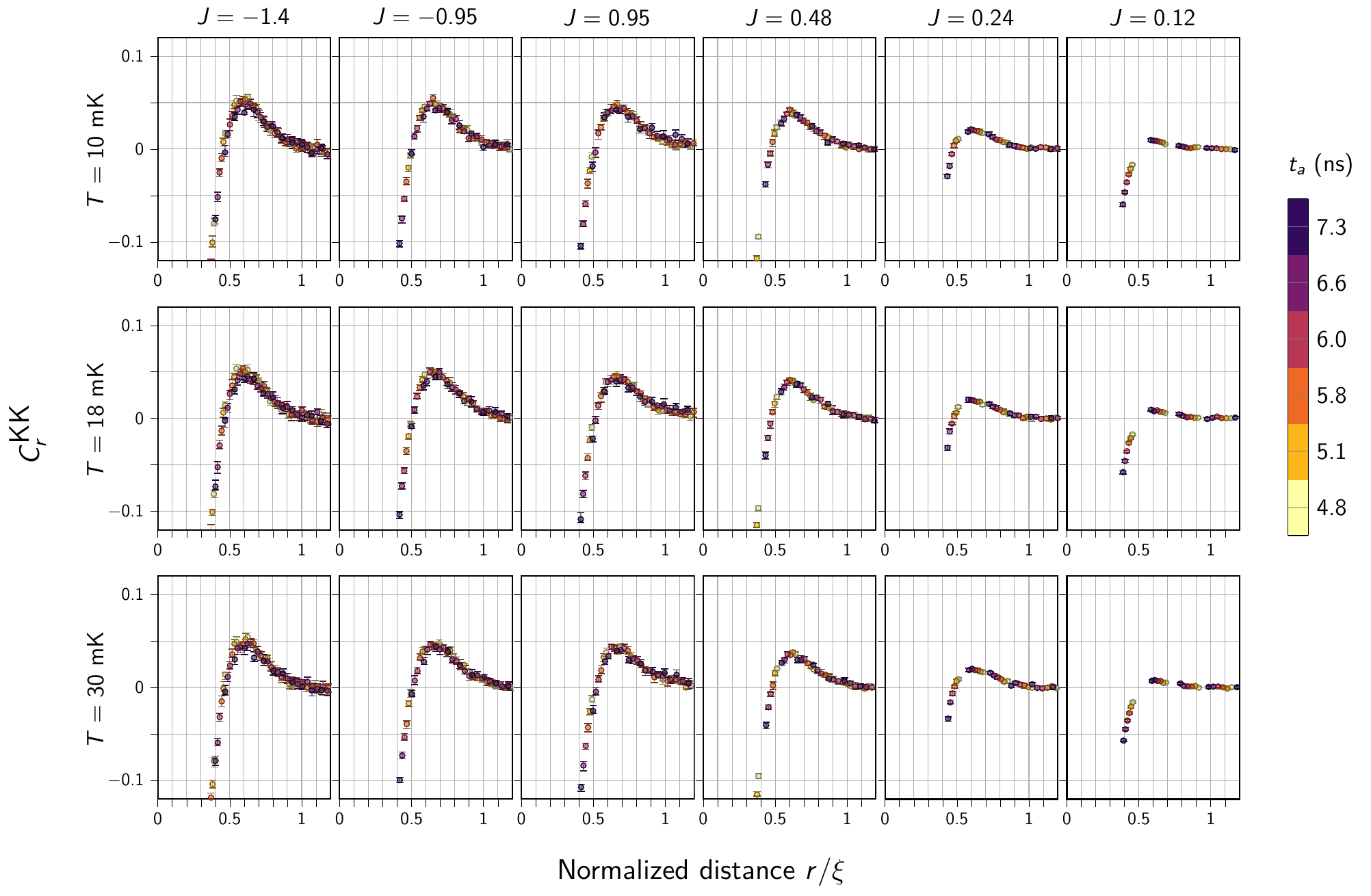}
  \caption{{\bf Additional kink-kink correlator data for QA.}  Temperatures between $\SI{10}{mK}$ and $\SI{30}{mK}$ are probed for multiple coupling values $J$.  In this regime of very fast anneals, temperature has no significant impact.  Low coupling strength, which is associated with increased relative disorder, suppresses the correlator peak.}\label{fig:kkc_backmatter}
\end{figure}

\begin{figure}
  \includegraphics[width=0.5\linewidth]{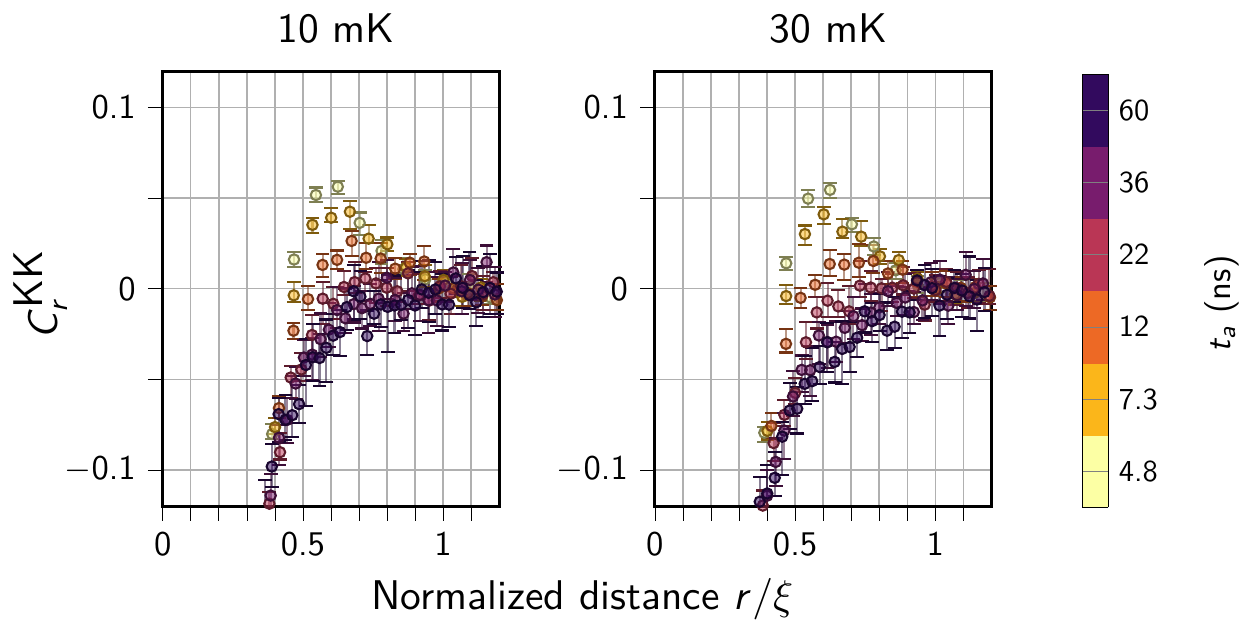}
  \caption{{\bf Kink-kink correlator data for QA at longer anneal times.}  Shown are correlators at anneal times up to $\SI{60}{ns}$ at temperatures $T=\SI{10}{mK}$ and $\SI{30}{mK}$, with $J=-1.4$.}\label{fig:kkc_longer}
\end{figure}

Fig.~\ref{fig:density_backmatter} shows kink densities $\bar n $ for a range of coupling strengths, temperatures, and anneal times.  Theoretical values for $T=0$ are calculated using Eq.~(\ref{eq:tauq2}).

Fig.~\ref{fig:cumulants_backmatter} shows the first three cumulants of the kink probability distribution, as in Fig.~\ref{fig:2}c, for a range of parameters.

Fig.~\ref{fig:kkc_backmatter} shows kink-kink correlators for QA at a range of $J$ and $T$, generalizing the data shown in Fig.~\ref{fig:3}.  Temperature has minimal impact on correlations, but coupling strength has a significant impact.  This suggests that disorder plays a major role in suppressing the correlator peak, while temperature does not.

Fig.~\ref{fig:kkc_longer}, with $J=-1.4$, shows correlators for longer anneal times, where the positive peak disappears.  Since temperature has very little impact on these results, it is unlikely that decoherence drives this peak suppression.  Among the likely contributing factors are disorder and diffusion.  Disorder will play an increasingly important role for slow quenches because correlation length grows.  Diffusion is likely to occur late in the anneal, between the critical point and the freezing of qubit dynamics. 

\begin{figure}
  \includegraphics[width=\linewidth]{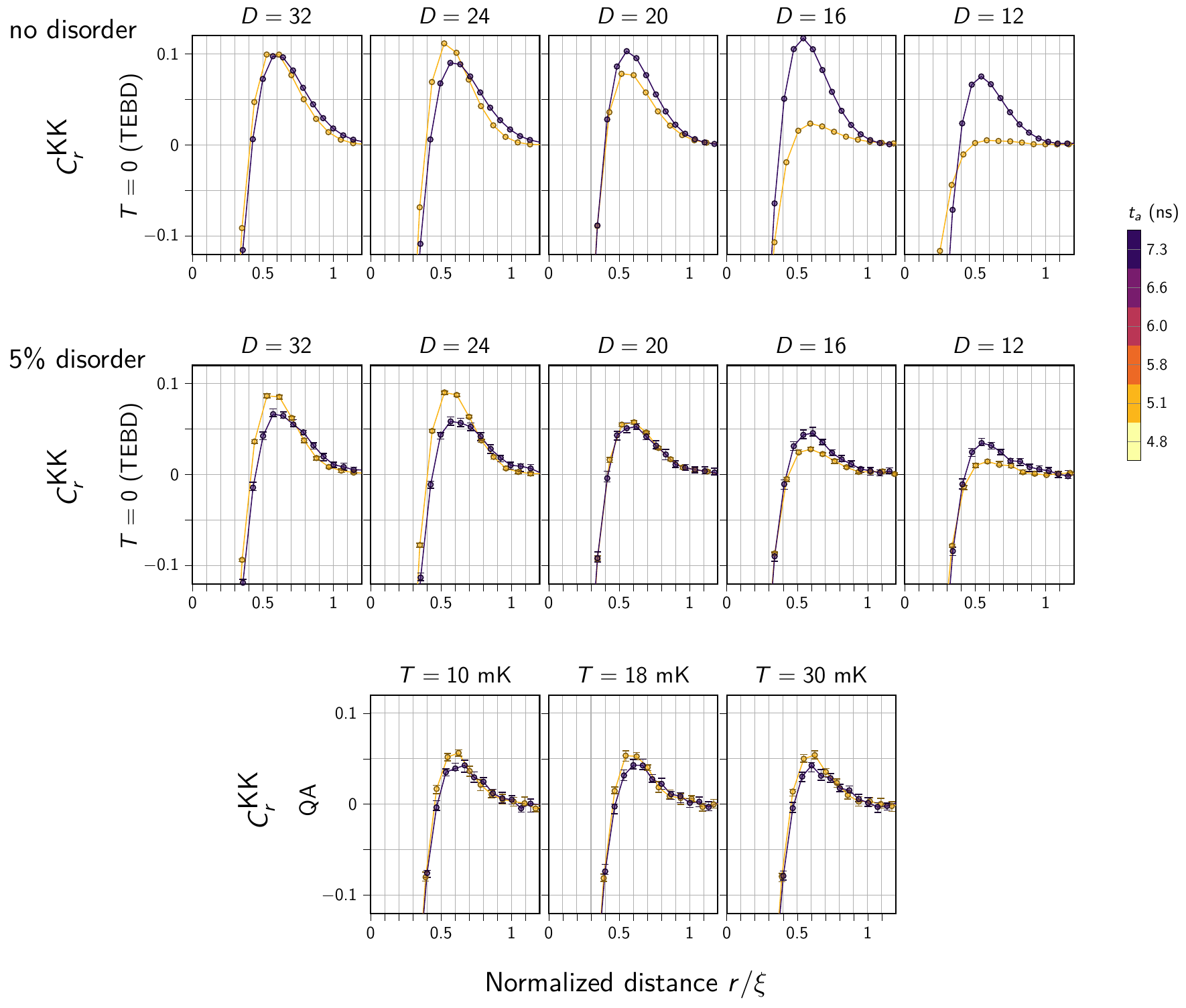}
  \caption{{\bf Kink-kink correlator data for TEBD and QA.}  Highlighted here are data only for $t_a = \SI{4.8}{ns}$ and $\SI{7.3}{ns}$, in TEBD with and without disorder, as well as QA at various temperatures, for $J=-1.4$.  The crossing of the QA data and the TEBD simulation results as a function of bond dimension $D$ at $D=20$ and 5\% disorder provides a possible measure of entanglement present in the experiment, since bond dimension is an upper bound on entanglement entropy~\cite{Schuch:2008ve}, as well as an estimate of the Schmidt number.}\label{fig:kkc_2rates}
\end{figure}

Fig.~\ref{fig:kkc_2rates} shows correlators for a range of bond dimensions at two short anneal times, as well as the QA data for these anneal times at different temperatures. The QA data at $t_a=\SI{4.8}{ns}$ are slightly above $t_a=\SI{7.3}{ns}$. The TEBD results exhibit a reversal in this trend at $D=20$ and 5\% disorder.  For $D<20$ the $t_a=\SI{4.8}{ns}$ TEBD curve is below the $t_a=\SI{7.3}{ns}$ curve, opposite from the QA data. Hence we conclude that $D=20$ is a lower bound on the bond dimension appropriate for describing the QA data. While bond dimension only provides an upper bound on the entanglement entropy, it is not unreasonable to conclude that this result also sets a lower bound on entanglement in the experiment, since the bond dimension is also an estimate of the Schmidt number of the state~\cite{Sanz:2016tb}.

\end{document}